\documentclass[prb,twocolumn,superscriptaddress,showpacs,preprintnumbers,floatfix,eqsecnum]{revtex4-2}
\usepackage{amsmath,amssymb}

\usepackage{graphicx}
\pdfoutput=1

\date{\today}

\begin{document}
\title{
Role of bias and tunneling asymmetries in  
nonlinear Fermi-liquid transport 
\\through an SU($N$) quantum dot
}

\author{Kazuhiko Tsutsumi}
\affiliation{Department of Physics, Osaka City University, 
Sumiyoshi-ku, Osaka 558-8585, Japan
}
\affiliation{
NITEP, Osaka Metropolitan University 
, Sumiyoshi-ku, Osaka 558-8585, Japan}

\author{Yoshimichi Teratani}
\affiliation{Department of Physics, Osaka City University, 
Sumiyoshi-ku, Osaka 558-8585, Japan
}
\affiliation{
NITEP, Osaka Metropolitan University 
, Sumiyoshi-ku, Osaka 558-8585, Japan}

\author{Kaiji Motoyama}
\affiliation{Department of Physics, Osaka City University, 
Sumiyoshi-ku, Osaka 558-8585, Japan
}

\author{Rui Sakano}
\affiliation{Department of Physics, Keio University, 3-14-1 Hiyoshi, Kohoku-ku, Yokohama, Kanagawa 223-8522, Japan}

\author{Akira Oguri}
\affiliation{Department of Physics, Osaka City University, 
Sumiyoshi-ku, Osaka 558-8585, Japan
}
\affiliation{
NITEP, Osaka Metropolitan University 
, Sumiyoshi-ku, Osaka 558-8585, Japan}
\begin{abstract}
We study how bias and tunneling asymmetries affect 
nonlinear current through a quantum dot with $N$ discrete levels    
in the Fermi liquid regime,    
 using an exact low-energy expansion of the current derived up to terms of order $V^3$ with respect to the bias voltage. 
The expansion coefficients are described in terms of 
the phase shift, the linear susceptibilities,
and the three-body correlation functions,
defined with respect to the equilibrium ground state  
of the Anderson impurity model.   
In particular, the three-body correlations 
play an essential role in the order $V^3$ term, 
and their coupling to the nonlinear current 
depends crucially on the bias and tunnel asymmetries.
The number of independent components 
of the three-body correlation functions  
increases with $N$ the internal degrees of the quantum dots,
and it gives a variety in the low-energy transport. 
We calculate the correlation functions 
over a wide range of electron fillings 
of the Anderson impurity model with the SU($N$) internal symmetry,   
using the numerical renormalization group. 
We find that the order $V^3$ nonlinear current through 
the SU($N$) Kondo state, which  occurs at electron fillings of $1$ and $N-1$ for strong Coulomb interactions,  
significantly varies with the three-body contributions 
as tunnel asymmetries increase.  
Furthermore,  in the valence fluctuation regime 
toward the empty or fully occupied impurity state,  
a sharp peak emerges in the coefficient of $V^3$ current 
 in the case at which bias and tunneling asymmetries 
cooperatively enhance the charge transfer from one of the electrodes.   
\end{abstract}
\maketitle

\section{Introduction}

The ground state and low-lying excited states of  
quantum-impurity systems \cite{Kondo2012,hewson_1993} 
such as dilute magnetic alloys and quantum dots  
can be described as a local Fermi liquid \cite{Nozieres1974,Yamada1975II,Yamada1975IV,Shiba1975,Yoshimori1976}, 
in which localized electrons with discrete energies are  
strongly coupled to conductions electrons in 
host metals or electrodes.
These low-energy states continuously evolve 
as the occupation number of the discretized states varies, 
and various interesting phenomena such as the Kondo effects 
and the valence fluctuations occur, 
depending on 
electron fillings and the configurations \cite{Wilson1975,KWW1,KWW2}.

After early observations 
 of the Kondo effect in quantum dots \cite{Goldhaber-Gordon1998nature,Goldhaber-Goldon1998PRL,Cronenwett-Oosterkamp-Kouwenhoven1998,vanderWiel2000, Schmid1998,Simmel1999,Sasaki2004}, 
universal Fermi-liquid behaviors were 
explored through highly sensitive measurements \cite{GrobisGoldhaber-Gordon,ScottNatelson,Heiblum,Delattre2009,KobayashiKondoShot,KondoCloud2020}
 and precise calculations \cite{Izumida2001,AO2001,Anders2008,WeichselbaumVonDelft}. 
Furthermore, various kinds of 
internal degrees of freedom 
bring an interesting variety to the Kondo effects in quantum dots, 
such as the one with the SU($4$) symmetry that 
can be realized in multiorbital dots and in carbon-nanotube (CNT) dots \cite{RMP-Kouwenhoven,Borda2003,Izumida1998,Sasaki2000,Pablo2005,Choi2005,Eto2005,Sakano2006,Sakano2007,Makarovski2007,Anders-SU4,SU4_Kondo_ferro_Weymann,MoraEtal2009,Mora2009,Cleuziou2013,MantelliMocaZarandGrifoni,Teratani2020PRL}.  
External magnetic fields also induce interesting crossover phenomena,  
such as the one between the SU(4) and SU(2) Kondo states 
observed in a CNT quantum dot \cite{Teratani2016,Ferrier2016,Ferrier2017,Teratani2020PRB}.

Recent development in the Fermi-liquid theory also  
reveals that the three-body correlations 
of electrons passing through the quantum dot 
play an essential role in 
the next-leading order terms of the transport coefficients 
\cite{MMvDZ2015,FMvDM2018,AO2017_I,AO2017_II,AO2017_III} 
when the system does not have both electron-hole and time-reversal symmetries.
This is because, 
in addition to the well-studied damping of order $\omega^2$, $T^2$, and $(eV)^2$, 
quasiparticles of the local Fermi liquid 
capture the energy shift of the same quadratic order 
which is induced by the three-body correlations,   
at low but finite frequencies $\omega$, temperatures $T$, and bias voltages $eV$.
The three-body contributions have also been confirmed experimentally  
in magnetoconductance and nonlinear thermocurrent spectroscopy 
measurements very recently \cite{Hata2021,HsuCosti2022}.

Here we focus on the effects of bias and tunneling asymmetries     
on the nonlinear current $I$ at low energies. 
Specifically,  we examine the bias asymmetry 
that can be described by the parameter 
$(\mu_L+\mu_R)/2 -E_F$,
with $\mu_L$ and $\mu_R$ 
the chemical potentials of the source ($L$) and drain ($R$) electrodes, 
applied such that  $eV \equiv \mu_L-\mu_R$,     
and $E_F$ the Fermi level at thermal equilibrium $eV=0$.     
The other one, tunneling asymmetry, occurs through the difference between 
the tunnel couplings $\Gamma_L$ and $\Gamma_R$  
for the source and drain electrodes, respectively.    
In the previous paper, we demonstrated how   
these asymmetries affect the transport through a quantum dot 
described by a spin-1/2 Anderson impurity model 
with no orbital degrees of freedom \cite{Tsutsumi2020}, 
and showed that the three-body correlations give 
significant contributions to the order $(eV)^3$ nonlinear current 
in the valence fluctuation regime,
using the numerical renormalization group (NRG) approach.

The purpose of this paper is to clarify how  
the conjunction of these asymmetries and 
the internal degrees of freedom, 
which increases the independent components of 
the three-body correlation functions,  
affects low-energy transport in the Fermi-liquid regime. 
The low-bias expansion of the conductance up to terms order $(eV)^2$ 
can be described by the Fermi-liquid theory with the three-body correlations, 
which at zero temperature $T=0$ takes the following form 
for a quantum dot 
 with $N$ discrete levels which include the spin components,  
\begin{align}
\frac{d I}{d V}\,=\,g_0\sum_{\sigma=1}^{N}\left[\,
\sin^2\delta_\sigma+c_{V,\sigma}^{(2)}\,eV-c_{V,\sigma}^{(3)}\,(eV)^2 
\,+\,\cdots \right].
\label{dif_cond_first}
\end{align}
Here,  $g_0=\frac{e^2}{h}\,4\Gamma_L\Gamma_R/(\Gamma_L+\Gamma_R)^2$, 
which depends on tunneling asymmetries.   
We present the exact formulas for the coefficients $c_{V,\sigma}^{(2)}$ 
and $c_{V,\sigma}^{(3)}$ of the multilevel Anderson impurity model
which are applicable to arbitrary impurity electron fillings and   
arbitrary level structures $\epsilon_{d\sigma}^{}$.  
These coefficients are determined by the phase shift $\delta_\sigma^{}$ 
and the other renormalized parameters,   
including the three-body correlation functions, 
and depend crucially on the bias and tunneling asymmetries. 
We show that, for $N \geq 3$, 
the three-body correlations between electrons in  
three different levels also couple to order $(eV)^3$ nonlinear current 
as well as the other components 
when there are some extents of bias and/or tunneling asymmetries.
Our formula includes  
the previous results of the other group as some special limiting cases:   
 $c_{V,\sigma}^{(2)}$ derived by Aligia for the  Anderson model \cite{Aligia2011,Aligia2014}, 
and $c_{V,\sigma}^{(3)}$ derived by Mora {\it et al.\ } 
for the SU($N$) Kondo model \cite{MoraEtal2009}.

We also calculate  
 $c_{V,\,\sigma}^{(2)}$ and $c_{V,\,\sigma}^{(3)}$ 
for the SU($N$) symmetric quantum dots with $N=4$ and $6$, 
using the NRG in a wide range of electron fillings, 
varying the parameter corresponding to the gate voltage. 
There emerge $(N-1)$ different Kondo states in the SU($N$) symmetric case, 
 which can be classified according to the occupation number
 $\langle n_d\rangle=1$, $2,$ $\ldots$, $N-1$. 
We find that  large Coulomb interaction suppresses charge fluctuations 
throughout the region of $1\lesssim\langle n_d\rangle\lesssim N-1$, 
and it makes the coefficients $c_{V,\sigma}^{(2)}$ and $c_{V, \sigma}^{(3)}$ 
less sensitive to the bias asymmetry,
whereas the tunneling asymmetry affects these coefficients. 
In particular,  in the SU($N$) Kondo states 
at electron fillings of $\langle n_d\rangle \simeq 1$ and $N-1$ for $N \geq 3$, 
the three-body contributions become sensitive to tunnel asymmetries, 
and it significantly varies the behavior of the order $(eV)^3$ nonlinear current.

In contrast, in the valence fluctuation regime  
toward the  empty or fully occupied impurity state,  i.e., 
at $0 \lesssim \langle n_d\rangle \lesssim 1$ or  
$N-1 \lesssim \langle n_d\rangle \lesssim N$, 
both the bias 
and tunneling asymmetries affect the nonlinear transport.
We find that, in these regions, a sharp peak 
emerges in the coefficient $c_{V, \sigma}^{(3)}$,  
 in the case at which the bias and tunneling asymmetries 
cooperatively enhance the charge transfer from one of the electrodes.

This paper is organized as follows. 
In Sec.\ \ref{Formulation}, 
we describe an outline of the microscopic Fermi-liquid theory 
for the multilevel Anderson model,   
and derive the low-energy asymptotic form of the differential conductance. 
Section \ref{NRGResultQuasiPara} shows the NRG results of quasiparticle parameters 
for SU($4$) and SU($6$)  quantum dots. 
In Secs.\  \ref{Cv2Results} and \ref{Cv3Results}, 
we show the NRG results for the coefficients 
$c_{V,\sigma}^{(2)}$ and $c_{V,\sigma}^{(3)}$, respectively.  
Summary is given in Sec.\ \ref{Conclusion}.

\section{Formulation}
\label{Formulation}
We consider a quantum dot with $N$-level coupled to two noninteracting leads by using the Anderson Hamiltonian:
\begin{align}
H\,=&\,H_d\,+\,H_c\, +\,H_T\, 
\\ \nonumber
\\
H_d\, =& \ \sum_{\sigma=1}^N\epsilon_{d\sigma}^{} n_{d\sigma}^{}
+ \frac{U}{2}\sum_{\sigma\neq\sigma'}\, n_{d\sigma}^{} n_{d\sigma'}^{}\,,
\qquad 
\\ 
H_c\, =&\sum_{\nu=L, \,R}\,\sum_{\sigma=1}^N\int_{-D}^{D}d\epsilon\,\,
\epsilon \,c_{\epsilon\nu\sigma}^{\dagger}c_{\epsilon\nu\sigma}^{\,}\,,
\\
H_T\,=  & \  \sum_{\nu=L,\,R}\,\sum_{\sigma=1}^N v_\nu^{}\,
(\psi_{\nu,\,\sigma}^{\dagger}d_\sigma^{}+d_\sigma^{\dagger}\,\psi_{\nu,\,\sigma}^{\,})\,,
\\
\psi_{\nu,\,\sigma}\equiv&\int_{-D}^{D}d\epsilon\,\sqrt{\rho_c}\,c_{\epsilon\nu\sigma}\,. 
\end{align}
Here, $d_\sigma^\dagger$ for $\sigma=1,2,\ldots, N$
creates an impurity electron with energy $\epsilon_{d\sigma}^{}$, 
$n_{d\sigma}^{} \equiv  d_\sigma^\dagger d_\sigma^{}$, 
and 
$U$ is the Coulomb interaction between electrons in the quantum dot. 
$c_{\epsilon\nu\sigma}^\dagger$  creates 
an electron with energy $\epsilon$ in the lead on the left or right ($\nu=L, R$), 
and it is normalized as 
 $\bigl\{c_{\epsilon\nu\sigma}\,,\,c_{\epsilon'\nu'\sigma'}^\dagger\bigr\}
=\delta(\epsilon-\epsilon')\delta_{\nu\nu'}\delta_{\sigma\sigma'}$. 
The linear combination of the conduction electron $\psi_{\nu,\sigma}$ 
couples to the dot via the tunneling matrix element $v_\nu^{}$.  
It determines the resonance width of the impurity level $\Delta=\Gamma_L+\Gamma_R$, 
with $\Gamma_\nu \equiv\pi\rho_cv_\nu^2$ 
the tunnel energy scale due to the lead $\nu$,
 and  $\rho_c=1/(2D)$ the density of state of the conduction band 
with a half-width $D$. 
We set  $k_B=1$ throughout this paper, and 
consider the parameter region where 
the half bandwidth $D$  
 is much greater than the other energy scales, 
$D \gg \max( U, \Delta, |\epsilon_{d\sigma}^{}|, T, |eV|)$. 

\subsection{Fermi-liquid parameters}

We describe here the definition of the correlation functions 
that play an essential role in the microscopic Fermi-liquid theory.
 
 The occupation number  and the linear susceptibilities 
of the impurity level can be derived from the free energy: 
\begin{align}
\bigl\langle n_{d\sigma}\bigr\rangle\,
=& \   \frac{\partial \Omega}{\partial \epsilon_{d\sigma}}\,,
\qquad \quad 
\Omega\,\equiv \,  
-T\log \,\bigl[\,\mathrm{Tr}\,e^{-\beta H}\,\bigr]\,, 
\label{ndFriedel}
\\
\chi_{\sigma\sigma'}\, \equiv 
&\ -\frac{\partial^2 \Omega}{\partial \epsilon_{d\sigma}\partial\epsilon_{d\sigma'}}
\,= \, 
\int_0^\beta \! d\tau\,
\bigl\langle \delta n_{d\sigma}(\tau)\,\delta n_{d\sigma'}\bigr\rangle\,,
\label{eq:linear_susceptibility}
\end{align}
where 
$\delta n_{d\sigma}=n_{d\sigma}-\langle n_{d\sigma}\rangle$,  
and  $\langle \cdots \rangle$ represents an equilibrium average, 
 with $\beta=1/T$ the inverse temperature.

In addition to linear susceptibilities, the nonlinear 
ones $\chi_{\sigma_1\sigma_2\sigma_3}^{[3]}$ 
also play an important role away from half filling:
\begin{align}
&\chi_{\sigma_1\sigma_2\sigma_3}^{[3]} \,\equiv\, 
-\frac{\partial^3\Omega}{\partial\epsilon_{d\sigma_1}
\partial\epsilon_{d\sigma_2}\partial\epsilon_{d\sigma_3}}
\,= \, 
\frac{\partial \chi_{\sigma_1\sigma_2}}{\partial \epsilon_{d\sigma_3}}
\nonumber
\\
=&-\int_0^\beta \! d\tau_1
\! \int_0^\beta \!d\tau_2\,
\bigl \langle T_\tau\delta n_{d\sigma_1}(\tau_1)\,\delta n_{d\sigma_2}(\tau_2)\,
\delta n_{d\sigma_3}\bigr\rangle\,.
\label{nonlinear}
\end{align}
Here, $T_\tau$ is the imaginary-time ordering operator. 
This  correlation function has the permutation symmetry: 
$\chi_{\sigma_1\sigma_2\sigma_3}^{[3]}
=\chi_{\sigma_2\sigma_1\sigma_3}^{[3]}
=\chi_{\sigma_3\sigma_2\sigma_1}^{[3]}
=\chi_{\sigma_1\sigma_3\sigma_2}^{[3]}=\cdots.$ 
We will use hereafter 
the ground-state values for $\langle n_{d\sigma}\rangle$,  
$\chi_{\sigma\sigma'}$ and $\chi_{\sigma_1\sigma_2\sigma_3}^{[3]}$,  
and thus the occupation number can be deduced from 
the phase shift $\delta_\sigma$ through the Friedel sum rule: 
$\langle n_{d\sigma}\rangle 
\xrightarrow{\,T\to 0\,}\delta_\sigma/\pi$ \cite{Shiba1975}.

The retarded Green's function also plays a central role 
in the microscopic description of the Fermi-liquid transport: 
\begin{align}
G_\sigma^r(\omega)=& \
 -i\int_{0}^{\infty}dt\,e^{i(\omega+i0^+) t}\,
\Bigl\langle\,\Bigl\{d_\sigma^{}(t),\,d_\sigma^\dagger\Bigr\}\Bigr\rangle_{eV},  
\label{eG}
 \\
A_{\sigma}^{}(\omega) \, \equiv & \  
-\frac{1}{\pi} \,\mathrm{Im}\,G_{\sigma}^{r}(\omega)\;.
\label{eq:spectral_function}
\end{align}
represents a nonequilibrium steady-state average 
taken with the statistical density matrix, 
which is constructed at finite bias voltages $eV$ and temperatures $T$, 
using the Keldysh formalism \cite{Hershfield1992,M_W1992}.

The phase shift $\delta_{\sigma}^{}$  is related to 
the value of the Green's function at the equilibrium ground state $T=eV=0$  
as  $G_\sigma^r(0) 
= -\left| G_\sigma^r(0)\right| e ^{i \delta_{\sigma}^{}}$.  
It also determines the behavior of the equilibrium spectral function  
$\rho_{d\sigma}^{}(\omega) \equiv 
\left. A_{\sigma}^{}(\omega)\right|_{T=eV=0}^{}$ in the low-frequency limit.  
At the Fermi level  $\omega=0$,   it takes the form 
\begin{align}
\rho_{d\sigma} \, \equiv& \  \rho_{d\sigma}^{}(0)  
\,= \, \frac{\sin^2\delta_\sigma}{\pi\Delta}  
\,. \label{Fermiparaorigin}
\end{align}
Furthermore, the first derivative of  $\rho_{d\sigma}(\omega)$  
is related to the diagonal susceptibility $\chi_{\sigma\sigma}$, as  
\begin{align}
\rho_{d\sigma}' 
 \equiv \,\left.\frac{\partial \rho_{d\sigma}^{}(\omega)}{\partial \omega}
\right|_{\omega=0}
\ =\ \frac{\chi_{\sigma\sigma}}{\Delta}\,\sin2\delta_{\sigma}\,.
 \label{DerivativeDOS}
 \end{align}
This is a result of a series of exact Fermi-liquid relations,
obtained by Yamada-Yosida [see Appendix A].

\subsection{Low-energy expansion of  nonlinear current}
Nonlinear current  through quantum dots can be calculated 
using a Landauer-type formula \cite{Hershfield1992,M_W1992}:  
\begin{align}
I\,=& \ \frac{e}{h}
\frac{4\Gamma_L\Gamma_R}{(\Gamma_L+\Gamma_R)^2}
\nonumber \\
& \times 
\sum_{\sigma=1}^{N}
\int_{-\infty}^{\infty}d\omega\,\bigl[f_L(\omega)-f_R(\omega)\bigr]\,
\pi\Delta A_\sigma(\omega) .
\label{eq:MWFormula}
\end{align}
Here, $f_\nu(\omega)= [e^{\beta(\omega-\mu_\nu)}+1]^{-1}$ is 
the Fermi distribution function for the conduction band on $\nu=L,\,R$. 
The chemical potentials of the left and right leads 
are driven from the Fermi energy at equilibrium $E_F=0$ 
by the bias voltage:  
$\mu_L=\alpha_L\,eV$ and $\mu_R=-\alpha_R\,eV$,  
with  $\alpha_L$ and $\alpha_R$ the parameters which satisfy $\alpha_L+\alpha_R=1$,
i.e., $\mu_L-\mu_R \equiv eV$. 
Asymmetries in tunnel couplings and 
that in bias voltages can be described, respectively, 
 by the following parameters, 
\begin{align}
\gamma_\mathrm{dif}^{}
\,\equiv & \  
\frac{\Gamma_L-\Gamma_R}{\Gamma_L+\Gamma_R}\,,  
\\
\alpha_\mathrm{dif}^{}\,\equiv & \ 
\frac{\mu_L+\mu_R}{\mu_L-\mu_R}
 \,=\,
\alpha_L-\alpha_R\,.
\end{align}

 In this work, we have derived the explicit expressions of  
the coefficients  $c_{V,\sigma}^{(2)}$ and  $c_{V,\sigma}^{(3)}$ 
for the first two nonlinear-response terms of $dI/dV$ in Eq.\ \eqref{dif_cond_first}, 
 using  the exact low-energy asymptotic form of the spectral 
 function $A_\sigma(\omega) $ of quantum dots embedded in asymmetric junctions 
 obtained up to terms of order $\omega^2$, $(eV)^2$, and $T^2$. 
The derivation is given in Appendix \ref{sec:asymptotic_form_of_A}.    
Specifically, we  will use Eqs.\ \eqref{cv2_general} and \eqref{cv3_general} 
 to clarify the roles of the tunnel and bias asymmetries,  
which enter through the parameters 
$\gamma_\mathrm{dif}^{}$ and $\alpha_\mathrm{dif}^{}$, 
in the nonlinear Fermi-liquid transport.

\subsubsection{$dI/dV$ of an SU($N$) dot  with tunnel and bias asymmetries }
In this paper, we consider the SU($N$) symmetric case, 
 at which the impurity level has the $N$-fold degeneracy: 
$\epsilon_{d\sigma}^{} \equiv \epsilon_d^{}$ for  $\sigma =1,2,\ldots,N$. 
For convenience, 
we use a shifted impurity energy level, 
defined by  $\xi_d^{}\equiv \epsilon_{d}^{} +(N-1)U/2$ in the following. 
Note that the system additionally has an electron-hole symmetry at $\xi_{d}^{}=0$.

The linear susceptibilities have two independent components 
in the SU($N$)  symmetric case, i.e., 
the diagonal component $\chi_{\sigma\sigma}$ 
and the off-diagonal one  $\chi_{\sigma\sigma'}$  for $\sigma \neq \sigma'.$ 
These two parameters  determine the essential properties of quasiparticles:  
\begin{align}
T^\ast \, \equiv \,  \frac{1}{4\chi_{\sigma\sigma}} \,, 
\qquad \quad 
R \,\equiv\, 1-\frac{\chi_{\sigma\sigma'}}{\chi_{\sigma\sigma}}
\,. 
\label{Fermiparaorigin}
\end{align}
Here,  $T^\ast $  is a characteristic energy scale  of the SU($N$) Fermi liquid, 
for instance, 
 the  $T$-linear specific heat of impurity electrons is scaled in the form  $\mathcal{C}_\mathrm{imp}^{} 
= \frac{N\pi^2}{12} T/T^{*}$.  
The Wilson ratio $R$ corresponds to  
a dimensionless residual interaction between quasiparticles
\cite{HewsonRPT2001}:
we will use the following rescaled Wilson ratio $\widetilde{K}$ 
which is bounded in the range  $0 \leq  \widetilde{K} \leq 1$,
\begin{align}
  \widetilde{K}\,\equiv\, (N-1)(R-1)\,.  
\end{align}

The differential conductance for SU($N$) symmetric quantum dots 
can be expressed in the following form, 
scaling the bias voltage  $eV$ by $T^\ast$, 
\begin{align}
\frac{d I}{d V}\,&=  \ 
N g_0^{} 
\left[\,\sin^2\delta\,+\,C_V^{(2)}\,\frac{eV}{T^\ast}
\,-\,C_V^{(3)}\left(\frac{eV}{T^\ast}\right)^2\,+\,\cdots \right], 
\label{DiffCond}
\\
g_0^{}\,&\equiv\  
\frac{e^2}{h}\,\frac{4\Gamma_L\Gamma_R}{(\Gamma_L+\Gamma_R)^2}
\ = \, \frac{e^2}{h} \bigl(1- \gamma_\mathrm{dif}^{2}\bigr) \,.
\label{eq:g0_gamma_dif}
\end{align}
The  dimensionless coefficients $C_V^{(2)}$ and $C_V^{(3)}$, 
can be deduced from the general formulas given in  
 Eqs.\ \eqref{cv2_general} and \eqref{cv3_general}, 
taking into account the SU($N$) symmetry: 
\begin{align}
C_V^{(2)} \,= &\ 
\frac{\pi}{4} \left[\, \alpha_\mathrm{dif}^{}
\left(1-\widetilde{K}\right)\,-\,\gamma_\mathrm{dif}^{}\,\widetilde{K}
\, \right] \,\sin2\delta \,. 
\label{Cv2}
\end{align}
For $N=2$, this reproduces the  
previous result obtained 
by Aligia for the spin-1/2 Anderson model \cite{Aligia2011,Aligia2014}.
$C_V^{(2)}$ consists of a linear combination 
of  $\alpha_\mathrm{dif}^{}$ and $\gamma_\mathrm{dif}^{}$, 
and  it identically vanishes   $C_V^{(2)}\to 0$ 
when both the tunnel couplings and the bias voltages are symmetrical  
 $\alpha_\mathrm{dif}^{}=\gamma_\mathrm{dif}^{}=0$. 
The magnitude is determined also by  $\widetilde{K}$
and  $\sin 2 \delta$ 
which can be related to the derivative of the spectral function $\rho_d'$,  
given in Eq.\ \eqref{DerivativeDOS}: $\sin2\delta/T^\ast=4\Delta\rho_d'$.  
Furthermore, $C_V^{(2)}$  is an odd function of $\xi_d^{}$. 
Note that  $C_V^{(2)}$  depends on the Coulomb interaction 
only through the real part of the self-energy.

In contrast,  
 $C_V^{(3)}$ given in the following depends 
on both the real part that determines the high-order energy shifts 
and imaginary part 
that destroy phase coherence \cite{Zarand2004,Kehrein2005},
specifically on the order $\omega^2$ and $(eV)^2$ terms. 
The exact formula of the coefficient $C_V^{(3)}$ for order $(eV)^3$ nonlinear current 
is composed of two parts: $W_V$ and $\Theta_V$ 
which represent the two-body contribution and three-body one, respectively, 
\begin{align}
C_V^{(3)} \,= & \ \frac{\pi^2}{64}\,
\bigl (W_V\,+\,\Theta_V \bigr) \,, 
\label{Cv3eq}
\\
W_V\equiv&\ -\cos2\delta 
\Biggl[1+3\alpha_\mathrm{dif}^2-6\Bigl(\alpha_\mathrm{dif}^2
+\alpha_\mathrm{dif}^{}\gamma_\mathrm{dif}^{}\Bigr)\,\widetilde{K} 
\nonumber
\\
& 
\!\!\!\!\!\!\!\!\!\!\!\!\!\!
 + \left\{\frac{5}{N-1}+3\alpha_\mathrm{dif}^2
+6\alpha_\mathrm{dif}^{}\gamma_\mathrm{dif}^{}
+\frac{3(N-2)}{N-1}\,\gamma_\mathrm{dif}^2\right\}\,\widetilde{K}^2\Biggr] ,
\label{Wv}
\\
\Theta_V\equiv&\Bigl[1+3\alpha_\mathrm{dif}^{2}\Bigr]\Theta_\mathrm{I}
+3\Bigl[1+3\alpha_\mathrm{dif}^{2}
+4\alpha_\mathrm{dif}^{}\gamma_\mathrm{dif}^{}\Bigr]\,
\widetilde{\Theta}_\mathrm{II} 
\nonumber
\\
& \ +  6\Bigl[\alpha_\mathrm{dif}^{2}
+2\alpha_\mathrm{dif}^{}\gamma_\mathrm{dif}^{}
+\gamma_\mathrm{dif}^{2}\Bigr]\,\widetilde{\Theta}_\mathrm{III}\,.
\label{THv}
\end{align}
 $C_V^{(3)}$ depends on 
 the asymmetry parameters through the quadratic terms, i.e.,    
 $\,\alpha_\mathrm{dif}^{2}$,  
$\,\alpha_\mathrm{dif}^{}\gamma_\mathrm{dif}^{}$,
 and $\,\gamma_\mathrm{dif}^{2}$.  
Effects of the  three-body corrections 
 enter through the dimensionless parameters:  
\begin{align}
\Theta_{\mathrm{I}}\, 
\equiv & \ 
\frac{\sin2\delta}{2\pi}\,
\frac{\chi_{\sigma\sigma\sigma}^{[3]}}{\chi_{\sigma\sigma}^2} 
\,,
\label{eq:ThetaI_def}
\\
\widetilde{\Theta}_{\mathrm{II}}\, 
\equiv& \ 
(N-1)\,\frac{\sin2\delta}{2\pi}\,
\frac{\chi_{\sigma\sigma'\sigma'}^{[3]}}{\chi_{\sigma\sigma}^2}\,,
\label{eq:ThetaII_def}
\\
\widetilde{\Theta}_{\mathrm{III}}\, \equiv& \ 
\frac{(N-1)(N-2)}{2}\,\frac{\sin2\delta}{2\pi}\,
\frac{\chi_{\sigma\sigma'\sigma''}^{[3]}}{\chi_{\sigma\sigma}^2}\,,
\label{eq:ThetaIII_def}
\end{align}
for $\sigma\neq\sigma'\neq\sigma''\neq\sigma$. 
In particular,  
$\widetilde{\Theta}_\mathrm{III}$ 
represents the correlation between three different levels which appears for $N \geq 3$. 
This component $\widetilde{\Theta}_\mathrm{III}$ 
 will give no contribution to $C_V^{(3)}$ through Eq.\ \eqref{THv} 
if there is no tunnel or bias asymmetry \cite{Teratani2020PRL,Oguri2022}: 
$W_V$ and $\Theta_V$ take the following form  
at  $\alpha_\mathrm{dif}^{}=\gamma_\mathrm{dif}^{}=0$,  
\begin{align}
W_V\,\xrightarrow{\,\alpha_\mathrm{dif}^{}=\gamma_\mathrm{dif}^{}=0\,}& 
\ -\cos2\delta \, \left( 1\,+\, \frac{5\widetilde{K}^2}{N-1} \right) , \label{WvALSymXLSym}
\\
\Theta_V\,\xrightarrow{\,\alpha_\mathrm{dif}^{}=\gamma_\mathrm{dif}^{}=0\,}& 
\ \,  \Theta_\mathrm{I}\,+\,3 \,\widetilde{\Theta}_\mathrm{II} \,. \label{THvALSymXLSym}
\end{align}
%

When both 
the bias and tunnel asymmetries 
are inverted 
such that   
$(\alpha_\mathrm{dif}^{}$, $\gamma_\mathrm{dif}^{}) 
\to 
 (-\alpha_\mathrm{dif}^{}$, $-\gamma_\mathrm{dif}^{})$,  
the coefficients  $C_V^{(2)}$ and $C_V^{(3)}$ 
exhibit odd and even properties,   
respectively: 
$C_V^{(2)} \to - C_V^{(2)}$ and $C_V^{(3)} \to C_V^{(3)}$ as shown in Appendix \ref{sec:asymptotic_form_of_A}.  
 These formulas of  $C_V^{(2)}$ and $C_V^{(3)}$ of the Anderson model 
for arbitrary level structures $\epsilon_{d\sigma}^{}$ 
are consistent, 
 in the  limit of strong interaction $U \to \infty$,   
with the corresponding formulas  for the SU($N$) Kondo model   
obtained by Mora {\it et al.\/} \cite{MoraEtal2009} 
at integer-filling points  $\bigl\langle n_{d}\bigr\rangle =1,2,\ldots,N-1$,  
with  $n_{d} \equiv \sum_{\sigma} n_{d\sigma}$.

\section{NRG results for Fermi-liquid parameters}
\label{NRGResultQuasiPara}

In this section, we summarize the basic properties of quasiparticles 
in the SU($4$) and SU($6$) symmetric cases.  
Specifically, we have calculated the Fermi-liquid parameters 
$\delta$, $\chi_{\sigma\sigma'}$,  and $\chi_{\sigma\sigma'\sigma''}^{[3]}$ 
with the NRG approach, 
dividing  $N$ channels into $N/2$ pairs, 
and using the SU($2$) spin and U($1$) charge symmetries for each pair, i.e., 
totally $\prod_{k=1}^{N/2}\left[\mathrm{SU}(2)\otimes\mathrm{U}(1)\right]_k$ 
symmetries. 
The discretization parameter $\Lambda$ and the number of low-lying energy 
states  $N_\mathrm{trunc}$  are chosen 
such that $(\Lambda, N_\mathrm{trunc})=(6, 10000)$ 
for $N=4$,   and $(20, 30000)$ for $N=6$ \cite{Teratani2020PRL}. 
The phase shift and renormalized parameters have been deduced 
from the energy flow of NRG \cite{Hewson2004,Nishikawa2010v1,Nishikawa2010v2,OguriSakanoFujii2011,Teratani2020PRB}.

\begin{figure}[t]
	\rule{1mm}{0mm}
	\begin{minipage}[r]{\linewidth}
	\centering
	\includegraphics[keepaspectratio,scale=0.21]{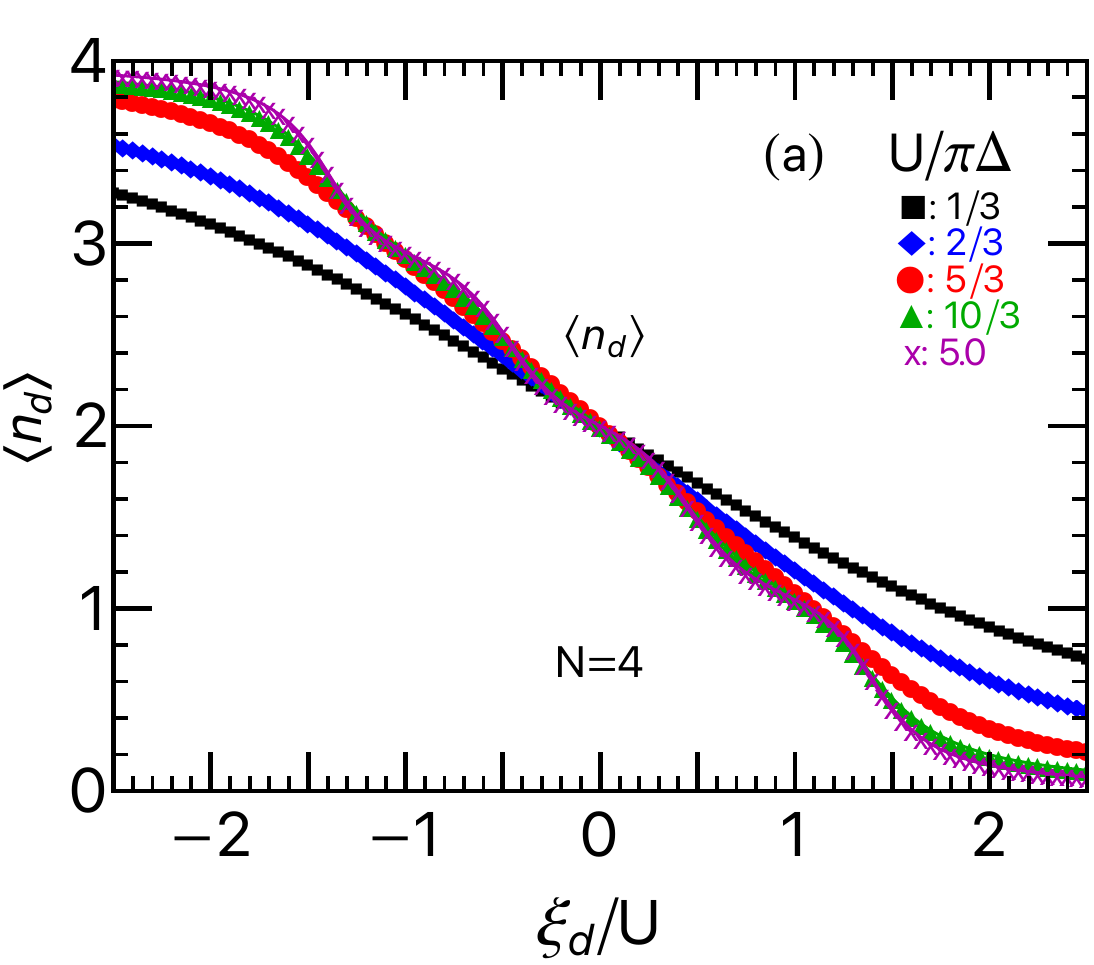}
	\centering
	\includegraphics[keepaspectratio,scale=0.22]{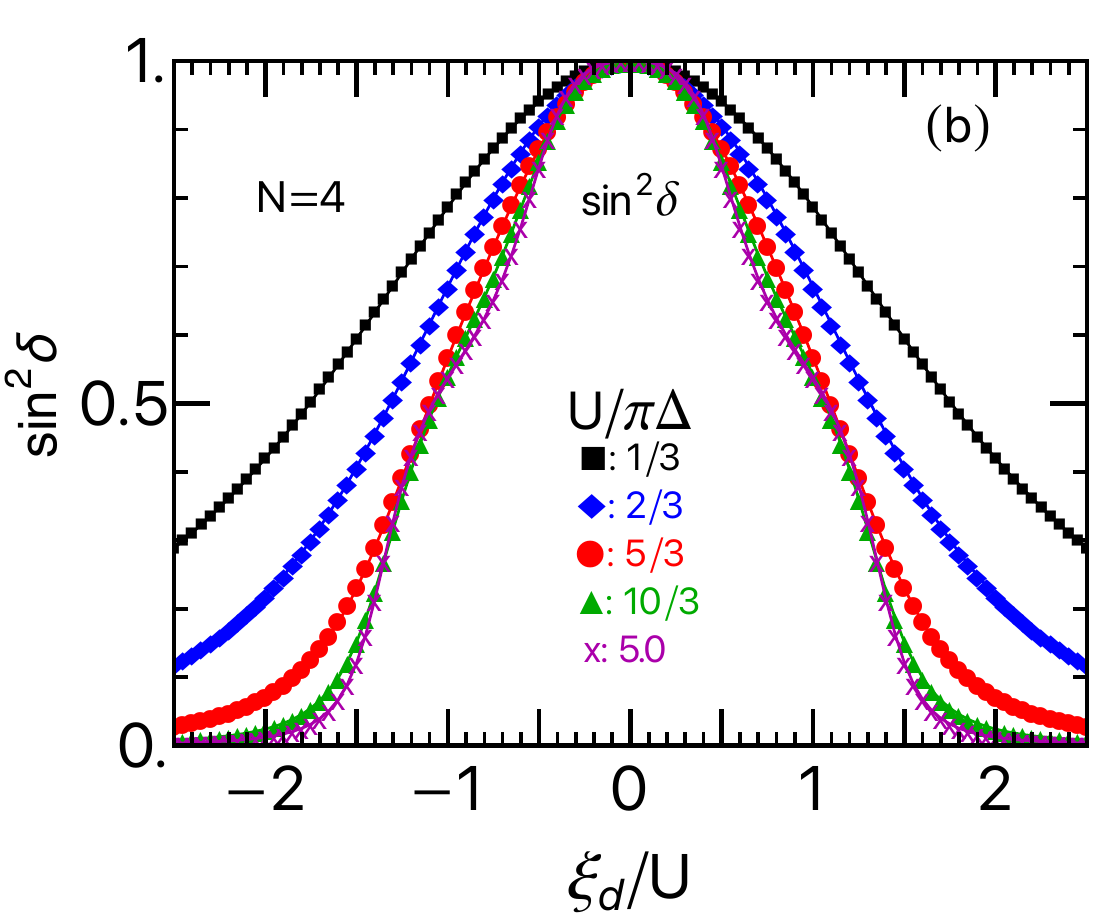}
	\end{minipage}
	\begin{minipage}[r]{\linewidth}
	\centering
	\includegraphics[keepaspectratio,scale=0.22]{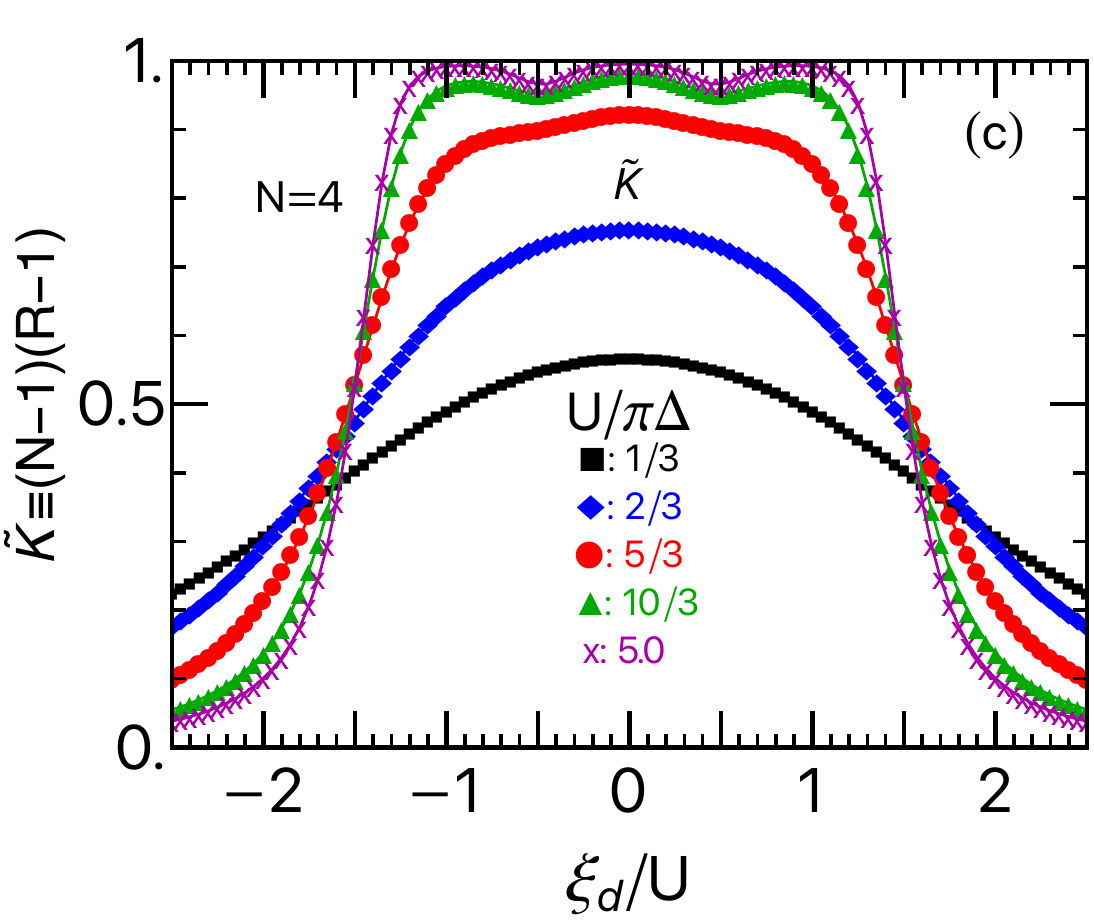}
	\centering
	\includegraphics[keepaspectratio,scale=0.22]{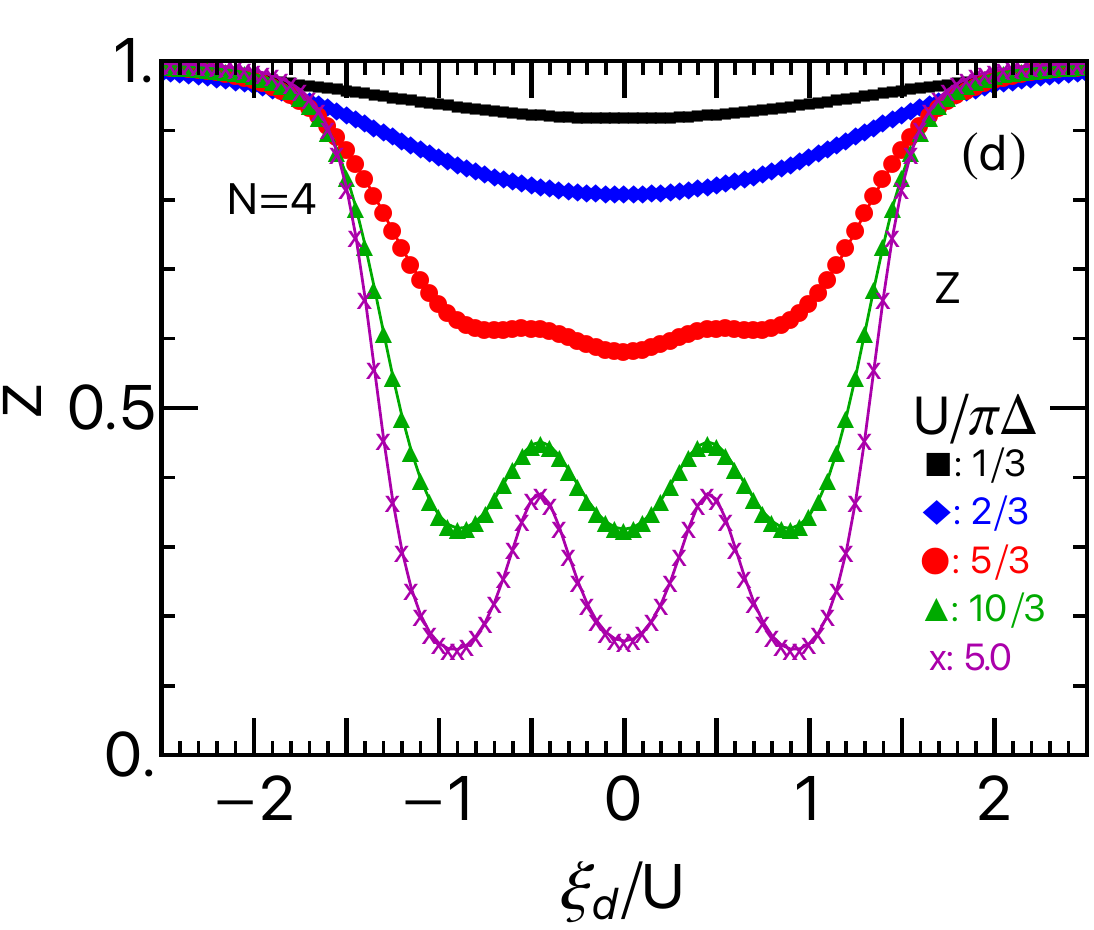}
	\end{minipage}
	\begin{minipage}[r]{\linewidth}
	\centering
	\includegraphics[keepaspectratio,scale=0.22]{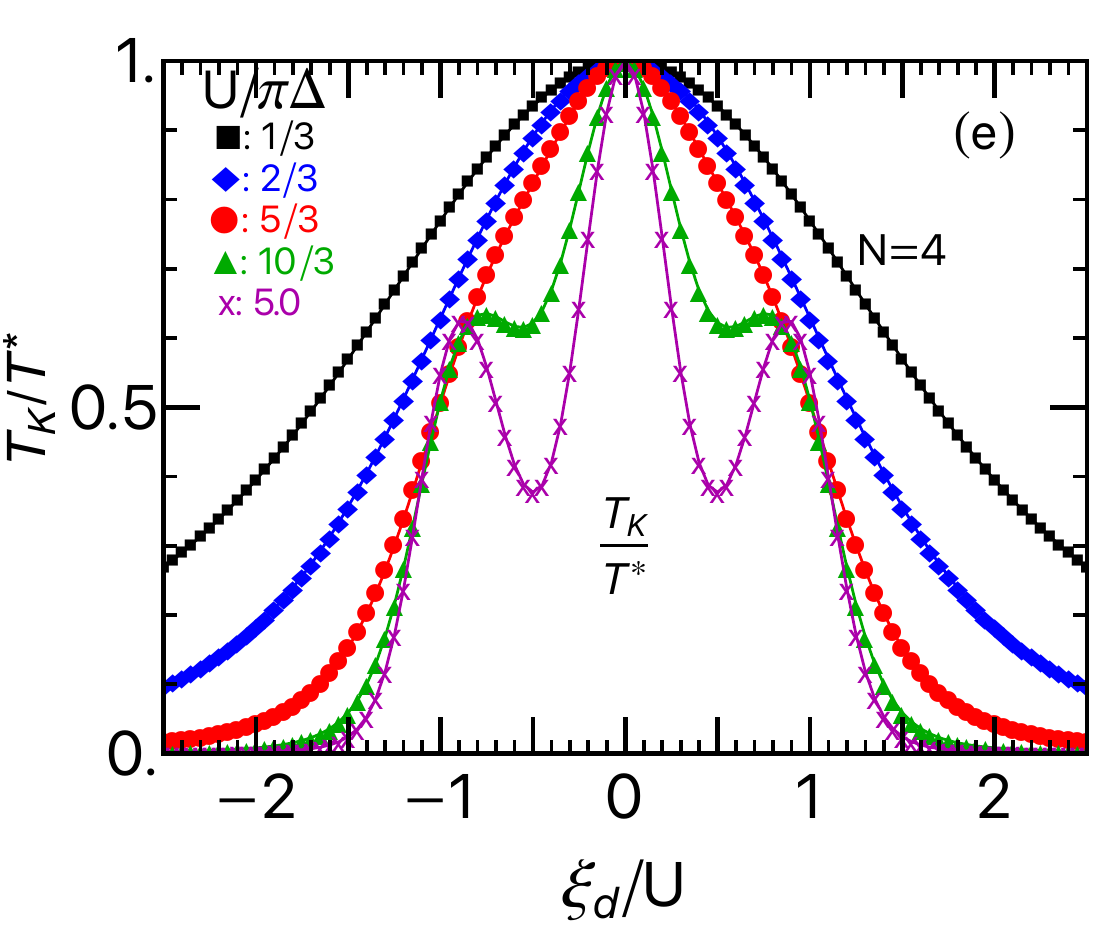}
	\centering
	\includegraphics[keepaspectratio,scale=0.22]{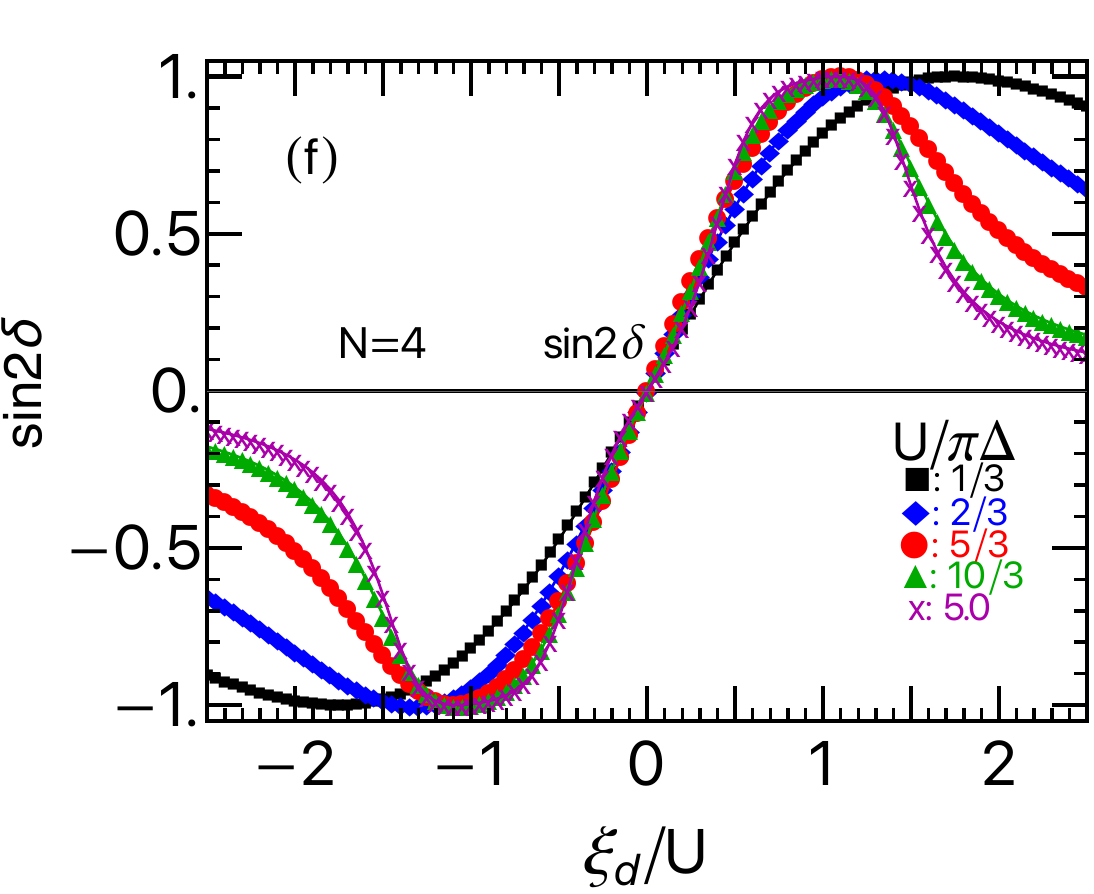}
	\end{minipage}
\caption{Fermi-liquid parameters of  an SU(4)  dot 
are plotted as a function of $\xi_d^{}/U$ 
for  $U/(\pi\Delta) =1/3$, $2/3$, $5/3$, $10/3$, $5.0$: 
(a) $\langle n_{d}\rangle$,\,  
(b) $\sin^2\delta$,\,
(c) rescaled Wilson ratio $\widetilde{K} \equiv  (N-1)(R-1)$,\,
(d) renormalized factor $z$,\,
(e) $T_K/T^\ast$, \, and  \,
(f) $\sin2\delta$. 
Here,  $T_K$ is defined as the value of 
$T^\ast  \equiv 1/(4\chi_{\sigma\sigma}^{})$ 
at half filling $\xi_d^{}=0$. 
}
  \label{FermiparaSU4}
\end{figure}

\subsection{ SU(4) Fermi-liquid properties}

\subsubsection{Two-body correlation functions of an SU(4) dot}

The Fermi-liquid parameters of an SU(4) quantum dot,   
 which can be deduced from the phase shift and the linear susceptibilities,  
are shown in Fig.\ \ref{FermiparaSU4}  
as a function of $\xi_d^{}$ for several different values of   
 $U/(\pi\Delta) =1/3$, $2/3$, $5/3$, $10/3$, $5$. 
For large  $U/(\pi\Delta)\gtrsim 2.0$,  
the occupation number in Fig.\ \ref{FermiparaSU4}(a) 
shows a Coulomb-staircase behavior with the plateaus of integer height 
 $\langle n_{d}\rangle \simeq 1.0$, $2.0$,  $3.0$ 
and  the steps at $\xi_d^{} \simeq -U$, $0$, $U$: 
 the structure becomes clearer for stronger interactions. 
Figure \ref{FermiparaSU4}(b) shows $\sin^2\delta$ 
corresponding 
to the linear term of the differential conductance in Eq.\ \eqref{dif_cond_first}. 
For  strong interactions $U/(\pi\Delta) \gtrsim 2.0$, 
the Kondo ridges which reflect
 the step structure of the occupation 
number of the values $\langle n_{d}\rangle\simeq 1,2, 3$ evolve at  $\xi_d\simeq U, 0, -U$, respectively, as $U$ increases.

Figure \ref{FermiparaSU4}(c) shows the rescaled Wilson ratio  $\widetilde{K} \equiv (N-1)(R-1)$. 
It has a wide plateau that reaches the strong-coupling limit value $\widetilde{K}\simeq 1$ in the region $-1.5U\lesssim\xi_d^{}\lesssim 1.5U$, for large interactions  $U/(\pi\Delta)\gtrsim3.0$. 
This is caused by the fact that 
charge fluctuations are  suppressed in this region 
and it makes the charge susceptibility vanish: 
$\chi_c \propto \chi_{\sigma \sigma}^{}
+(N-1)\chi_{\sigma \sigma'}^{} \to 0$.  
The shallow dips of $\widetilde{K}$ at  $\xi_d^{}\simeq  \pm0.5U$ 
is caused by the charge fluctuations at the steps of the Coulomb staircase structure 
of $\langle n_d^{} \rangle$.

Correspondingly, the renormalization factor $z$ in Fig.\ \ref{FermiparaSU4}(d)  
 exhibits a broad valley structure at $-1.5U\lesssim\xi_d^{}\lesssim 1.5U$, 
which becomes deeper as $U$ increases. 
 It has  a clear  local minimum  for $U/(\pi\Delta)\gtrsim2.0$ at $\xi_d^{}\simeq$  $0$ and  $\pm U$, 
where $\langle n_d\rangle$ takes integer values: 
it also  has  a local maximum 
at intermediate valence states in between 
the two adjacent minima. 
Figure \ref{FermiparaSU4}(e) shows 
the gate voltage $\xi_d^{}$ dependence of 
the inverse of the characteristic energy scale,  $1/T^\ast$,  
scaled by $T_K$ that is defined as the value of $T^\ast$ 
at the elelctron-hole symmetric point $\xi_d^{}=0$. 
It shows an oscillatory behavior for strong interactions
 $U/(\pi\Delta)\gtrsim3.0$, 
reflecting the dependence of $z$ on $\xi_d^{}$. 
In particular, $1/T^\ast$ reaches a local maximum 
at the integer-filling points where the SU($4$) Kondo effect occurs.

Some of the two-body correlation functions contribute to the low-energy transport through
 the derivative of the impurity density of state:    
$\rho_d' = (\chi_{\sigma \sigma}^{}/\Delta)\sin2\delta$ 
given in Eq.\ \eqref{DerivativeDOS}.
For instance, the coefficient $C_V^{(2)}$ is proportional to $\rho_d' $. 
Figure \ref{FermiparaSU4}(f) shows $\sin2\delta$. 
It is an odd  function of $\xi_d^{}$ and vanishes at $\xi_d^{}=0$ 
where the phase shift takes the value  $\delta=\pi/2$. 
For strong interactions $U/(\pi\Delta)\gtrsim3.0$,  
it has a wide maximum  (minimum) at $\xi_d^{}\simeq U$  ($-U$), 
where the $1/4$-filling ($3/4$-filling)  SU($4$) Kondo occurs.

\begin{figure}[t]
	\rule{1mm}{0mm}
	\begin{minipage}[r]{\linewidth}
	\centering
	\includegraphics[keepaspectratio,scale=0.22]{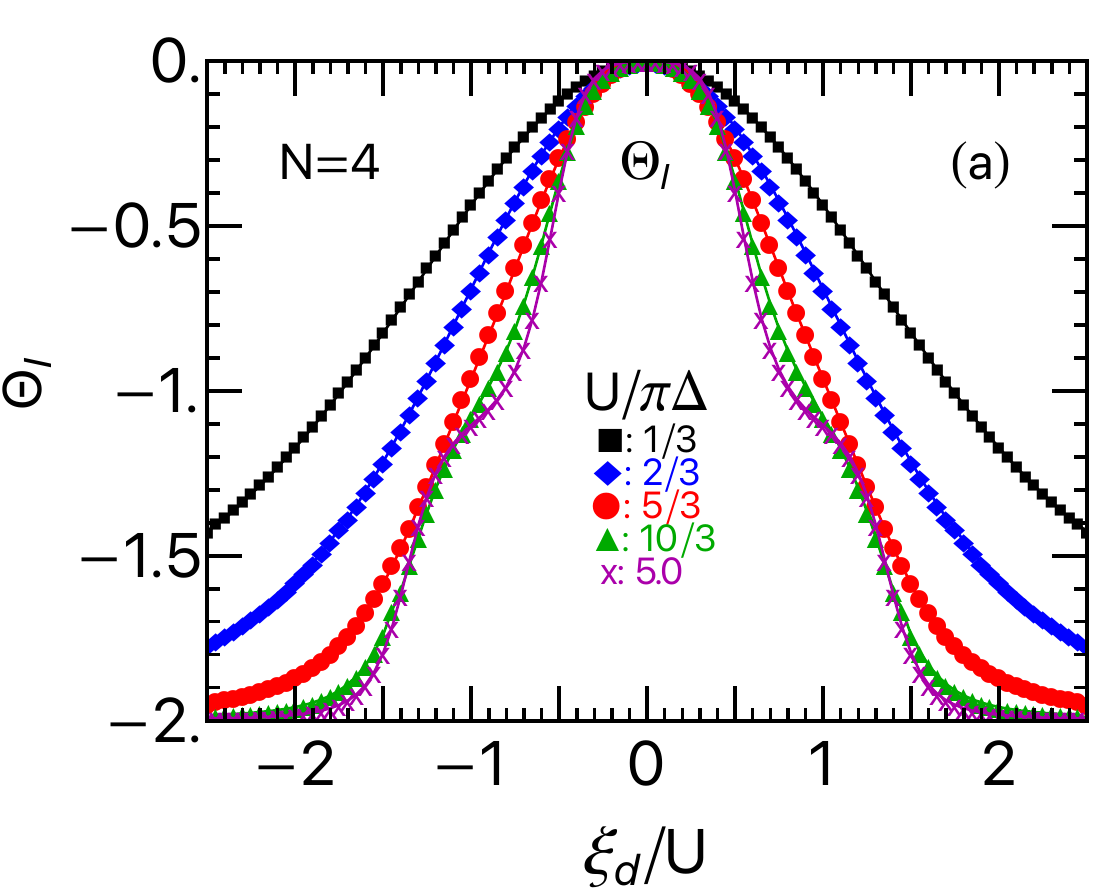}
	\centering
	\includegraphics[keepaspectratio,scale=0.21]{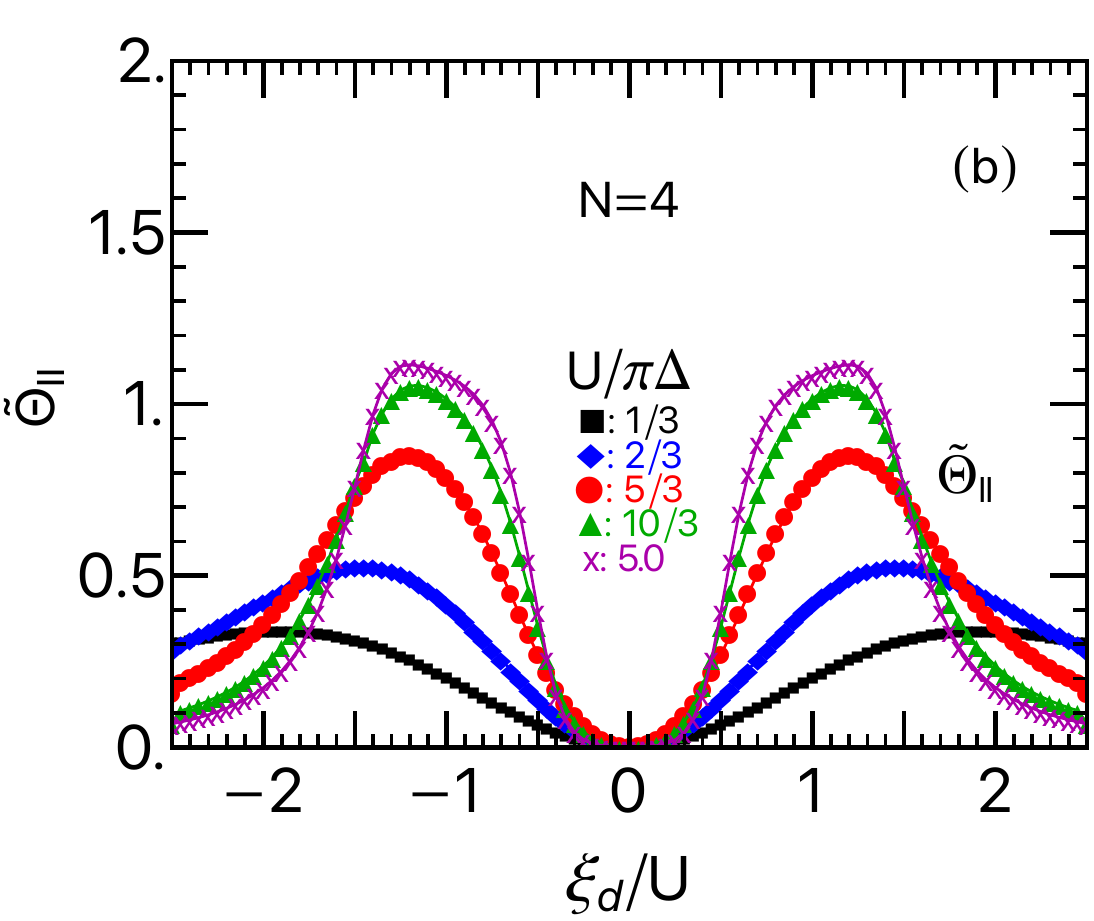}
	\end{minipage}
	\begin{minipage}[r]{\linewidth}
	\centering
	\includegraphics[keepaspectratio,scale=0.22]{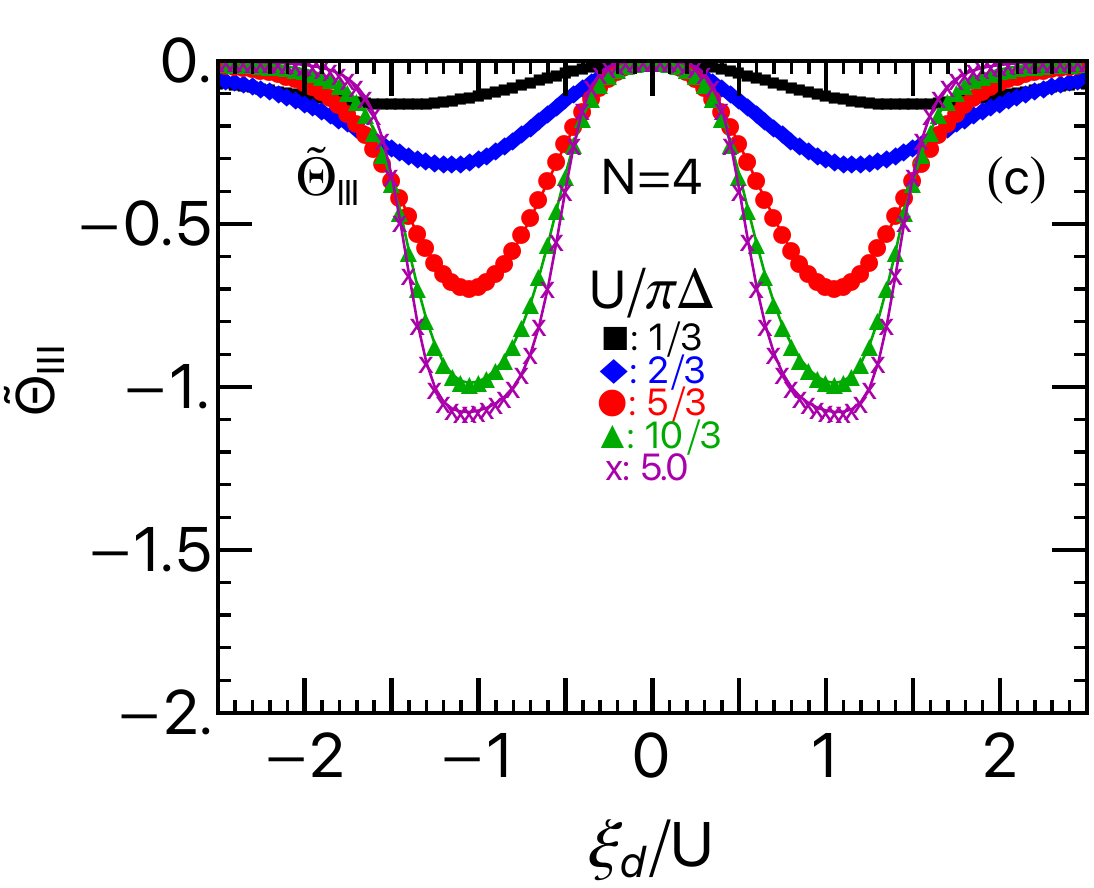}
	\centering
	\includegraphics[keepaspectratio,scale=0.22]{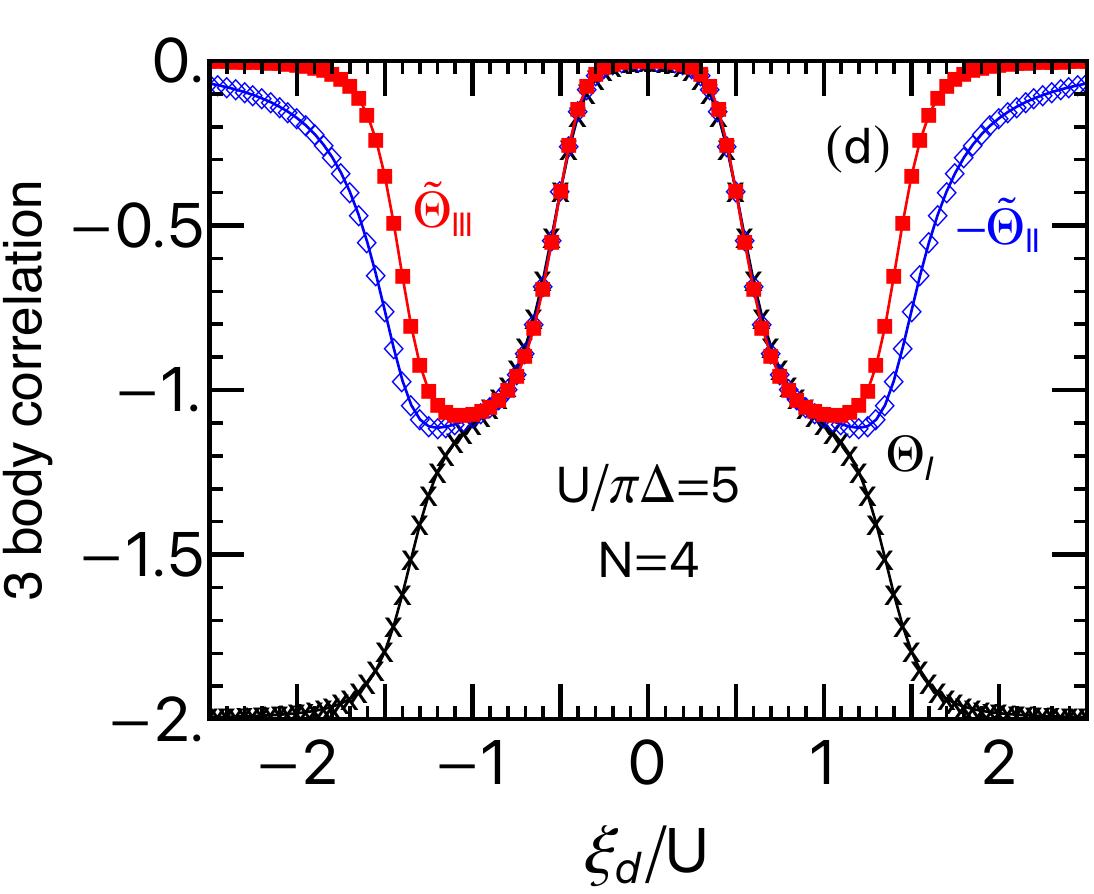}
	\end{minipage}
\caption{
Three-body correlation functions of an SU(4) dot plotted vs $\xi_d^{}/U$
 for $U/\pi\Delta=1/3$, $2/3$, $5/3$, $10/3$, $5.0$: 
(a) $\Theta_\mathrm{I}$, (b) $\widetilde{\Theta}_\mathrm{II}$,
 (c) $\widetilde{\Theta}_\mathrm{III}$.  
 (d)  comparison of 
$\Theta_\mathrm{I}$, $-\widetilde{\Theta}_\mathrm{II}$, 
and $\widetilde{\Theta}_\mathrm{III}$ 
at a large interaction $U/(\pi\Delta) = 5.0$. 
It indicates that $\Theta_\mathrm{I} \simeq 
-\widetilde{\Theta}_\mathrm{II} 
\simeq \widetilde{\Theta}_\mathrm{III}$ 
over a wide range of $-U\lesssim\xi_d^{}\lesssim U$.
 }
  \label{TH123SU4}
\end{figure}

\subsubsection{Three-body correlation functions of an SU(4) dot}

We next consider the three-body correlations between electrons 
passing through an SU(4) symmetric dot.
Figure \ref{TH123SU4} shows  
$\Theta_\mathrm{I}$, $\widetilde{\Theta}_\mathrm{II}$, 
and $\widetilde{\Theta}_\mathrm{III}$ as a function of $\xi_d/U$ 
for several different values of $U$. 
These dimensionless three-body correlation functions 
 vanish in the electron-hole symmetric case  $\xi_d^{}=0$.  
For strong interactions $U/(\pi\Delta)\gtrsim3.0$, 
 a plateau of the width $U$ emerges 
at  $\xi_d^{}\simeq 0$ and  $\pm U$, 
i.e., at integer filling points  $\langle n_d \rangle \simeq 1,$ $2$, $3$.  
The plateau structure evolves further as interaction $U$ increases. 
Specifically,
the correlation function between three different levels 
  $\widetilde{\Theta}_\mathrm{III}$ appears 
for SU($N$) quantum dots with $N \geq 3$,  
and contributes to the nonlinear conductance 
when  there are some asymmetries 
in the tunnel couplings or the bias voltages.

A comparison of the three independent components
$\Theta_\mathrm{I}$, $-\widetilde{\Theta}_\mathrm{II}$,  
and $\widetilde{\Theta}_\mathrm{III}$ 
is made in 
Fig.\ \ref{TH123SU4}(d) for a large interaction $U/(\pi\Delta)=5.0$. 
It shows that these three components approach each other very closely 
over a wide range of gate voltages  $-1.5U\lesssim\xi_d^{} \lesssim 1.5 U$,  
\begin{align}
\Theta_\mathrm{I} \,\simeq\, 
-\widetilde{\Theta}_\mathrm{II}
\,\simeq \,
\widetilde{\Theta}_\mathrm{III}\,.
\label{TH1simTH2simTH3}
\end{align}
This is caused by the fact that the derivatives 
of the two independent components of linear susceptibilities, 
$|\frac{\partial \chi_{\sigma\sigma}^{}}{\partial \epsilon_d^{}}|$ 
and $|\frac{\partial \chi_{\sigma\sigma'}^{}}{\partial \epsilon_d^{}}|$,  
become  much smaller than an  inverse of the 
 characteristic energy scale $(T^\ast)^{-2}$  in a wide range  
of electron fillings $1 \lesssim \langle n_d^{} \rangle \lesssim N-1$, 
in addition to the suppression of the charge fluctuations 
$\chi_c^{} \simeq  0$, mentioned above
[see Appendix \ref{ThreeBodySame}] \cite{Teratani2020PRL}.

\begin{figure}[t]
	\rule{1mm}{0mm}
	\begin{minipage}[r]{\linewidth}
	\centering
	\includegraphics[keepaspectratio,scale=0.21]{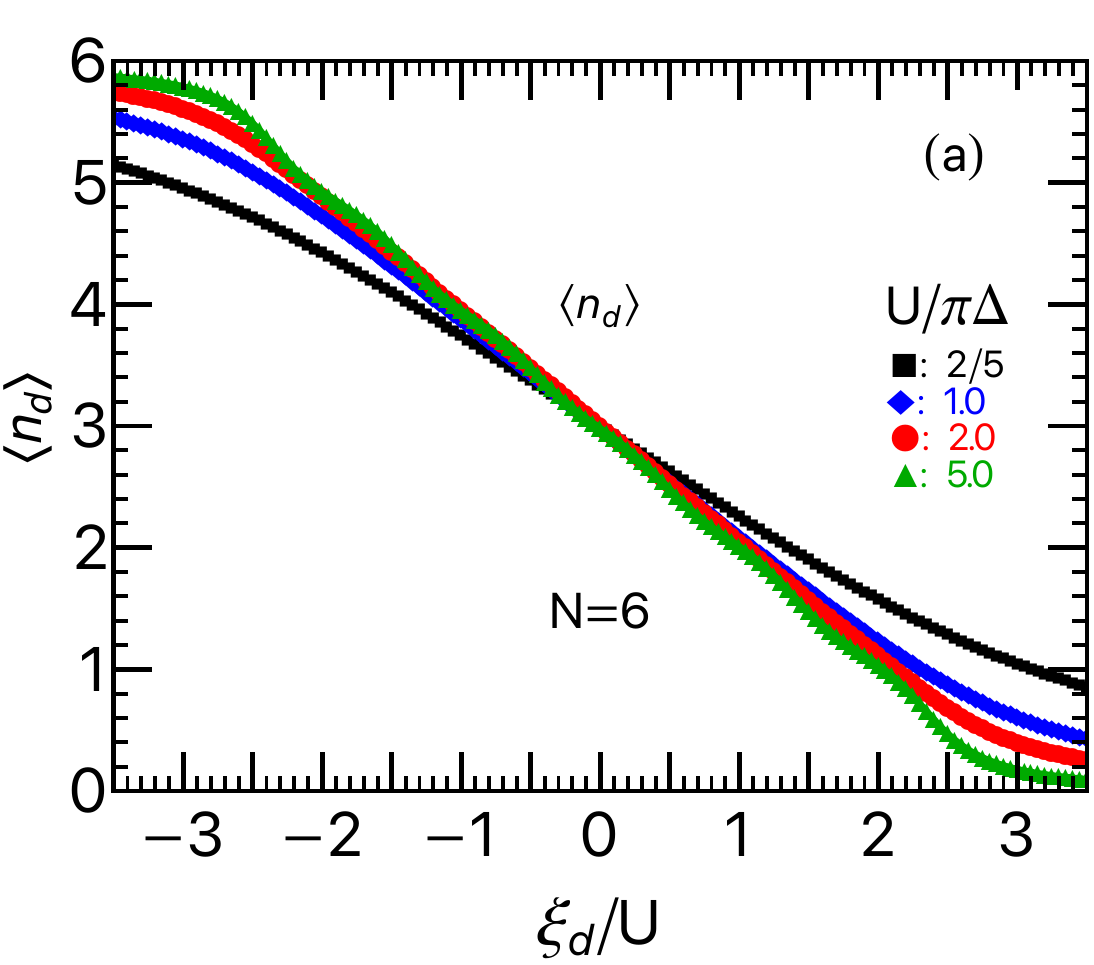}
	\centering
	\includegraphics[keepaspectratio,scale=0.22]{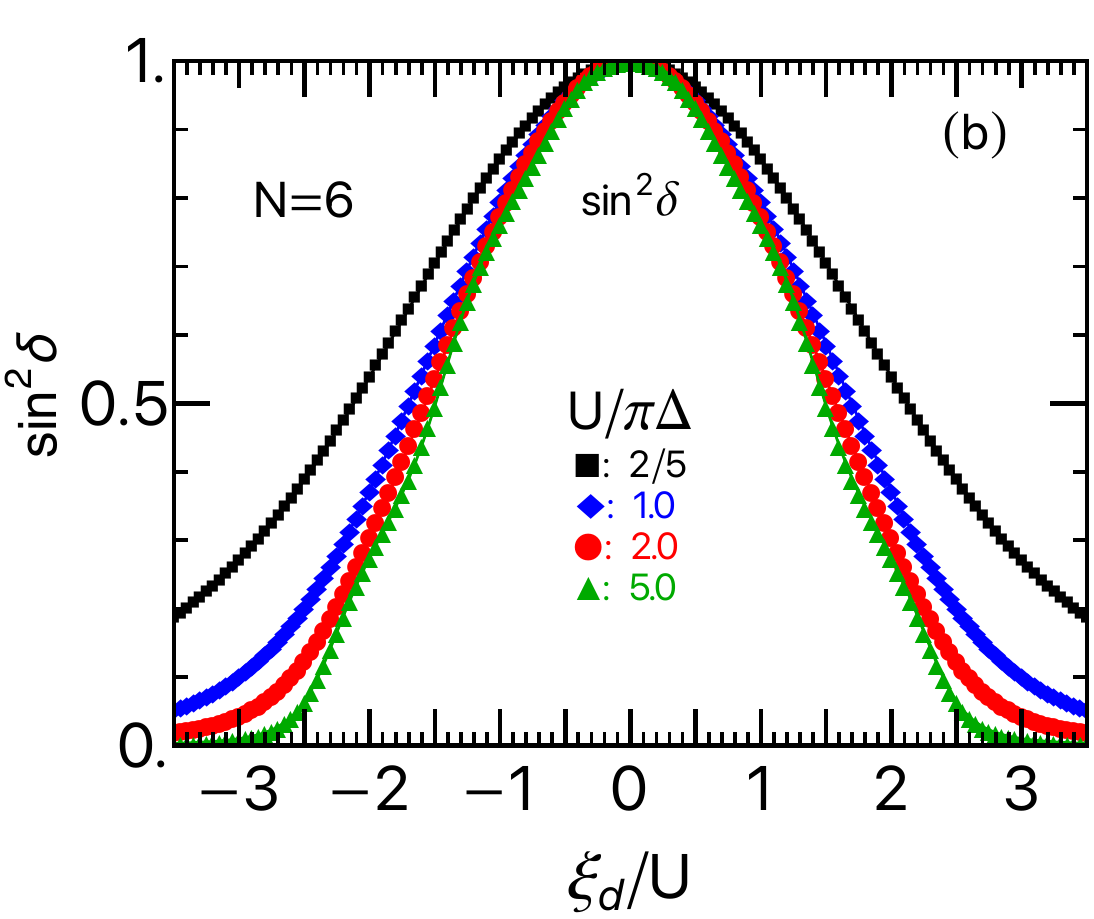}
	\end{minipage}
	\begin{minipage}[r]{\linewidth}
	\centering
	\includegraphics[keepaspectratio,scale=0.22]{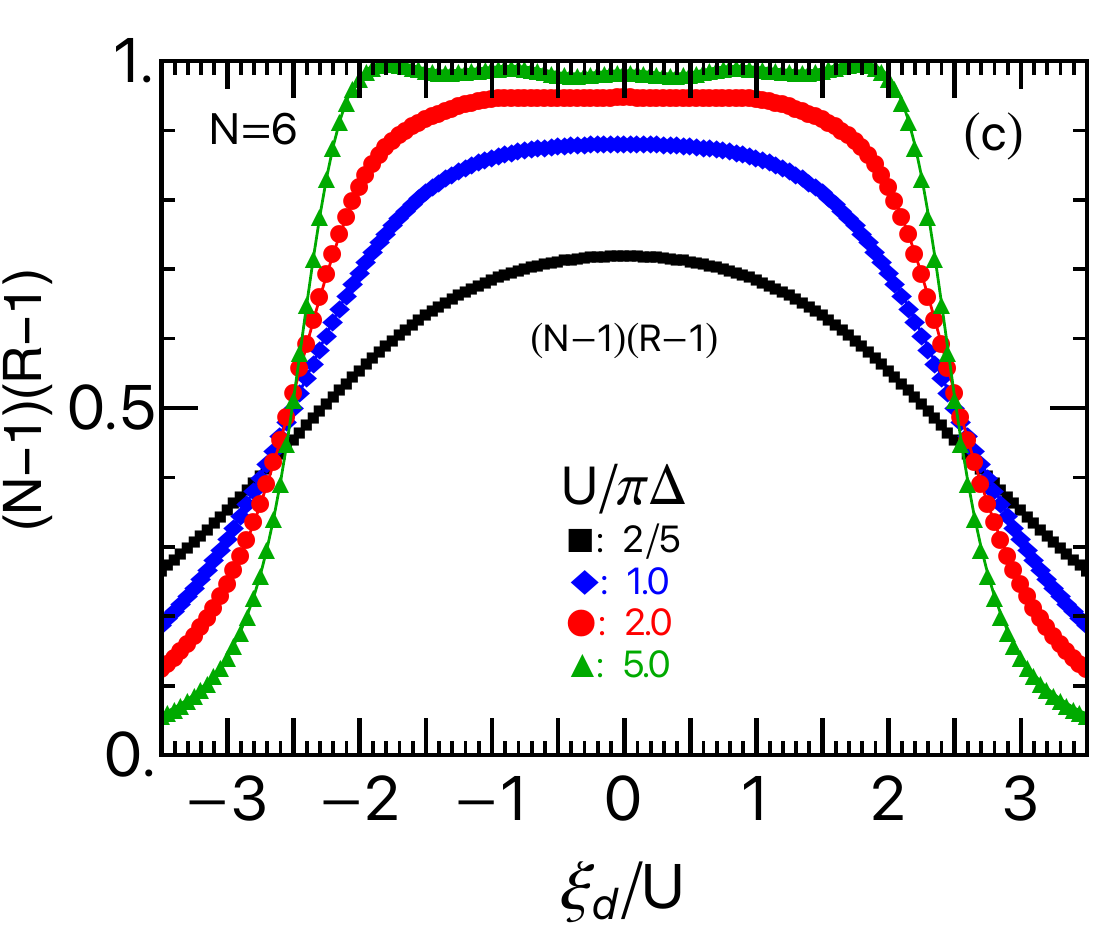}
	\centering
	\includegraphics[keepaspectratio,scale=0.22]{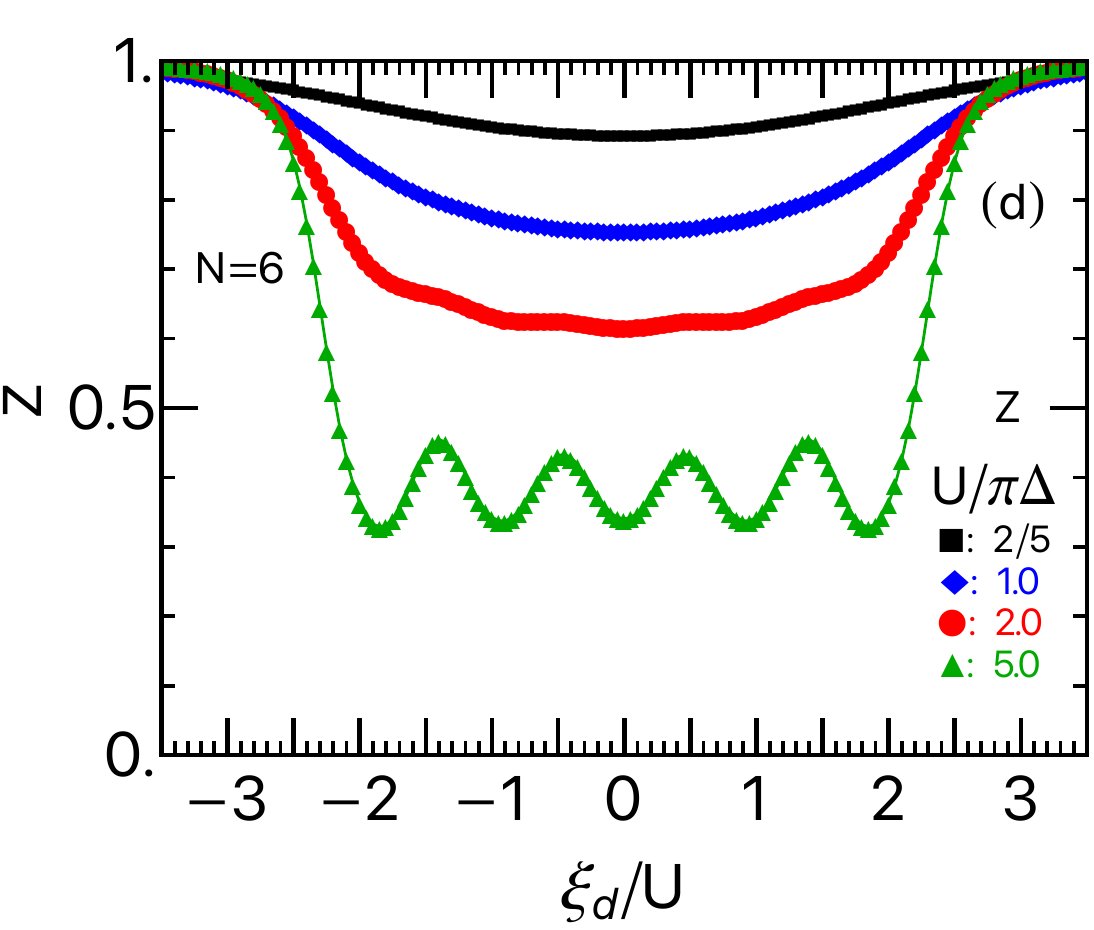}
	\end{minipage}
	\begin{minipage}[r]{\linewidth}
	\centering
	\includegraphics[keepaspectratio,scale=0.22]{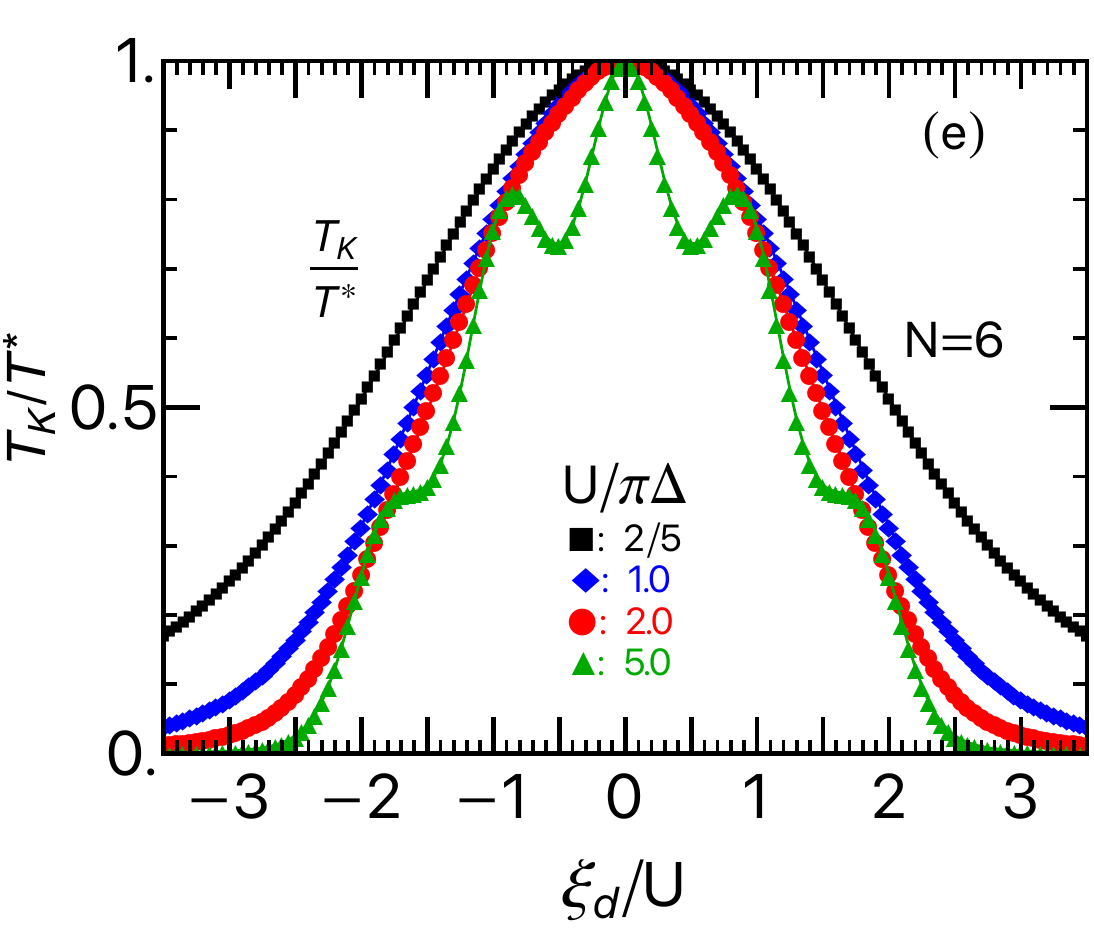}
	\centering
	\includegraphics[keepaspectratio,scale=0.22]{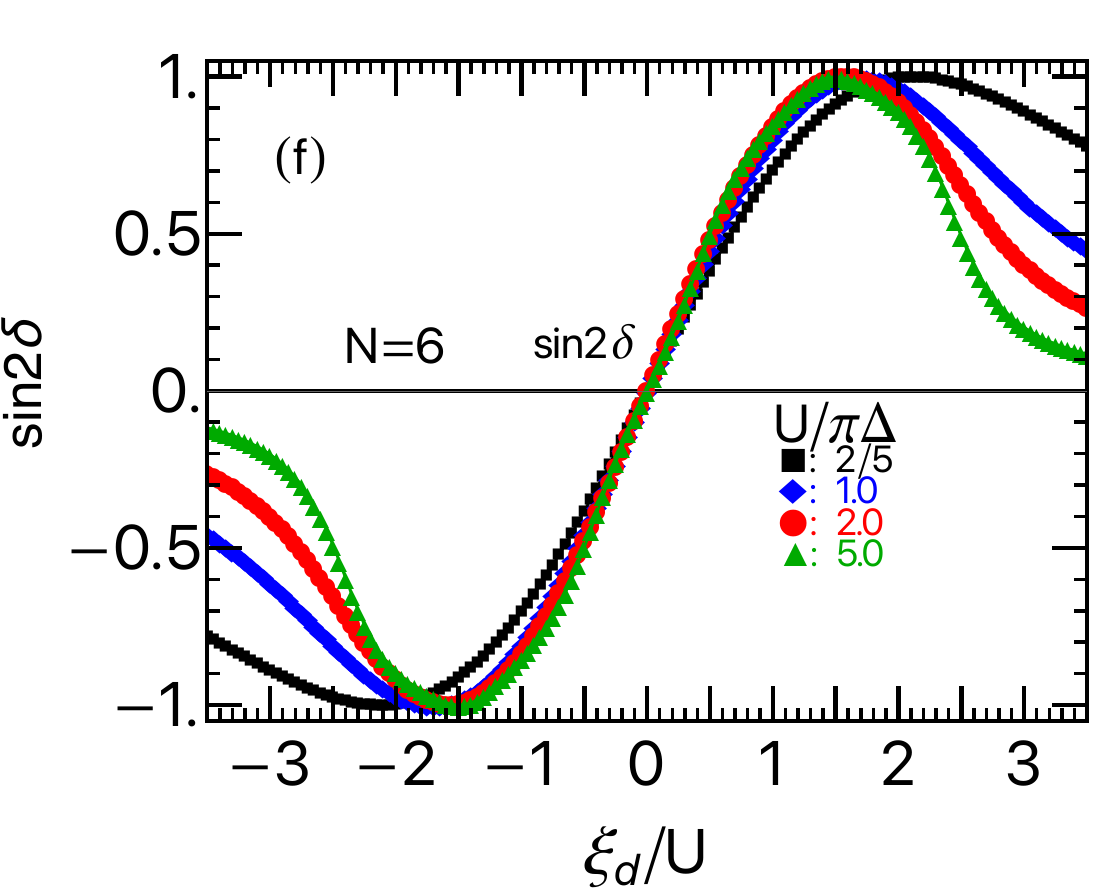}
	\end{minipage}
\caption{
Fermi-liquid parameters of an SU(6)  dot 
are plotted as a function of $\xi_d^{}/U$ 
for $U/(\pi\Delta) =2/5$,\,$1.0$,\,$2.0$,\,$5.0$. 
(a) $\langle n_{d}\rangle$,\,  
(b) $\sin^2\delta$,\,
(c) rescaled Wilson ratio $\widetilde{K} \equiv (N-1)(R-1)$,\,
(d) renormalized factor $z$,\,
(e) $T_K/T^\ast$,\, and \,
(f) $\sin2\delta$. 
Here,  $T_K$ is defined as the value of 
$T^\ast  \equiv 1/(4\chi_{\sigma\sigma}^{})$ 
at half filling $\xi_d^{}=0$.
}
  \label{FermiparaSU6}
\end{figure}

\begin{figure}[t]
	\rule{1mm}{0mm}
	\begin{minipage}[r]{\linewidth}
	\centering
	\includegraphics[keepaspectratio,scale=0.22]{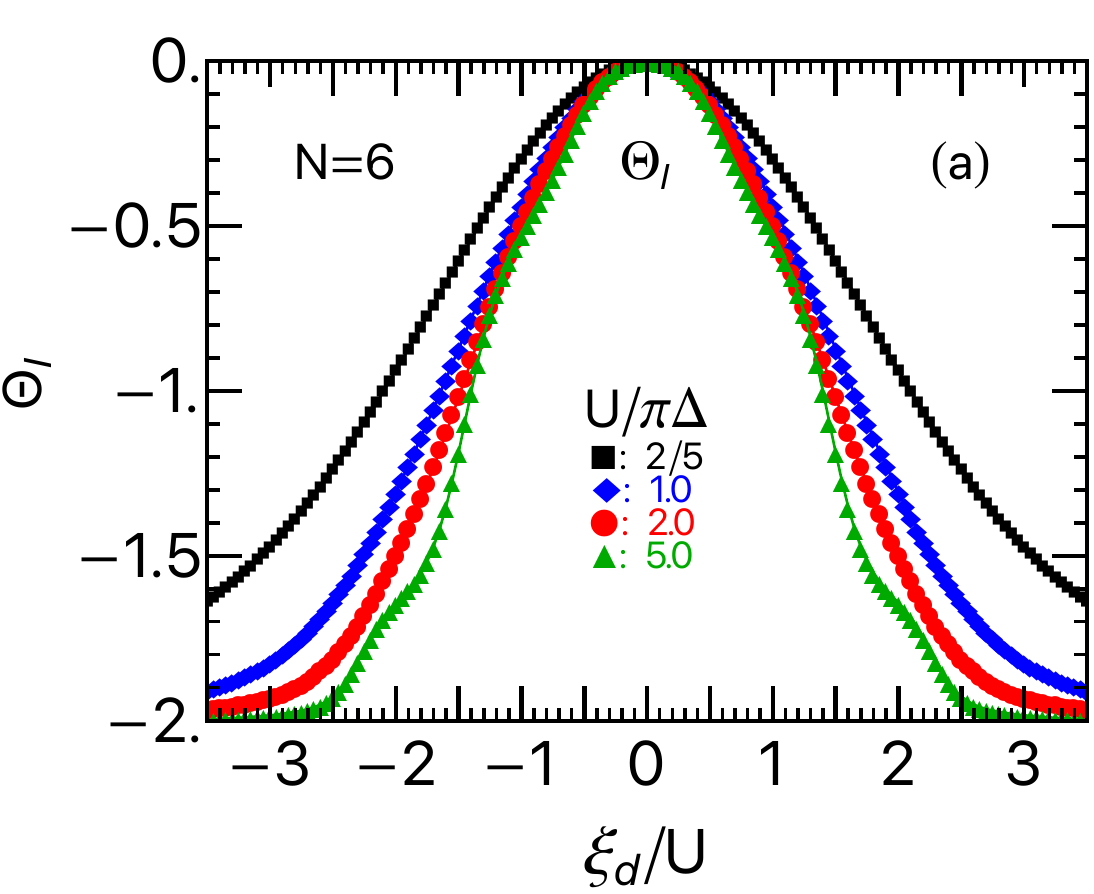}
	\centering
	\includegraphics[keepaspectratio,scale=0.21]{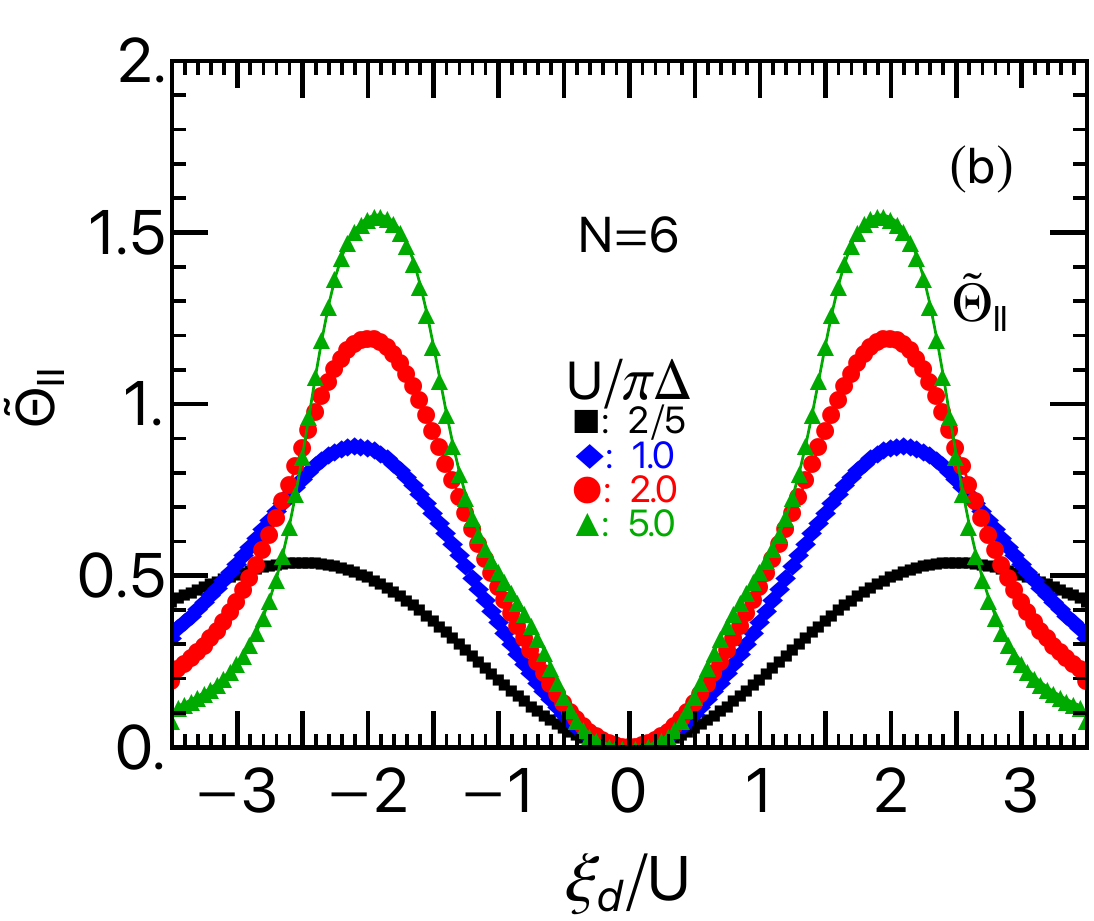}
	\end{minipage}
	\begin{minipage}[r]{\linewidth}
	\centering
	\includegraphics[keepaspectratio,scale=0.22]{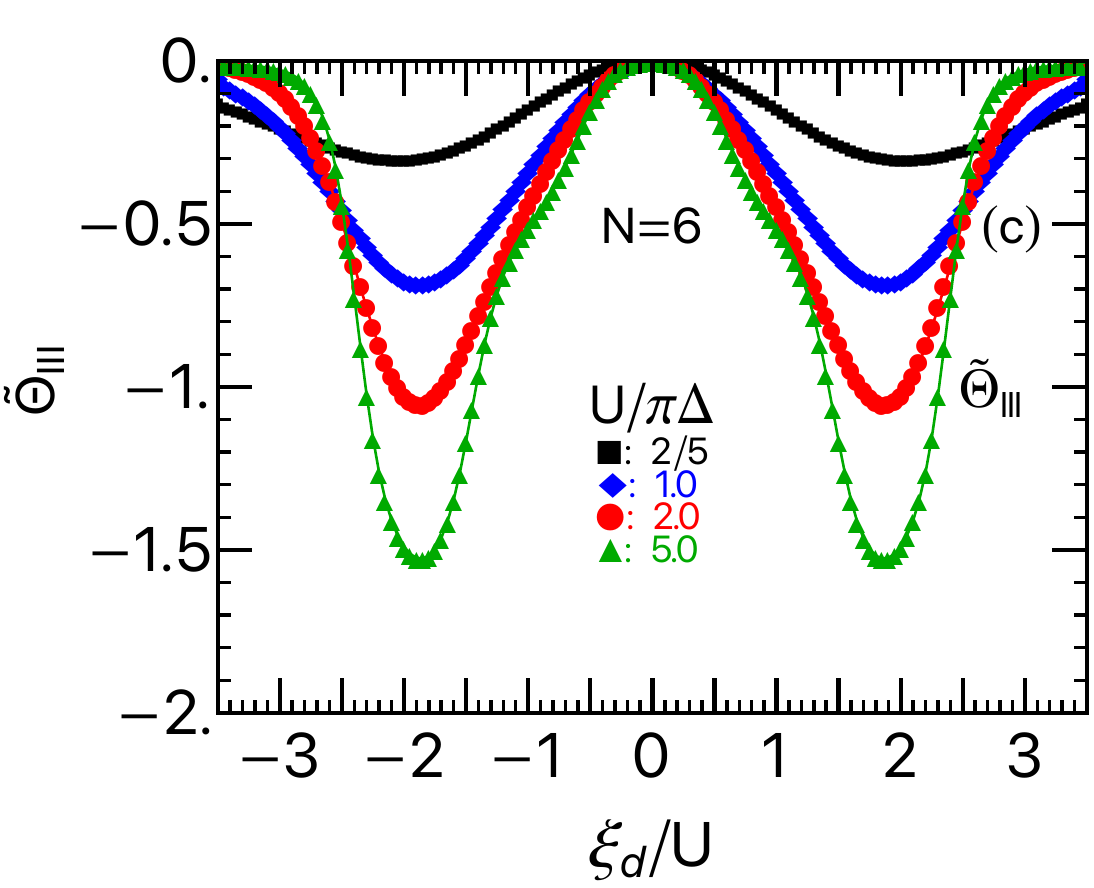}
	\centering
	\includegraphics[keepaspectratio,scale=0.22]{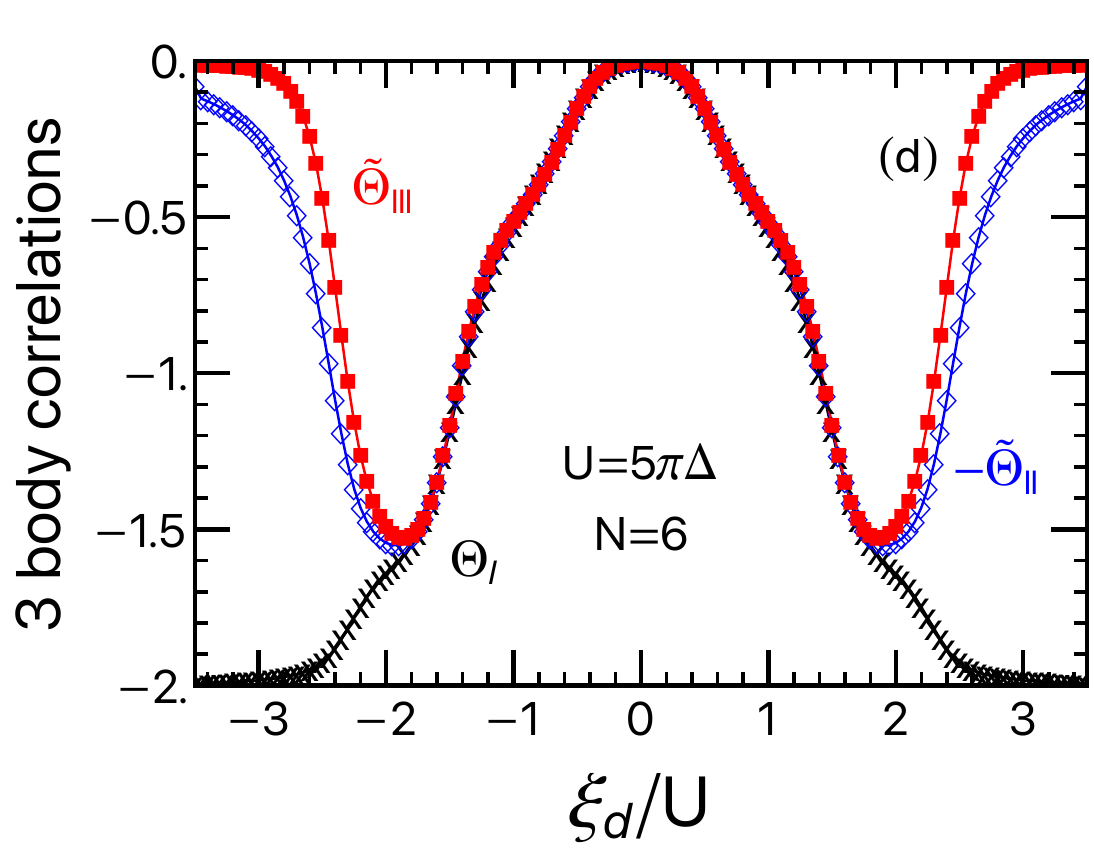}
	\end{minipage}
\caption{
Three-body correlation functions of an SU(6) dot 
 plotted vs $\xi_d/U$ for $U/\pi\Delta=2/5$,\,$1.0$,\,$2.0$,\,$5.0$: 
(a) $\Theta_\mathrm{I}$, (b) $\widetilde{\Theta}_\mathrm{II}$,
 (c) $\widetilde{\Theta}_\mathrm{III}$.  
 (d)  comparison of 
$\Theta_\mathrm{I}$, $-\widetilde{\Theta}_\mathrm{II}$, 
and $\widetilde{\Theta}_\mathrm{III}$ 
at a large interaction $U/(\pi\Delta) = 5.0$. 
It indicates that $\Theta_\mathrm{I} \simeq 
-\widetilde{\Theta}_\mathrm{II} 
\simeq \widetilde{\Theta}_\mathrm{III}$ 
over a wide range of $-(N-1)/2 U\lesssim\xi_d^{}\lesssim (N-1)/2$ for $N=6$.
}
  \label{TH123SU6}
\end{figure}

\subsection{SU(6) Fermi-liquid properties }

\subsubsection{Two-body correlation functions for an SU(6) dot}

We next summarize the low-energy Fermi-liquid properties 
of SU($6$) quantum dots. 
The phase shift and the renormalized parameters 
that can be deduced from the two-body correlations 
are plotted as a function of the gate voltage $\xi_d^{}/U$ 
in Fig.\ \ref{FermiparaSU6},  
for several different values of  $U/(\pi\Delta) =2/5$, 
$1.0$, $2.0$, $5.0$, i.e.,  from weak to strong interactions. 
The behaviors of these parameters are similar to those of the SU($4$) quantum dots. 
However,  the number of different SU($N$) Kondo states 
occurring at  integer fillings increases with $N$, i.e.,  
  $\langle n_d^{} \rangle=1$, $2$, $\ldots$, $N-1$.
It takes place at  $\xi_d^{} \simeq 0.0$, $\pm U$, \ldots,  $\pm \frac{N-2}{2}U$, 
and gives an interesting variety in low-energy transport.  
As $N$ increases,  
quantum fluctuations caused by the Coulomb interaction $U$ is suppressed, 
and in particular, the mean-field theory becomes  exact  in the large $N$ limit of  
the finite-$U$ Anderson impurity model \cite{Oguri2012}.
Hence for larger $N$,  
electron-correlation effects occur for stronger interactions $U$.

In Fig.\ \ref{FermiparaSU6}(a),  we can see that 
the Coulomb staircase behavior of the occupation number 
$\langle n_{d}\rangle$ emerges for a large interaction $U/(\pi\Delta) = 5.0$, 
which has steps 
at $\xi_d^{} \simeq 0.0$, $\pm U$,  $\pm 2 U$  for  SU($6$) quantum dots.  
The transmission probability $\sin^2\delta$,  
shown in Fig.\ \ref{FermiparaSU6}(b),   
 reaches the unitary limit value $\sin^2\delta=1.0$ 
at half filling $\xi_d^{}=0$.   
We can see that  the plateau structure develops as $U$ increases, 
and it becomes visible at a strong interaction $U/(\pi\Delta) = 5.0$.

Figure \ref{FermiparaSU6}(c) shows that the rescaled  
Wilson ratio $\widetilde{K}$ for $N=6$ 
has a wide flat peak at  $-(N-1)U/2\lesssim \xi_d\lesssim (N-1)U/2$, 
the height of which increases with  $U$.
In particular,  it approaches saturation value $\widetilde{K}\simeq 1$ 
at  $U/(\pi\Delta) = 5.0$ due to the suppression of charge fluctuations, 
 as mentioned.

The renormalization factor $z$ for SU($6$) quantum dots, 
shown in Fig.\  \ref{FermiparaSU6}(d), 
 exhibits a broad valley structure similar to the one for the SU(4) symmetric case. 
The valley becomes deeper as $U$ increases, 
 and the local minima emerge at the integer-filling points,  
reflecting the electron correlations due to the SU(6) Kondo effects. 
Figure \ref{FermiparaSU6}(e) shows that 
the inverse of the characteristic energy $1/T^\ast$, 
which is scaled by the value  $T_K$ defined at half filling $\xi_d^{}=0$.  
It has peaks situated  at  $\xi_d^{}\simeq 0$,  $\pm U$, $\pm 2U$ 
for a strong interaction $U/(\pi\Delta)=5.0$,  
 although the ones at $\pm 2U$ are still developing.  
These peaks correspond to the SU(6) Kondo temperature 
at each of the integer-filling points. 

Figure \ref{FermiparaSU6}(f) shows $\sin2\delta$ for $N=6$,
which  is proportional to  the derivative of the density of states  
$\rho_d' = (\chi_{\sigma\sigma}^{}/\Delta) \sin 2 \delta$, 
and  determines the magnitude of the coefficient 
$C_V^{(2)}$ of the nonlinear current,  as mentioned. 
This factor $\sin2\delta$  is an odd function of $\xi_d^{}$, 
and at a strong interaction $U/(\pi \Delta)=5.0$ 
it takes a broad peak (dip) at $\delta= \pi/4$ ($3\pi/4$) 
corresponding to the half-integer filling 
 $\langle n_{d}\rangle=1.5$ ($4.5$) 
where charge fluctuations are not fully suppressed.

\subsubsection{Three-body correlation functions of an  SU(6) dot}
  
Figure \ref{TH123SU6} shows  three-body correlations between 
electrons passing through an SU(6) quantum dot, 
calculated as a function of $\xi_d^{}$  
for several values of interactions  $U/\pi\Delta=2/5$, $1.0$, $2.0$, $5.0$. 
These dimensionless correlation functions  $\Theta_\mathrm{I}$, 
$\widetilde{\Theta}_\mathrm{II}$, and $\widetilde{\Theta}_\mathrm{III}$ 
vanish at half filling $\xi_d^{}=0$, 
and away from half filling, they are enhanced as $U$ increases. 
We can see that for a strong  interaction $U/(\pi \Delta) =5.0$   
there emerges either a wide peak,  a wide dip, or a plateau    
at  $\xi_d^{} \simeq \pm U$,   $\pm 2U$, i.e., 
at integer filling points. 
The component between three different levels 
$\widetilde{\Theta}_\mathrm{III}$ also contributes 
to the order $(eV)^3$ term of nonlinear current 
through an SU($6$)  quantum dot   
when there are some asymmetries in the tunnel couplings 
or the bias voltages.

In Fig.\ \ref{TH123SU6}(d) 
the three independent components 
 $\Theta_\mathrm{I}$, 
$-\widetilde{\Theta}_\mathrm{II}$,  and $\widetilde{\Theta}_\mathrm{III}$ 
are  compared in a strong interaction case  $U/(\pi\Delta)=5.0$. 
It shows that these three components approach 
each other very closely 
$\Theta_\mathrm{I} \simeq -\widetilde{\Theta}_\mathrm{II}
\simeq\widetilde{\Theta}_\mathrm{III}$  
over a wide range of the gate voltage 
 $-(N-1)U/2\lesssim\xi_d\lesssim (N-1)U/2$ with $N=6$. 
This is due to  the fact that 
the diagonal component  $\chi_{\sigma\sigma\sigma}^{[3]}$ 
dominates the three-body correlation,  
and the derivatives  
$|\frac{\partial \chi_{\sigma\sigma}}{\partial \epsilon_d}|$ 
and $|\frac{\partial \chi_{\sigma\sigma'}}{\partial \epsilon_d}|$ 
 become much smaller than 
 $(T^\ast)^{-2}$  in a wide range 
of electron fillings $1 \lesssim \langle n_d^{} \rangle \lesssim N-1$ 
[see Appendix \ref{ThreeBodySame}] \cite{Teratani2020PRL}.

\section{Order $(eV)^2$ nonlinear current for SU(4) and SU(6) cases}
\label{Cv2Results}

The coefficient $C_V^{(2)}$, 
defined in Eqs.\ \eqref{DiffCond} and \eqref{Cv2}, emerges 
when the tunnel coupling or the bias voltage is not symmetric.  
Its magnitude is determined by the Wilson ratio $\widetilde{K}$ 
and the derivative of the density of states:
$\rho_d' = (\chi_{\sigma\sigma}^{}/\Delta) \sin 2 \delta$.

In this section, we first of all describe 
behavior of $C_V^{(2)}$ in some limiting cases, 
and then discuss the NRG results  
to show how the coefficient evolves with  
the tunnel and bias asymmetries.


\subsection{Behavior of $C_V^{(2)}$ in some limiting cases}

We have  seen in  Figs.\ \ref{FermiparaSU4} and \ref{FermiparaSU6} 
that the Wilson ratio reaches the saturation value $\widetilde{K}\simeq 1$ 
over a wide range of gate voltages 
$-(N-1)U/2\lesssim\xi_d^{}\lesssim(N-1)U/2$ for large $U$, 
where the quantum dot is partially filled 
$1\lesssim\langle n_d\rangle\lesssim N-1$. 
This is caused by the fact that 
the charge fluctuations are significantly suppressed in this region, 
as mentioned. 
In the limit of strong interactions,  
$C_V^{(2)}$ becomes independent of the bias asymmetry $\alpha_\mathrm{dif}^{}$:   
\begin{align}
C_V^{(2)}\xrightarrow{\,\widetilde{K} \to 1\,}&\,-\frac{\pi}{4}\,
\gamma_\mathrm{dif}^{}\,\sin2\delta\,,
\label{Cv2Kondo}
\end{align}
%

In contrast, in the limit of  $|\xi_d^{}|\to\infty$ 
where  $\langle n_d\rangle$ approaches $0$ or $N$, 
the Wilson ratio approaches the noninteracting value $\widetilde{K}\to 0$,   
 and $C_V^{(2)}$ becomes independent of tunnel asymmetry $\gamma_\mathrm{dif}$: 
\begin{align}
C_V^{(2)}\xrightarrow{\,\widetilde{K}\to 0\,}\,\frac{\pi}{4}\,\alpha_\mathrm{dif}\,\sin2\delta\,. 
\label{eq:Cv2K=0}
\end{align}

\begin{figure}[t]
	\begin{minipage}[r]{\linewidth}
	\centering
	\includegraphics[keepaspectratio,scale=0.22]{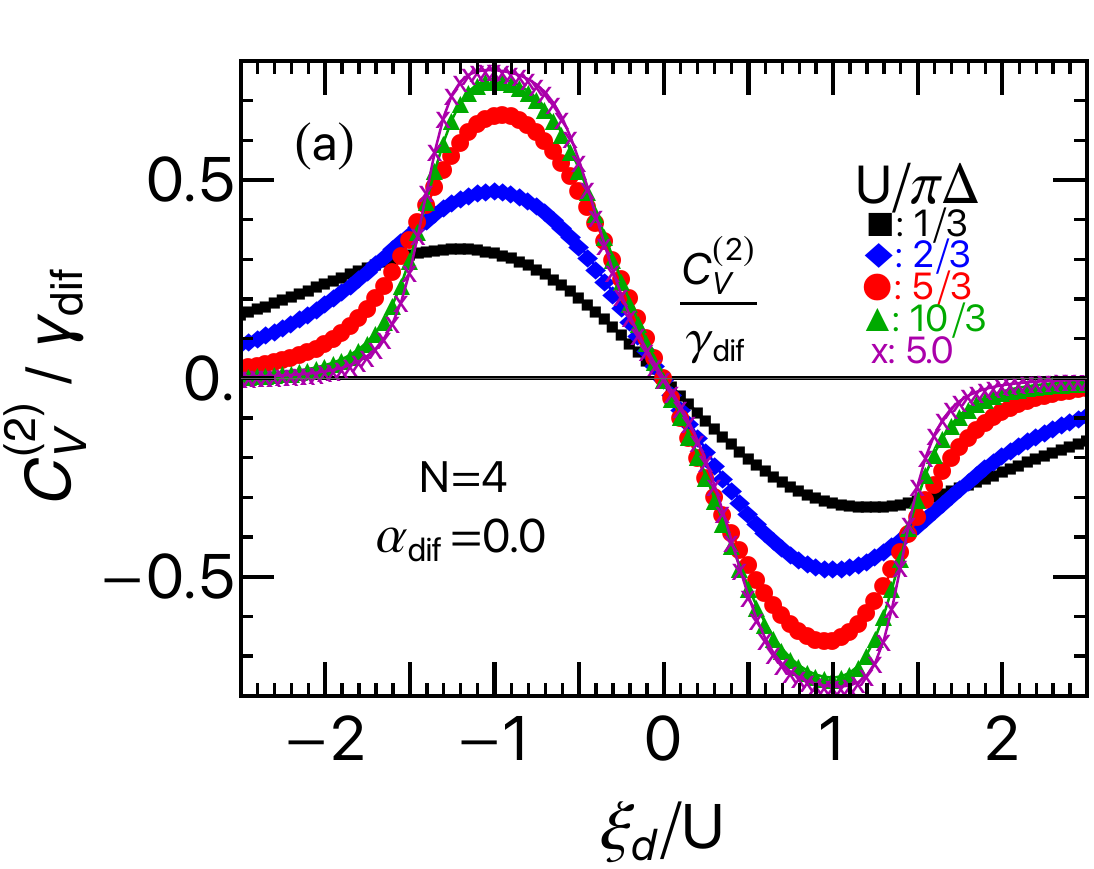}
	\centering
	\includegraphics[keepaspectratio,scale=0.22]{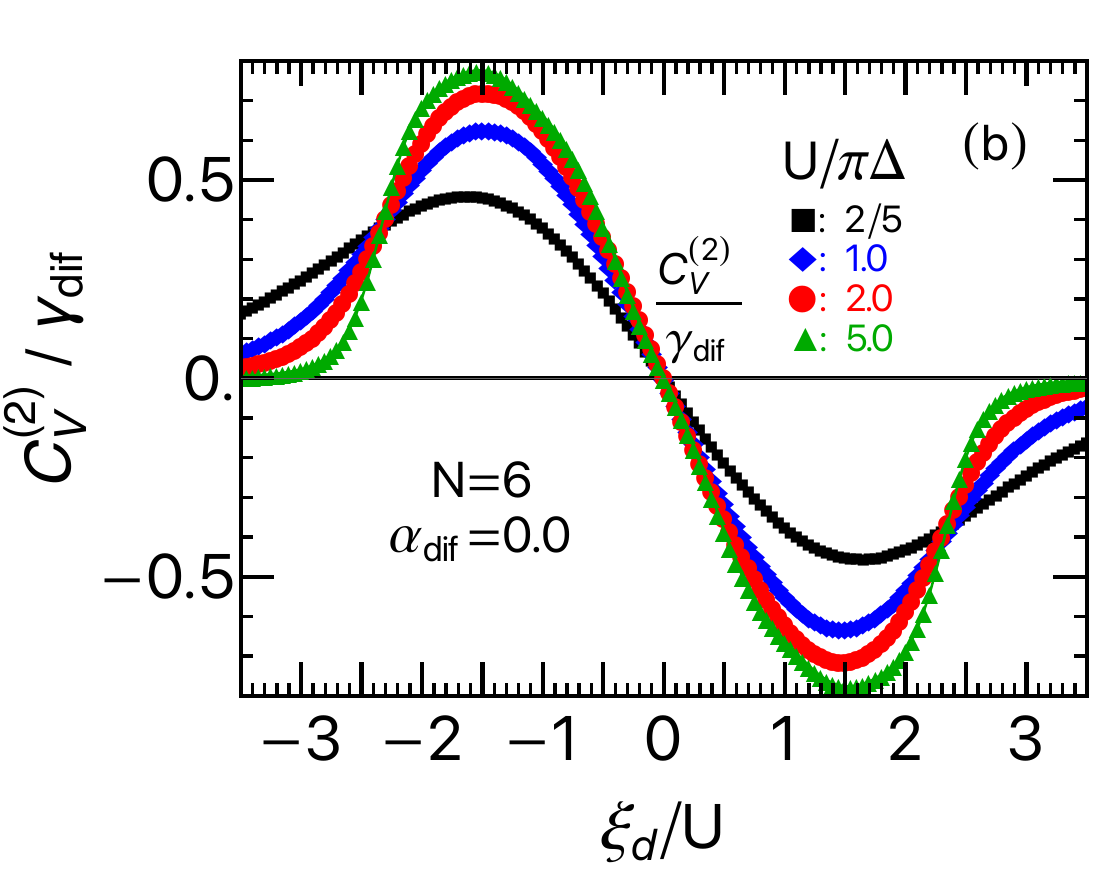}
	\end{minipage}
\caption{
Effects of tunnel asymmetry  $\gamma_\mathrm{dif}^{}$ 
on order $(eV)^2$ nonlinear current:   
$C_V^{(2)}/\gamma_\mathrm{dif}^{}$ given by Eq.\ \eqref{Cv2symeq} 
for symmetrical bias voltages $\alpha_\mathrm{dif}^{}=0.0$  
is plotted vs  $\xi_d^{}/U$, varying interactions from weak to strong.     
(a) for SU(4) quantum dots with  $U/(\pi\Delta)=1/3$, $2/3$, $5/3$, $10/3$, $5.0$. 
(b) for SU(6) quantum dots with  $U/(\pi\Delta)=2/5$, $1.0$, $2.0$, $5.0$.   
}
  \label{Cv2sym}
\end{figure}

\subsection{
 Effects of tunnel asymmetry $\gamma_\mathrm{dif} \neq 0$ on  $C_V^{(2)}$  \\ 
for symmetric bias voltages $\alpha_\mathrm{dif}=0$  
}

We first of all consider   
effects of tunneling asymmetries $\gamma_\mathrm{dif}^{}$ on  $C_V^{(2)}$, 
taking  bias voltages to be symmetric $\alpha_\mathrm{dif}=0$: 
\begin{align}
C_V^{(2)}\xrightarrow{\,\alpha_\mathrm{dif}=0\,}\,-\frac{\pi}{4}\,\gamma_\mathrm{dif}\,\widetilde{K}\,\sin2\delta \,. 
\label{Cv2symeq}
\end{align}
In this case, $C_V^{(2)}$ is proportional
to $\gamma_\mathrm{dif}^{}$, 
and is determined by $\widetilde{K}$ and $\sin 2 \delta$.
Figure  \ref{Cv2sym} shows 
$C_V^{(2)}/\gamma_\mathrm{dif}^{}$ 
as a function of $\xi_d^{}/U$ for symmetric 
bias voltages  $\alpha_\mathrm{dif}^{}=0.0$. 
We have examined the behaviors from weak to strong interactions: 
(a) for  SU(4) quantum dots 
with $U/(\pi\Delta)=1/3$, $2/3$, $5/3$, $10/3$, $5.0$, 
and (b) for SU(6) quantum dots with    
$U/(\pi\Delta)=2/5$, $1.0$, $2.0$, $5.0$.

As $U$ increases, 
 a wide peak and a wide dip of  $C_V^{(2)}$ evolve  
at $\xi_d^{}\simeq U/4$ and $-U/4$, 
where  the occupation number reaches the value of    
 $\langle n_d\rangle\simeq N/4$ and $3N/4$, respectively.  
It corresponds  to phase shifts of $\delta \simeq \pi/4$ and $3\pi/4$, 
at which $\sin2\delta$ takes an extreme value as seen   
in Figs.\ \ref{FermiparaSU4}(f)  and \ref{FermiparaSU6}(f). 
The peak and dip structures also reflect the fact that 
the Wilson ratio is almost saturated     
 $\widetilde{K}\simeq 1.0$   in a wider range of 
$ |\xi_d^{}| \lesssim (N-1)U/2$  
for large $U$ in Figs.\ \ref{FermiparaSU4}(c)  and \ref{FermiparaSU6}(c). 
The Kondo effect of an integer filling occurs 
at the flat peak and the flat dip for $N=4$ quantum dots.  
In contrast, for $N=6$,
it is an intermediate valence state that occurs at the peak and dip, 
and thus the structures become round rather than flat 
since charge fluctuations remain active. 

Outside the correlated region, 
 the absolute value of $C_V^{(2)}$ decreases as $U$ increases,  
for both  $N=4$ and $6$  in Figs.\ \ref{Cv2sym} (a) and (b), respectively.
In particular, at  $|\xi_d^{}| \gg (N-1)U/2$, 
it vanishes asymptotically  $C_V^{(2)}\to 0$ 
as the occupation number approaches  $\langle n_d\rangle \to 0$ or  $N$.

\begin{figure}[t]
	\begin{minipage}[r]{0.49\linewidth}
	\centering
	\includegraphics[keepaspectratio,scale=0.23]{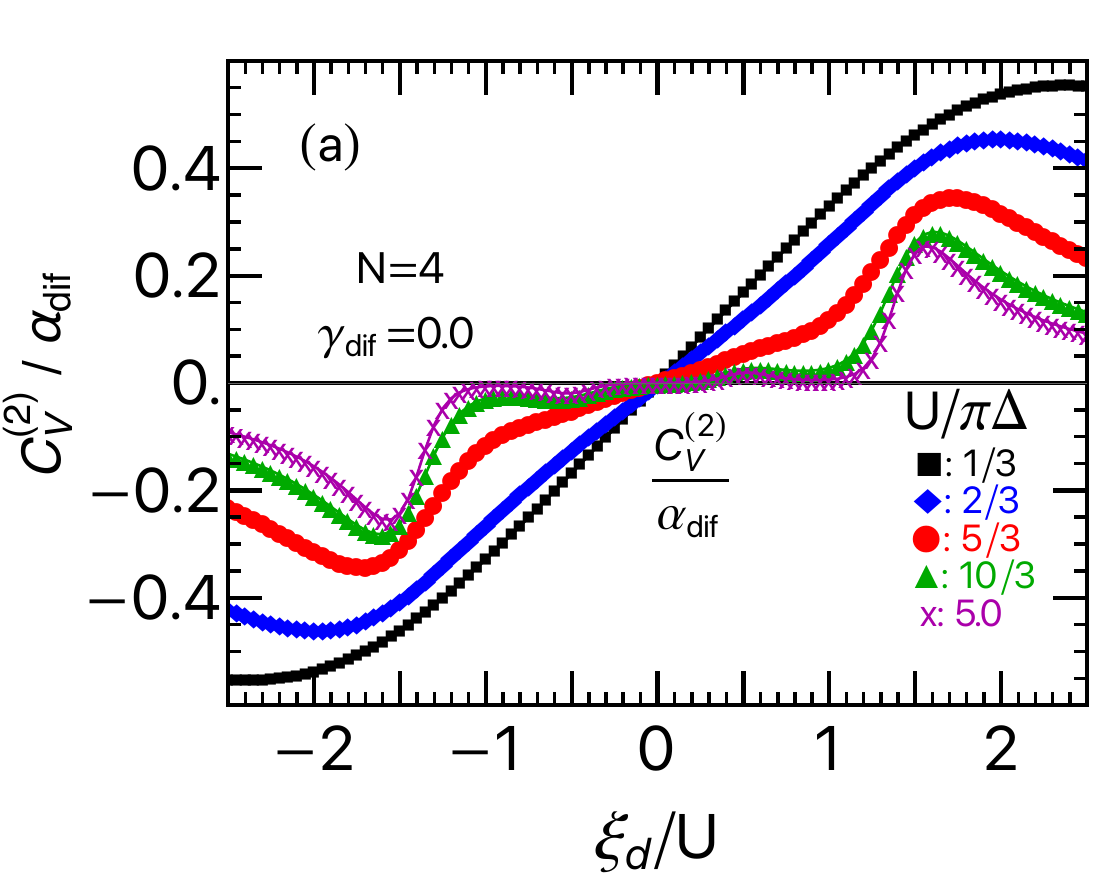}
	\end{minipage}
	\begin{minipage}[r]{0.49\linewidth}
	\centering
	\includegraphics[keepaspectratio,scale=0.23]{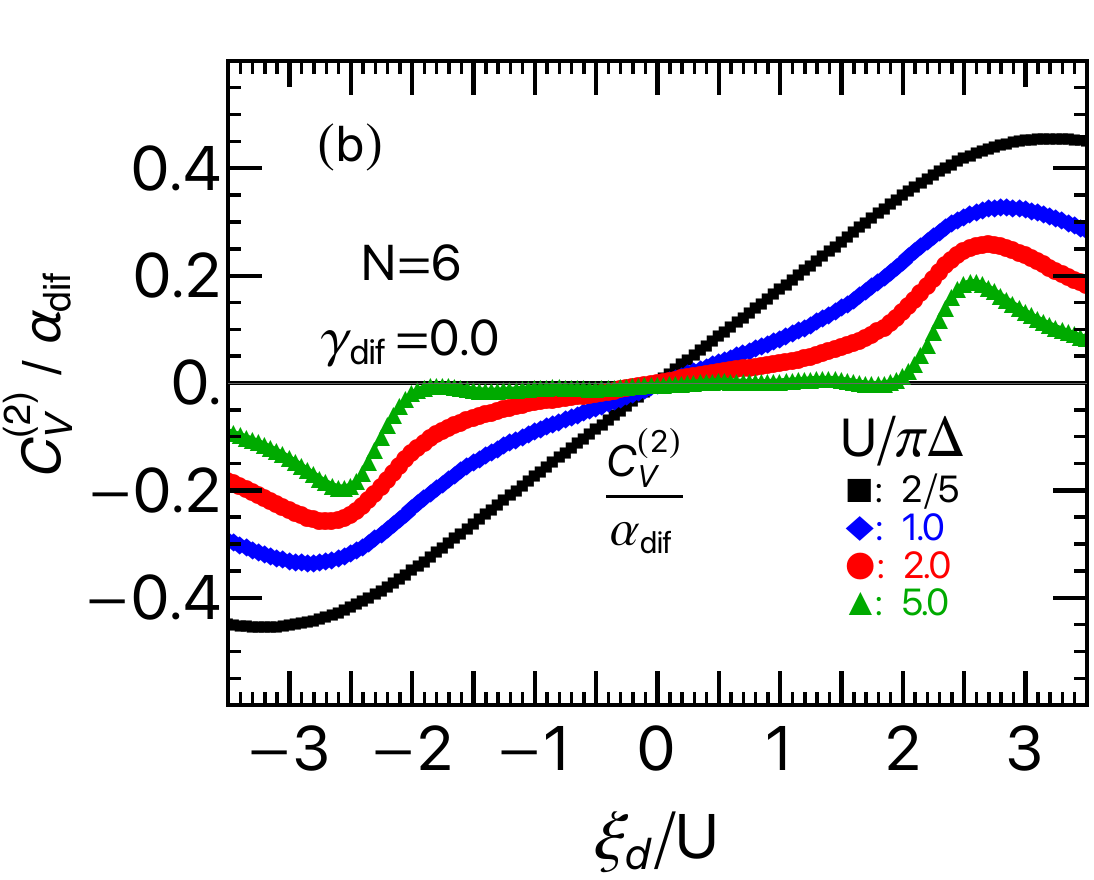}
	\end{minipage}
\caption{
Effects of bias-voltage asymmetry  $\alpha_\mathrm{dif}^{}$ 
on order $(eV)^2$ nonlinear current: 
$C_V^{(2)}/\alpha_\mathrm{dif}^{}$ given by Eq.\ \eqref{Cv2GdifSym}
for symmetric junctions with $\gamma_\mathrm{dif}^{}=0.0$  
is plotted vs  $\xi_d^{}/U$, varying interactions from weak to strong.     
(a) for SU(4) quantum dots with  $U/(\pi\Delta)=1/3$, $2/3$, $5/3$, $10/3$, $5.0$. 
(b) for SU(6) quantum dots with  $U/(\pi\Delta)=2/5$, $1.0$, $2.0$, $5.0$.  
}
  \label{Cv2aLvaryxL05N4N6}
\end{figure}

\subsection{Effects of bias asymmetry $\alpha_\mathrm{dif} \neq 0$ on 
 $C_V^{(2)}$ \\ 
for symmetric tunnel junctions $\gamma_\mathrm{dif}=0$
}

We next consider 
the effects of bias asymmetries $\alpha_\mathrm{dif}^{}$  
on $C_V^{(2)}$, 
setting tunnel junctions to be symmetric  $\gamma_\mathrm{dif}=0$.  
In this case,  $C_V^{(2)}$ becomes proportional to  $\alpha_\mathrm{dif}^{}$, 
as 
\begin{align}
C_V^{(2)}\xrightarrow{\,\gamma_\mathrm{dif}^{}=0\,}\,
\frac{\pi}{4}\,\alpha_\mathrm{dif}\,
\Bigl(1-\widetilde{K}\Bigr)\,\sin2\delta\,. 
\label{Cv2GdifSym}
\end{align}
The ratio $C_V^{(2)}/\alpha_\mathrm{dif}^{}$ for this case 
 is plotted vs $\xi_d^{}/U$ in Fig.\ \ref{Cv2aLvaryxL05N4N6}. 
We have also examined
 from weak to strong interactions 
 for symmetric tunnel couplings $\gamma_\mathrm{dif}^{}=0.0$:  
(a) for  SU(4) quantum dots 
with $U/(\pi\Delta)=1/3$, $2/3$, $5/3$, $10/3$, $5.0$, 
and (b) for SU(6) quantum dots with    
$U/(\pi\Delta)=2/5$, $1.0$, $2.0$, $5.0$. 

At $|\xi_d^{}|\lesssim(N-1)\,U/2$, 
where the  localized levels of quantum dots are partially filled 
 $1\lesssim\langle n_d\rangle\lesssim N-1$,  
the factor  $1-\widetilde{K}$ in Eq.\ \eqref{Cv2GdifSym} becomes  
very small as $U$ increases  
since the Wilson ratio approaches the saturation value $\widetilde{K}\to 1.0$. 
Therefore,  $C_V^{(2)}$ almost vanishes in this region 
for strong interactions  $U/(\pi\Delta)\gtrsim 3.0$.  
However, weak oscillatory behavior survives  
at  $\xi_d^{}/U\simeq \pm(N-1-2m)/2$ for $m=1,\ldots ,N/2 -1$, 
and it  is caused by charge fluctuations 
in the intermediate valence states 
between two adjacent integer filling points.

 In contrast outside the correlated region,  
at  $|\xi_d^{}|\gg (N-1)\,U/2$,  
the rescaled Wilson ratio  decreases and it eventually vanishes 
 in the limit of  $|\xi_d^{}| \to \infty$ 
as shown in Figs.\ \ref{FermiparaSU4}(c)  and \ref{FermiparaSU6}(c), 
so that $1-\widetilde{K} \to 1.0$  in Eq.\ \eqref{Cv2GdifSym}. 
Therefore, the  behavior of  $C_V^{(2)}$ in this region 
 is mainly determined by the other factor $\sin 2 \delta$. 
The localized levels of quantum dots become almost 
 empty or fully occupied in the limit of $|\xi_d^{}| \to \infty$. 
Correspondingly, at the crossover region  $|\xi_d^{}|\simeq (N-1)\,U/2$, 
the phase shift takes the value around  
 $\delta \simeq \pi/N$ at  $\xi_d^{} \simeq (N-1)U/2$,    
and  
$\delta \simeq \pi(N-1)/N$ at  $\xi_d^{} \simeq -(N-1)U/2$.  
Thus, the peak height, or the dip depth, of $C_V^{(2)}$  decreases  as $N$ increases:  
the amplitude becomes larger for the SU(4) quantum dots than the SU(6)  
 in Fig.\ \ref{Cv2aLvaryxL05N4N6}.

For weak interactions, effects of bias asymmetries $\alpha_\mathrm{dif}^{}$  
appear in the whole region of $\xi_d^{}$, particularly 
 $C_V^{(2)}$ takes finite values  in the region of $|\xi_d^{}| \simeq (N-1)U/2$. 

\subsection{$C_V^{(2)}$ 
under both asymmetries $\alpha_\mathrm{dif}^{} \neq 0$ and 
$\gamma_\mathrm{dif}^{} \neq 0$ 
}

We next consider the behavior of $C_V^{(2)}$ in the 
presence of both asymmetries, i.e.,  
 $\alpha_\mathrm{dif}^{} \neq 0$ and 
$\gamma_\mathrm{dif}^{} \neq 0$.  
We have seen  above  that 
in the strongly correlated region at  $|\xi_d^{}|\lesssim(N-1)U/2$ for large $U$, 
the coefficient $C_V^{(2)}$ is given by Eq.\ \eqref{Cv2Kondo} 
and it becomes  almost independent of  bias asymmetries $\alpha_\mathrm{dif}^{}$,  
since the Wilson ratio approaches $\widetilde{K} \simeq 1.0$. 
Conversely,  $C_V^{(2)}$  is given by Eq.\ \eqref{eq:Cv2K=0},  
which  does not depend on tunnel asymmetries  $\gamma_\mathrm{dif}^{}$, 
for small $U$ or outside the correlated region as $\widetilde{K} \simeq 0.0$.  
These properties provide the key to clarify 
overall characteristics of $C_V^{(2)}$ in a wide parameter range.

\subsubsection{Effects of bias asymmetry $\alpha_\mathrm{dif}^{}\neq 0$ 
on $C_V^{(2)}$ \\
at large tunnel asymmetry $\gamma_\mathrm{dif}^{}$ 
($\Gamma_L^{}\gg \Gamma_R^{}$)}

\begin{figure}[t]
\begin{minipage}[r]{0.49\linewidth}
\centering
\includegraphics[keepaspectratio,scale=0.23]{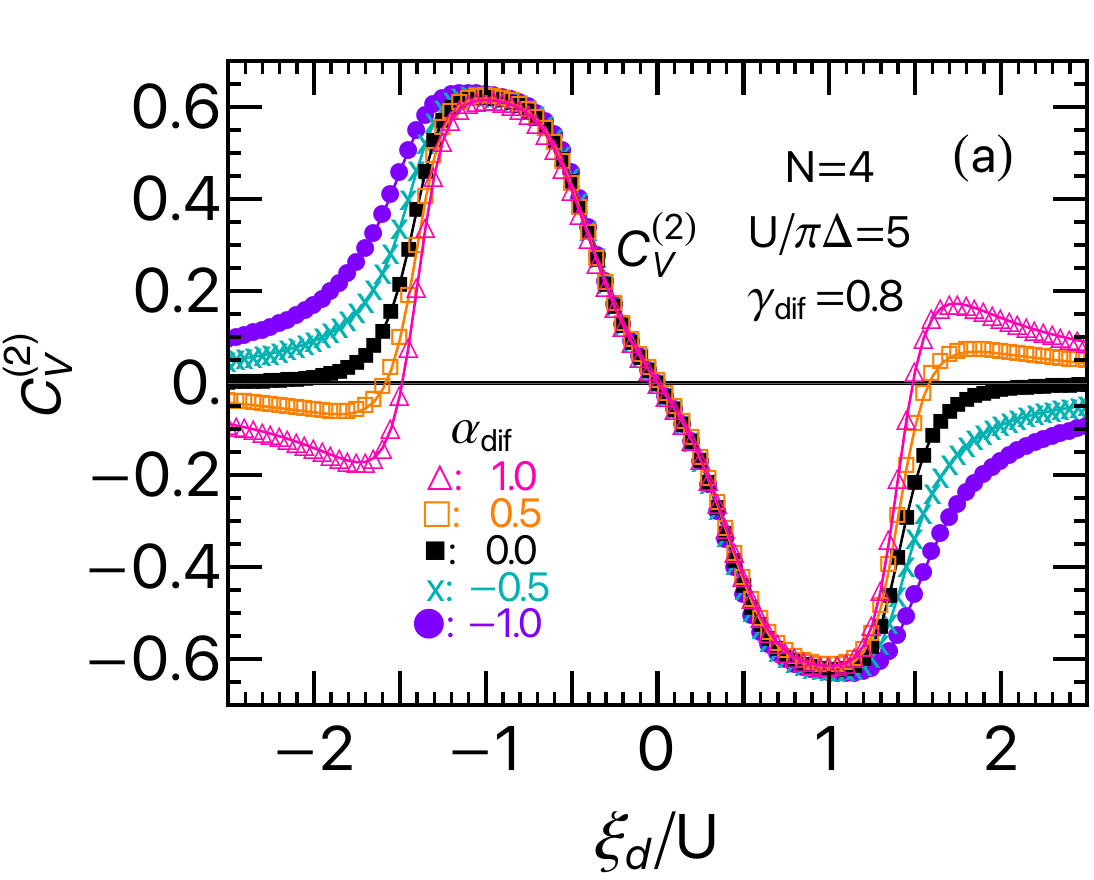}
\end{minipage}
\begin{minipage}[r]{0.49\linewidth}
\centering
\includegraphics[keepaspectratio,scale=0.23]{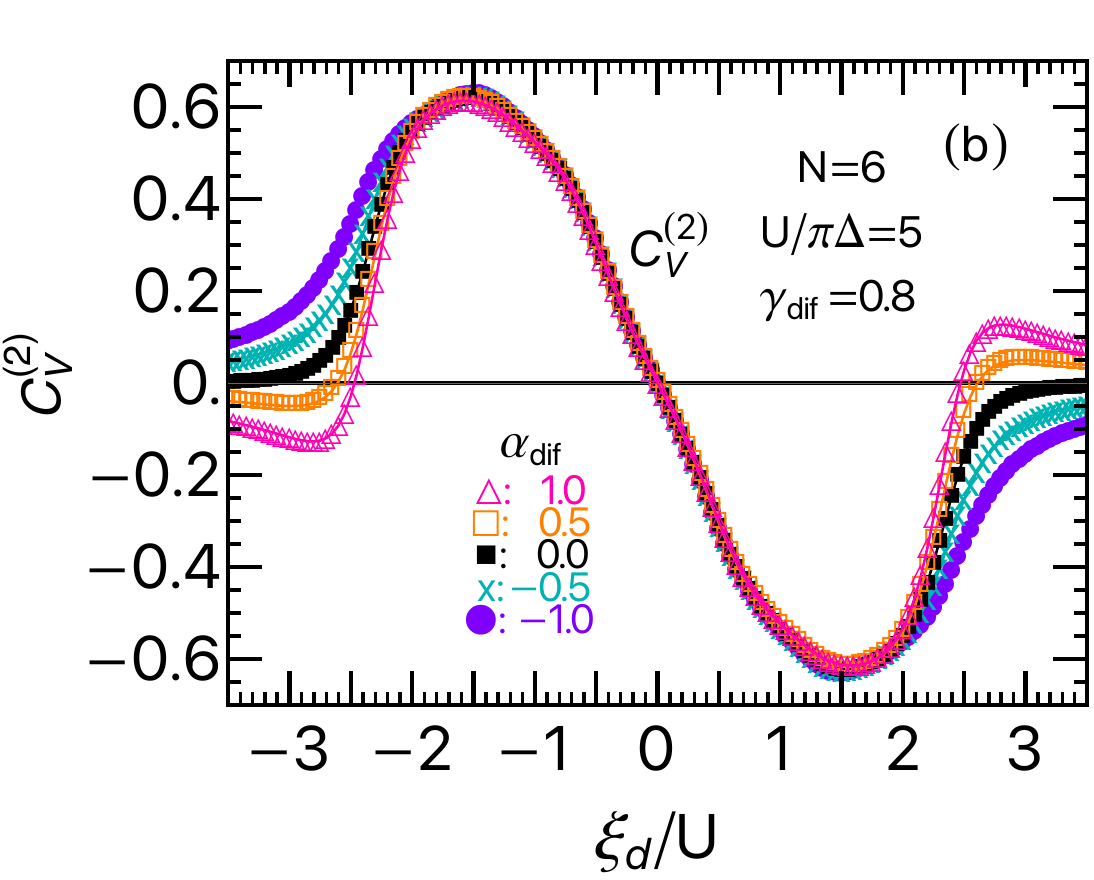}
\end{minipage}
\begin{minipage}[r]{0.49\linewidth}
\centering
\includegraphics[keepaspectratio,scale=0.23]{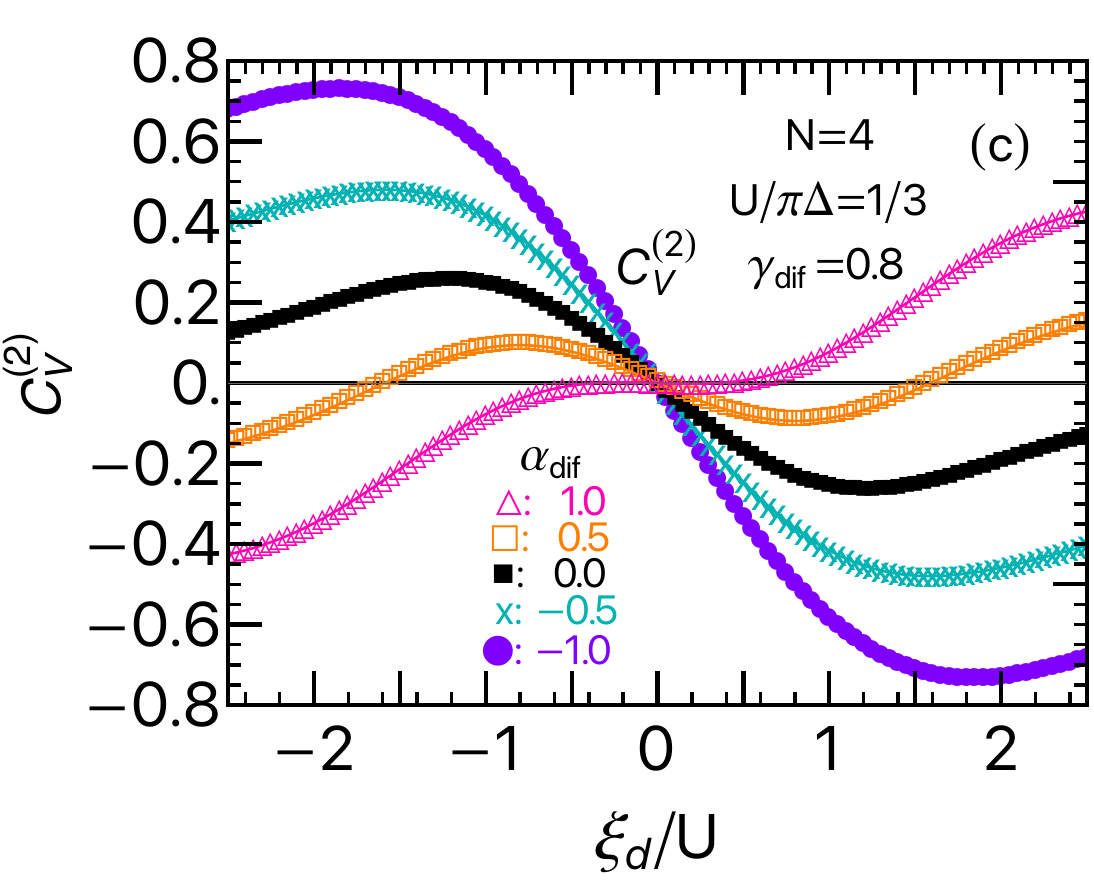}
\end{minipage}
\begin{minipage}[r]{0.49\linewidth}
\centering
\includegraphics[keepaspectratio,scale=0.23]{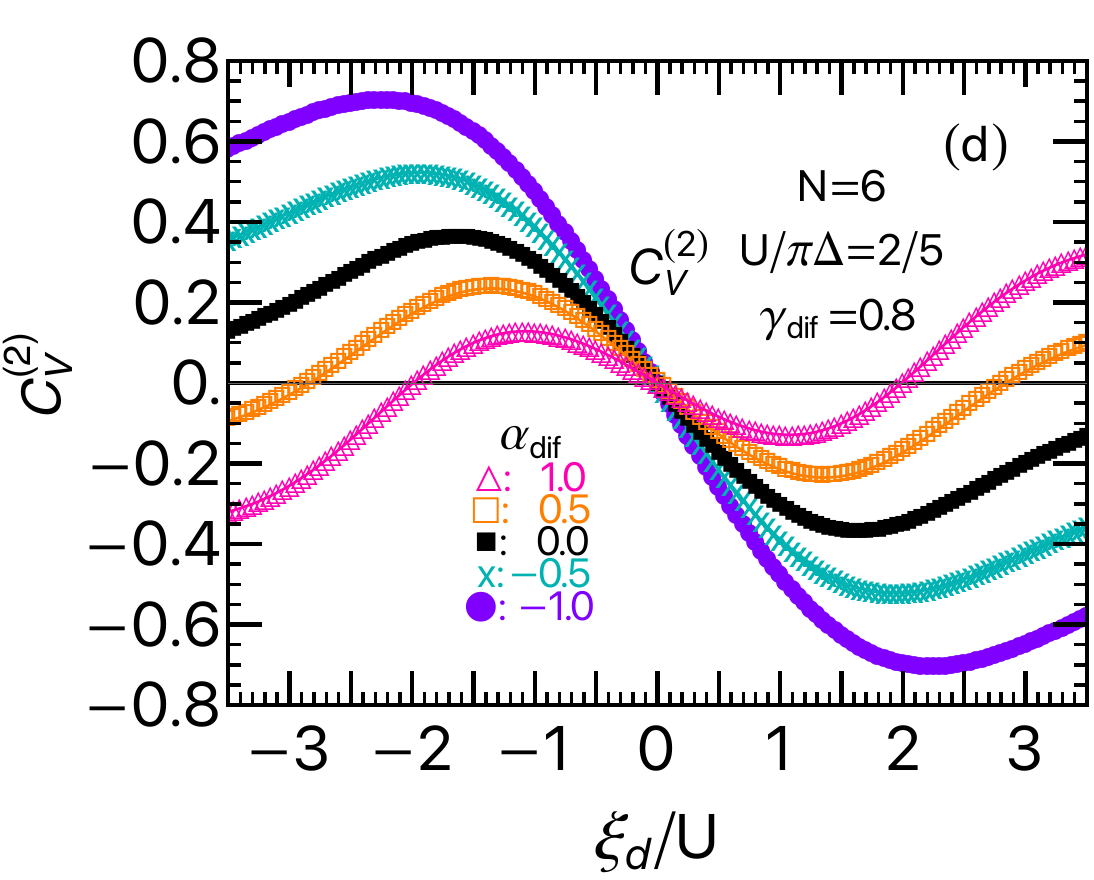}
\end{minipage}
\caption{$C_V^{(2)}$ for a junction with 
large tunnel asymmetry $\gamma_\mathrm{dif}^{}=0.8$ 
is plotted vs $\xi_d^{}/U$,  varying bias-voltage asymmetries, as       
 $\alpha_{\mathrm{dif}}=-1.0$ ($\bullet$), $-0.5$ ($\times$), 
 $0.0$ ($\blacksquare$), 
$0.5$ ($\square$), $1.0$ ($\triangle$).  
 (Left panels) for SU(4) quantum dots and (Right panels) for SU(6) ones,    
with (top panels) 
a strong interaction $U/(\pi\Delta)=5.0$  
and with (bottom panels) weak interactions 
$U/(\pi\Delta)=1/3$ for SU($4$) and $U/(\pi\Delta)=2/5$ for SU($6$). 
  }
  \label{Cv2aLvaryxL09N4N6}
\end{figure}

The coefficient  $C_V^{(2)}$ 
for a  large fixed tunnel asymmetry $\gamma_\mathrm{dif}^{} = 0.8$, 
 is plotted vs  $\xi_d^{}$ in  Fig.\ \ref{Cv2aLvaryxL09N4N6},  
varying  bias asymmetries,  
$\alpha_\mathrm{dif}^{}= 0.0$, $\pm 0.5$, $\pm 1.0$. 
The upper panels describe the behavior 
 for a strong interaction $U/(\pi\Delta)=5.0$,   
and the lower ones describe that for weak interactions, i.e.,  
$U/(\pi\Delta)=1/3$ for SU($4$),
 and $U/(\pi\Delta)=2/5$ for SU($6$).

The results  in Figs.\ \ref{Cv2aLvaryxL09N4N6} (a) and (b) show
that the coefficient  $C_V^{(2)}$ becomes almost independent of  bias asymmetries 
in the strong coupling region $|\xi_d^{}|\lesssim(N-1)U/2$ for large $U$ 
since the Wilson ratio approaches the saturated value  $\widetilde{K} \simeq 1.0$.
In this region, 
 $C_V^{(2)}$  is proportional to  $\gamma_\mathrm{dif}^{}$, 
and shows the same behavior as that in Figs.\ \ref{Cv2sym}(a) and (b). 
For $N \geq 4$,  $C_V^{(2)}$ has a peak and a dip 
which correspond to the points $\sin 2\delta \simeq \pm 1.0$ 
inside the correlated region, and thus the coefficient takes the value   
 $C_V^{(2)} \simeq \pm (\pi/4) \gamma_\mathrm{dif}^{}$  
at the extreme points. 
This kind of extreme points in the correlated region 
do not 
take place for SU(2) quantum dots 
since, for $N=2$, the phase shift takes the values $\delta =\pi/4$ and $3\pi/4$
in the valence fluctuation regime,  instead of the Kondo regime \cite{Tsutsumi2020}. 
Around the extreme points, 
 $C_V^{(2)}$  takes a typical flat structure of the Kondo state for $N=4$, 
while it takes a round structure typical to the intermediate valence state for $N=6$, 
as mentioned above.

The coefficient $C_V^{(2)}$ has an additional zero point 
other than the one at $\xi_d^{}=0$ 
 in the case   $\alpha_\mathrm{dif}^{}\gamma_\mathrm{dif}^{}>0$, 
 at which bias and tunneling asymmetries 
cooperatively enhance the charge transfer from one of the electrodes,       
and the Wilson ratio satisfies  the following condition,
\begin{align}
\widetilde{K}=\frac{\alpha_\mathrm{dif}^{}}
{\alpha_\mathrm{dif}^{}+\gamma_\mathrm{dif}^{}}\,.
\label{WilsonRatioRestrict}
\end{align}
It represents the condition that 
the first and second terms in Eq.\ \eqref{Cv2} cancel each other out. 
It takes place valence fluctuation regime  $|\xi_d^{}|\gtrsim (N-1)U/2$ 
in  Fig.\ \ref{Cv2aLvaryxL09N4N6}.  
Conversely, effects of tunnel and bias asymmetries 
become constructive for  $\alpha_\mathrm{dif}^{}\gamma_\mathrm{dif}^{}<0.0$. 
The behavior at  $|\xi_d^{}|\gg (N-1)U/2$,  is determined by 
the factor $\alpha_\mathrm{dif}^{} \sin 2\delta $.

\subsubsection{Effects of tunnel asymmetry $\gamma_\mathrm{dif}^{}\neq 0$ 
on $C_V^{(2)}$ \\
at large bias asymmetry $\alpha_\mathrm{dif}^{} = 1$}

\begin{figure}[t]
	\begin{minipage}[r]{\linewidth}
	\centering
	\includegraphics[keepaspectratio,scale=0.22]{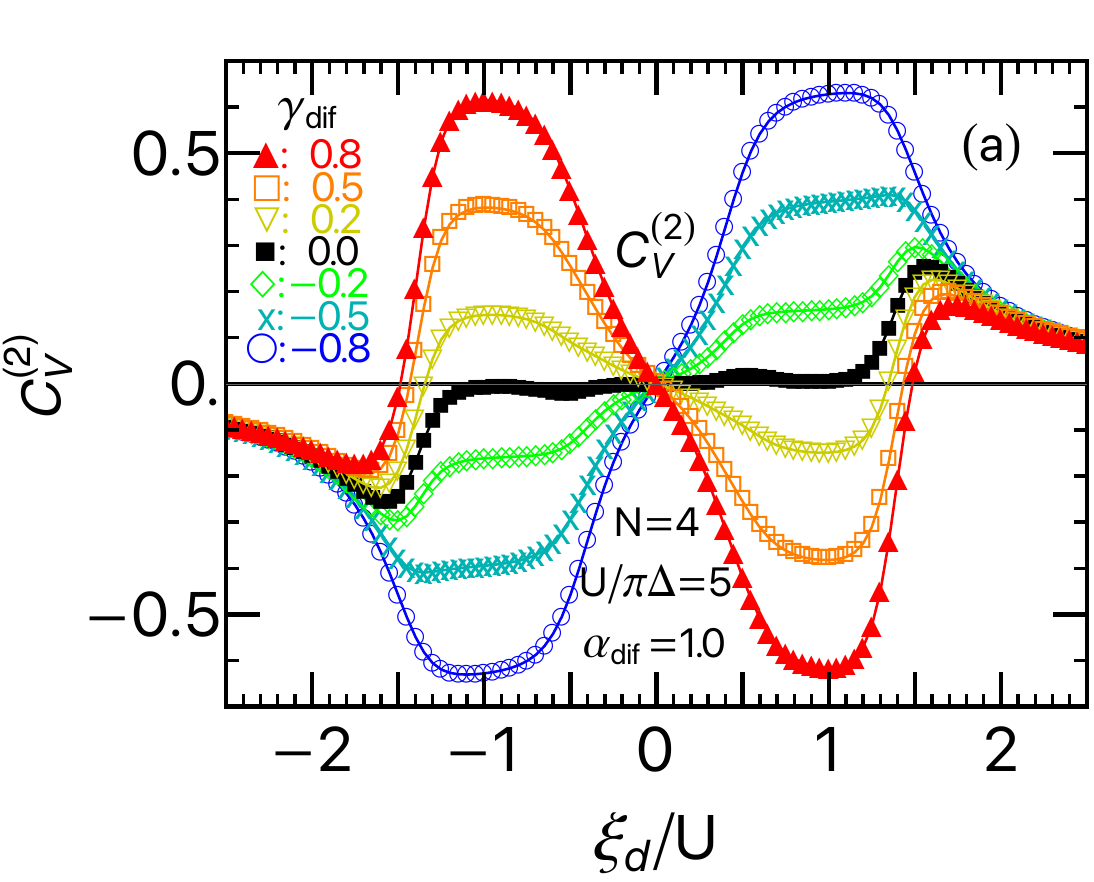}
	\centering
	\includegraphics[keepaspectratio,scale=0.22]{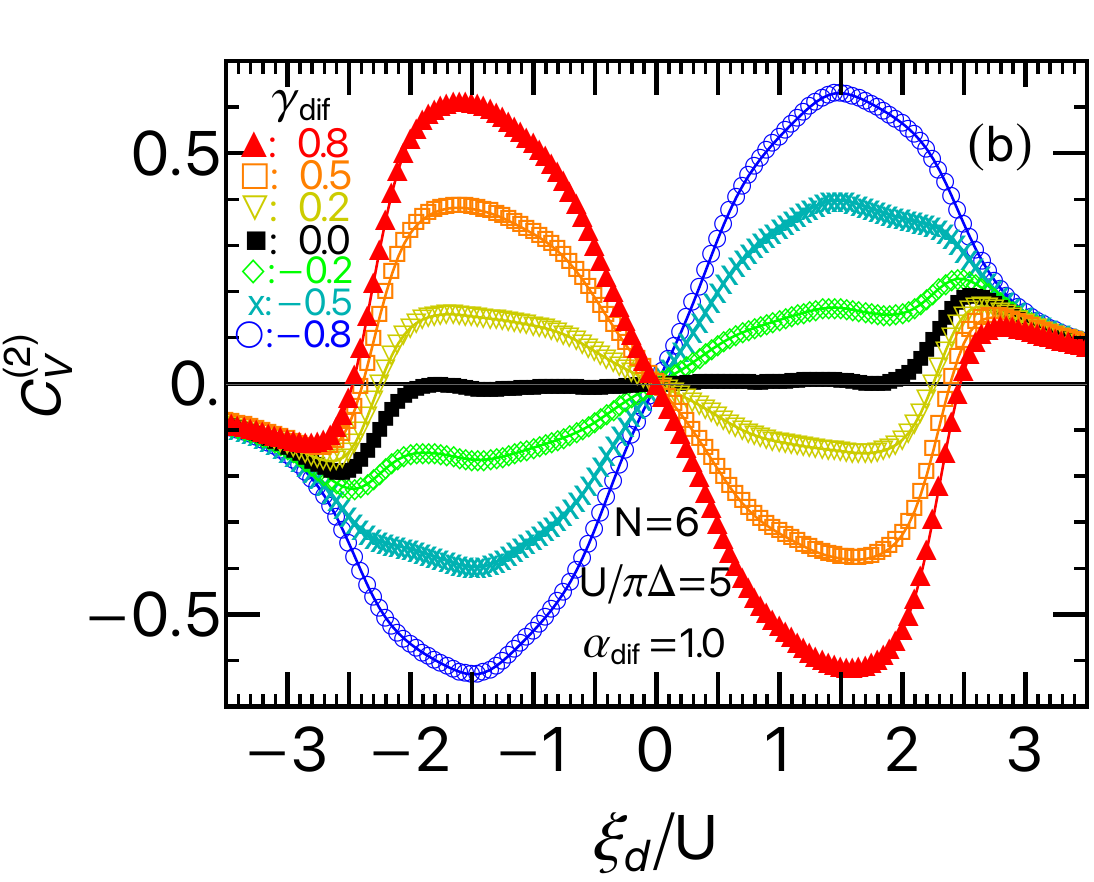}
	\end{minipage}
		\begin{minipage}[r]{\linewidth}
	\centering
	\includegraphics[keepaspectratio,scale=0.22]{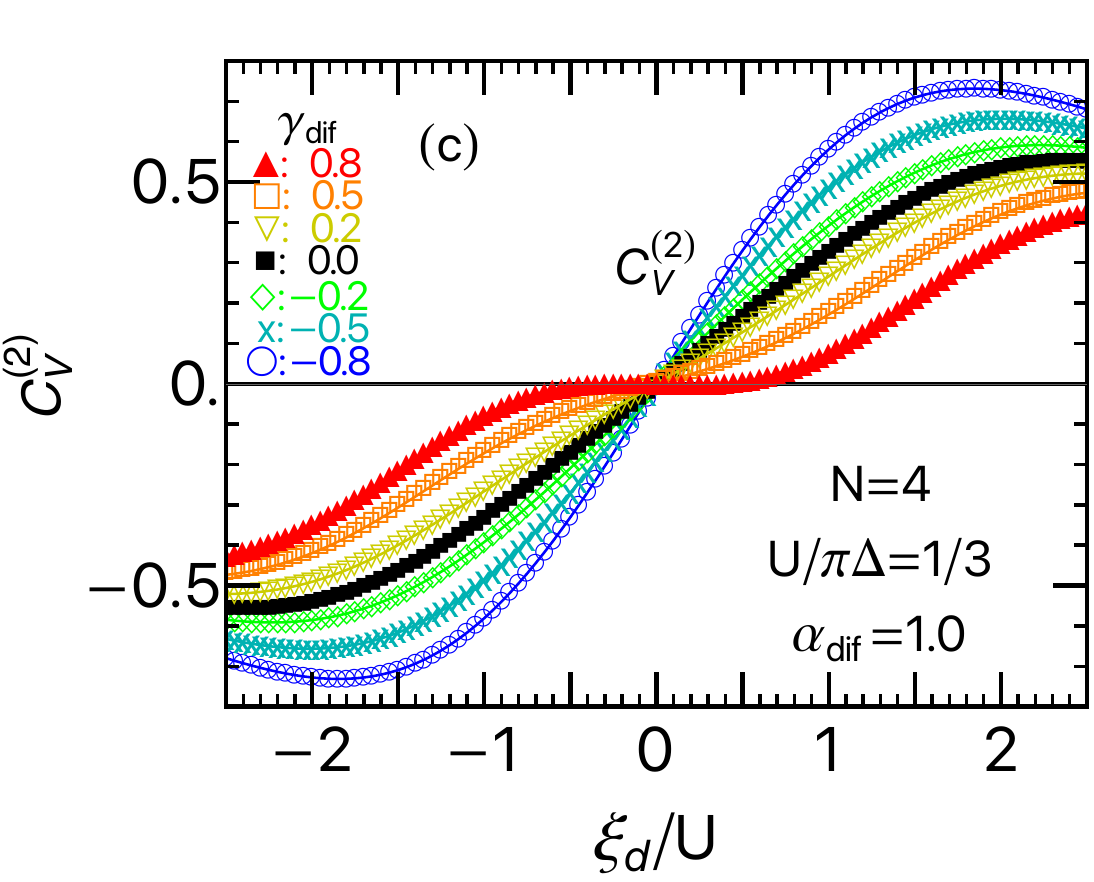}
	\centering
	\includegraphics[keepaspectratio,scale=0.22]{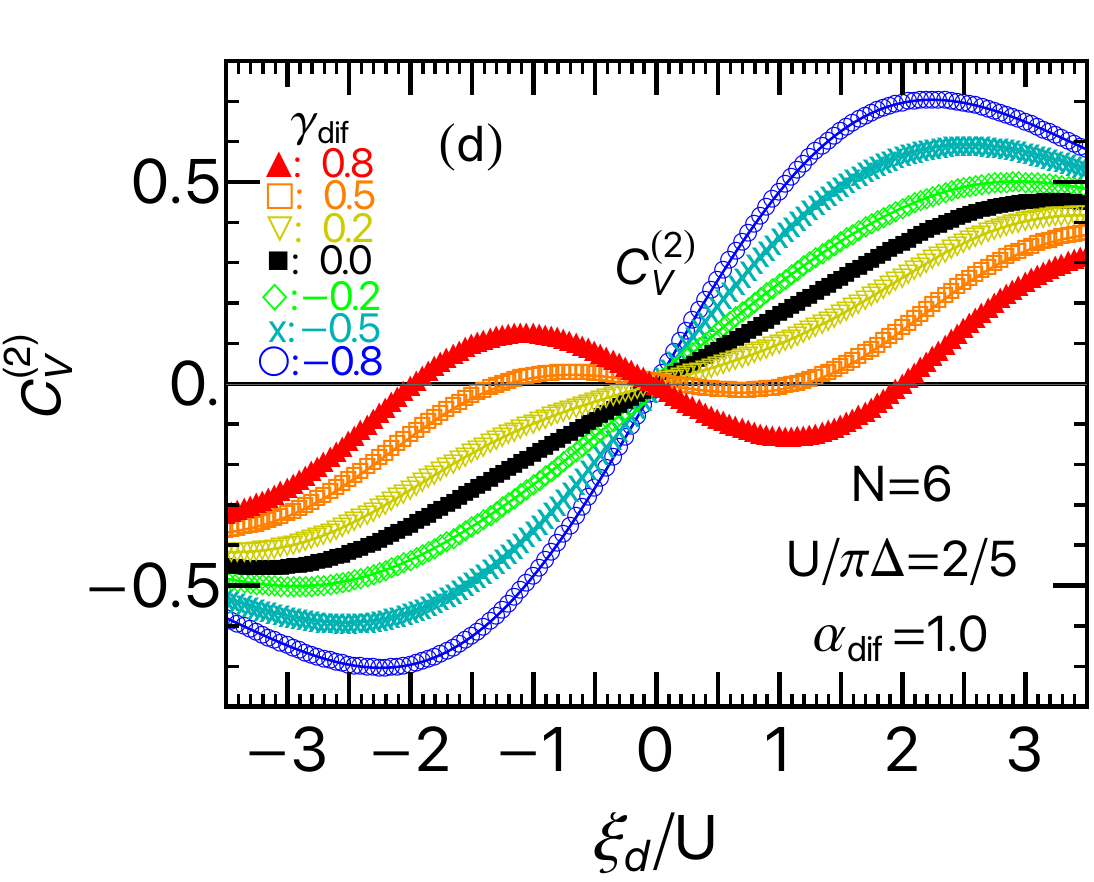}
	\end{minipage}
\caption{$C_V^{(2)}$ for a large bias-voltage asymmetry 
 $\alpha_\mathrm{dif}^{}=1.0$ is plotted vs $\xi_d^{}$, 
varying tunnel asymmetries, 
as 
 $\gamma_\mathrm{dif}^{}=-0.8$ ($\circ$),
$-0.5$ ($\times$), $-0.2$ ($\diamond$), $0.0$ ($\rule{2mm}{2mm}$),
$0.2$ ($\triangledown$), $0.5$ ($\square$) and $0.8$ ($\triangle$). 
 (Left panels) for SU(4) quantum dots and (Right panels) for SU(6) ones,    
with (top panels) a strong interaction $U/(\pi\Delta)=5.0$   
and (bottom panels) weak  interactions $U/(\pi\Delta)=1/3$ for SU($4$) 
and $U/(\pi\Delta)=2/5$ for SU($6$). 
}
  \label{Cv2asym}
\end{figure}

We next consider 
the effects of  tunnel asymmetry $\gamma_\mathrm{dif}^{}$ on  $C_V^{(2)}$ 
at large bias asymmetry $\alpha_\mathrm{dif}^{}=1.0$, 
which describes the situation that the right lead is grounded. 
Figure \ref{Cv2asym}
shows the result of $C_V^{(2)}$, plotted 
as a function  $\xi_d^{}$ varying tunnel asymmetries, 
as $\gamma_\mathrm{dif}^{}=0.0$, $\pm 0.2$, $\pm 0.5$, $\pm 0.8$: 
(top panels) for a strong $U/(\pi\Delta) =5.0$, and 
(bottom panels) for weak interactions with   
(left panels) $U/(\pi\Delta)=1/3$ for SU($4$) quantum dots 
and (right panels)  $U/(\pi\Delta)=2/5$ for SU($6$) ones.

In the strong-coupling limit region $|\xi_d|\lesssim (N-1)U/2$, 
the results in Figs.\ \ref{Cv2asym} (a) and (b) 
show almost the same behavior as that for symmetric bias voltage  
$\alpha_\mathrm{dif}^{}=0.0$ given in Figs.\ \ref{Cv2sym} 
(a) and (b), respectively, except for the ones for  $\gamma_\mathrm{dif}^{}=0.0$. 
This is because  the Wilson ratio reaches  saturated $\widetilde{K}\simeq 1.0$  
and $C_V^{(2)}$ becomes independent of 
$\alpha_\mathrm{dif}^{}$, 
 as shown in Eq.\ \eqref{Cv2Kondo}.

In the valence fluctuation region  $|\xi_d| \simeq  (N-1)U/2$,  
$C_V^{(2)}$ has an extra zero point 
other than the one at $\xi_d^{}=0$ 
 in the case at which $\alpha_\mathrm{dif}^{}\gamma_\mathrm{dif}^{}>0$, 
mentioned above at  Eq.\ \eqref{WilsonRatioRestrict}.
As the impurity level deviates further away from the electron-hole symmetric point, 
i.e., at $|\xi_d| \gtrsim  (N-1)U/2$,  
the coefficient $C_V^{(2)}$ becomes less sensitive to $\gamma_\mathrm{dif}^{}$ 
and its behavior is described by Eq.\ \eqref{eq:Cv2K=0}  
as the Wilson ratio approaches $\widetilde{K} \simeq 0.0$.   
We also see   in Fig.\ \ref{Cv2asym}(c) 
that  $C_V^{(2)}$ does not have an extra zero point 
for SU(4) symmetric quantum dots with a weak interaction.  
This is because the Wilson ratio 
in this case takes values in the range $\widetilde{K} \lesssim 0.54$ 
as shown in Fig.\ \ref{FermiparaSU4}(c) 
which does not satisfy the condition 
Eq.\ \eqref{WilsonRatioRestrict} 
as $\alpha_\mathrm{dif}^{}/(\alpha_\mathrm{dif}^{}
+\gamma_\mathrm{dif}^{}) \simeq 0.56$  
for $\alpha_\mathrm{dif}^{}=1.0$ and $\gamma_\mathrm{dif}^{}=0.8$. 
Nevertheless,   
extra zero points will  emerge even in this situation 
if tunnel asymmetries are slightly larger,  i.e.,  
  $\gamma_\mathrm{dif}^{}\gtrsim 0.85$. 
An example is shown in Fig.\ \ref{Cv2asym}(d) for SU($6$) quantum dots:
$C_V^{(2)}$ has an extra zero point clearly 
  for $\gamma_\mathrm{dif}^{}=0.8$, 
whereas  it does not for $\gamma_\mathrm{dif}^{} \lesssim 0.5$. 
In this case,  the Wilson ratio is bounded in the range of 
$\widetilde{K}\lesssim 0.72$ 
as shown in Fig.\ \ref{FermiparaSU6}(c), 
and thus the condition Eq.\ \eqref{WilsonRatioRestrict} 
is satisfied for $\gamma_\mathrm{dif}^{}=0.8$  
at which  $\alpha_\mathrm{dif}^{}/(\alpha_\mathrm{dif}^{}
+\gamma_\mathrm{dif}^{}) \simeq 0.56$, 
whereas it is not, for instance, for  $\gamma_\mathrm{dif}^{}=0.2$ 
 as  $\alpha_\mathrm{dif}^{}/(\alpha_\mathrm{dif}^{}
+\gamma_\mathrm{dif}^{}) \simeq 0.83$.

 \section{Order $(eV)^3$ nonlinear current for SU(4) and SU(6) cases}
\label{Cv3Results}

The coefficient $C_V^{(3)}$ 
of the order $(eV)^2$ term of the nonlinear conductance, 
defined  in Eqs.\ \eqref{Cv3eq}--\eqref{THv} 
has a quadratic dependence on the bias and tunnel asymmetries 
of the form,  
 $\alpha_\mathrm{dif}^{2}$, 
$\alpha_\mathrm{dif}^{}\gamma_\mathrm{dif}^{}$, 
 and $\gamma_\mathrm{dif}^{2}$. 
In particular, 
the order $\gamma_\mathrm{dif}^{2}$ term, 
which is absent in the SU(2) case \cite{Tsutsumi2021},  
emerges for multilevel quantum dots with $N \geq 3$ 
and plays an important role  
 in the Kondo states with no elelctron-hole symmetry.
In this section, we describe some limiting cases of $C_V^{(3)}$, 
and then discuss the NRG results for SU(4) and SU(6)  quantum dots.

\subsection{Behavior of $C_V^{(3)}$ in some limiting cases}

The two-body and three-body parts of 
$C_V^{(3)}$ defined in Eqs.\ \eqref{Cv3eq}--\eqref{THv} 
takes the following values 
 in the strong-coupling case 
where the Wilson ratio reaches the saturated value $\widetilde{K}\to 1.0$ 
and the three-body correlations functions show the properties 
$\Theta_\mathrm{I}\simeq-\widetilde{\Theta}_\mathrm{II}
\simeq\widetilde{\Theta}_\mathrm{III}$ 
shown in Figs.\ \ref{FermiparaSU4}--\ref{TH123SU6}, 
\begin{align}
W_V&\xrightarrow{\,\widetilde{K}\to1\,}
\,-\cos2\delta\left[1+\frac{5}{N-1}
+\frac{3(N-2)}{N-1}\gamma_\mathrm{dif}^{2}\right]\,, 
\label{WvStrongLimit}
\\
\Theta_V&\xrightarrow{\,\Theta_\mathrm{I}\simeq
-\widetilde{\Theta}_\mathrm{II}\simeq\widetilde{\Theta}_\mathrm{III}\,}
\,-2\,\Bigl[1-3\gamma_\mathrm{dif}^{2}\Bigr]\,\Theta_\mathrm{I}\,.  
\label{THvStrongLimit}
\end{align}
Our result in this limit  is consistent with the corresponding result 
 for the SU($N$) Kondo model, given 
 in Eqs.\ (22) and (32) of Ref.\ \onlinecite{MoraEtal2009}. 
Note that their notation and our one in the strong-interaction limit 
 correspond to each other such that 
$
\alpha_{1}^{}/(\pi T_K^{}) \Leftrightarrow    
  \chi_{\sigma\sigma}^{}$, and 
$\alpha_{2}^{}/(\pi T_K^2) 
 \Leftrightarrow   -\frac{1}{2}\, \chi_{\sigma\sigma\sigma}^{[3]}$.

In the limit of  $|\xi_d^{}|\to\infty$, 
the occupation number reaches $\langle n_d\rangle\to 0$ or $N$, 
and the correlation functions take the noninteracting values, 
and thus  
\begin{align}
W_V&\xrightarrow{\,|\xi_d^{}|\to\infty\,}\,
-\bigl(1+3\alpha_\mathrm{dif}^{2}\bigr)\,, \label{WvNonInt}
\\
\Theta_V&\xrightarrow{\,|\xi_d^{}|\to\infty\,}\,
-2\bigl(1+3\alpha_\mathrm{dif}^{2}\bigr)\,,\label{THvNonInt}
\\
C_V^{(3)}&\xrightarrow{\,|\xi_d^{}|\to\infty\,}\,
-\frac{3\pi^2}{64}\,\bigl(1+3\alpha_\mathrm{dif}^{2}\bigr)\,. \label{Cv3NonInt}
\end{align}
Note that in this limit, the three-body correlations are given by  
$\Theta_\mathrm{I} \to  -2$, 
$\widetilde{\Theta}_\mathrm{II}\to 0$,
and  $\widetilde{\Theta}_\mathrm{III}\to 0$  
 [see Appendix B in Ref.\ \onlinecite{Tsutsumi2021}].

\begin{figure}[t]
	\begin{minipage}[r]{0.49\linewidth}
	\centering
	\includegraphics[keepaspectratio,scale=0.23]{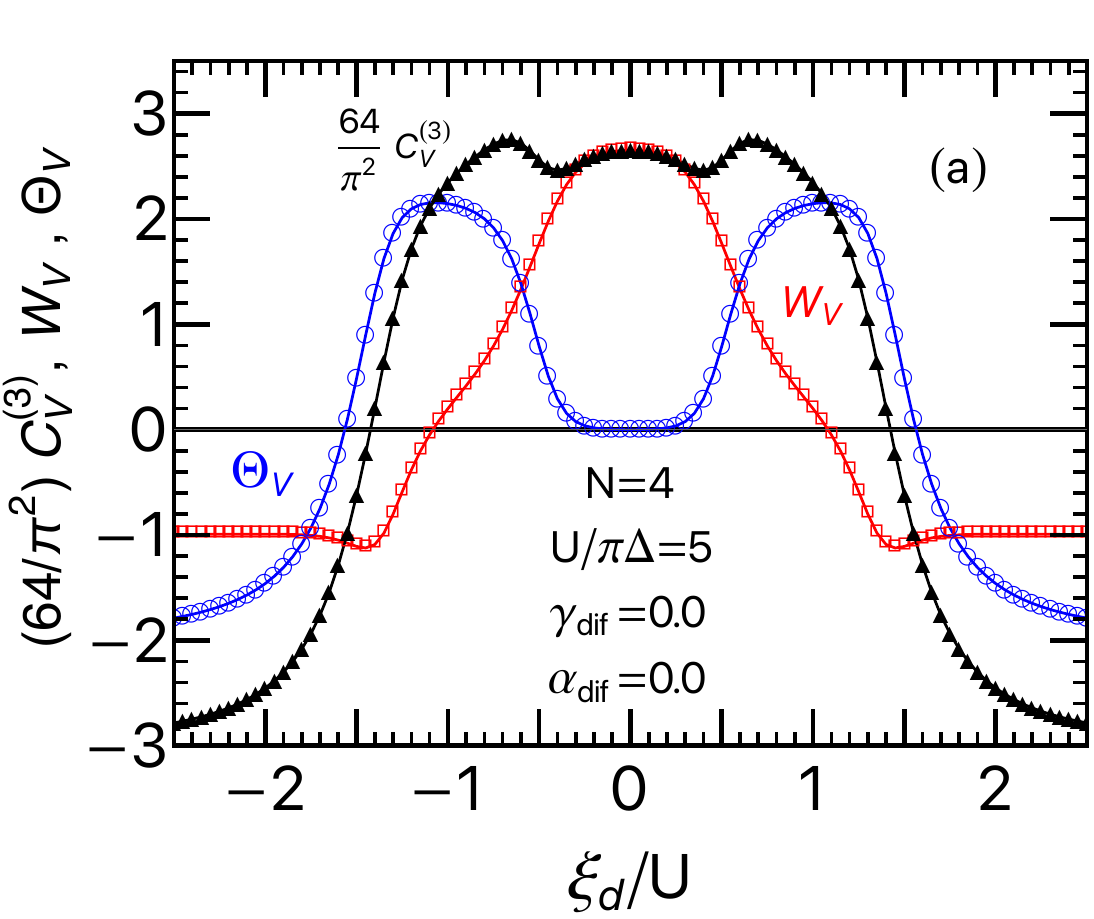}
	\end{minipage}
	\begin{minipage}[r]{0.49\linewidth}
	\centering
	\includegraphics[keepaspectratio,scale=0.23]{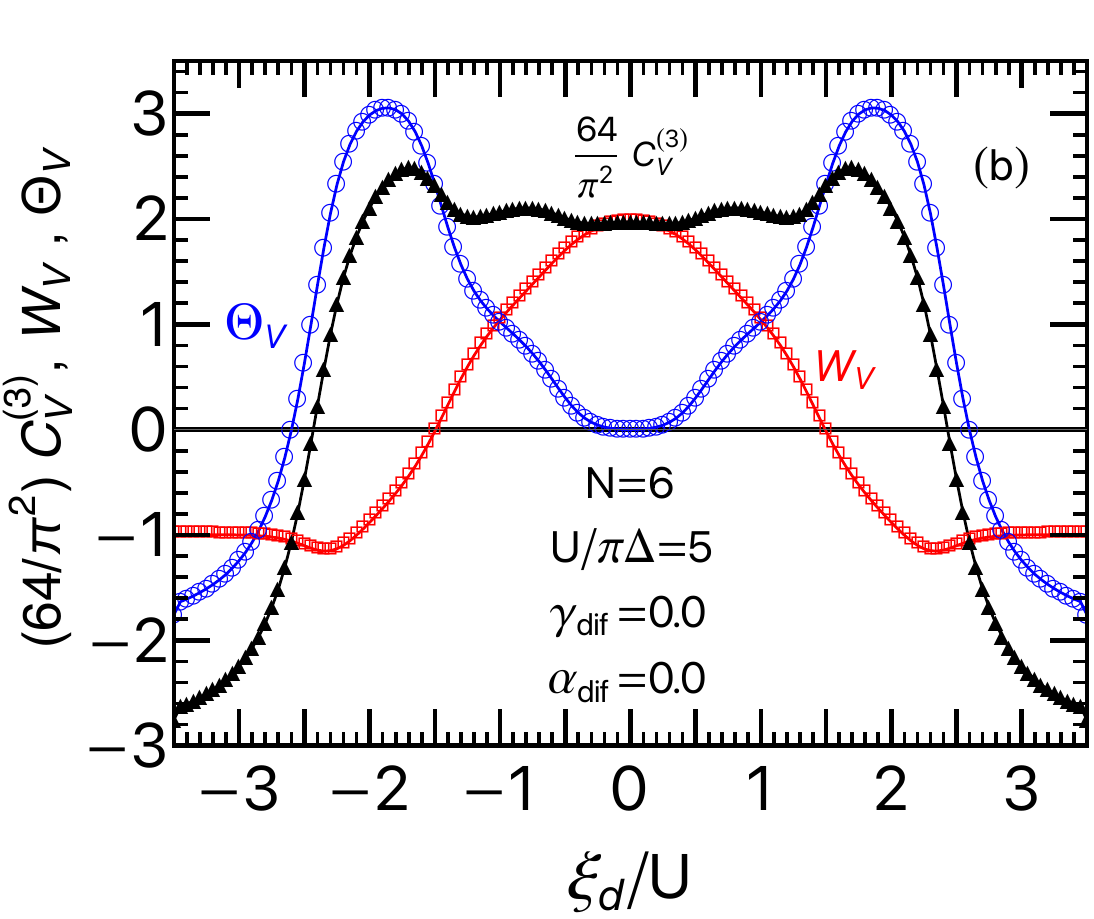}
	\end{minipage}
	\begin{minipage}[r]{0.49\linewidth}
	\centering
	\includegraphics[keepaspectratio,scale=0.23]{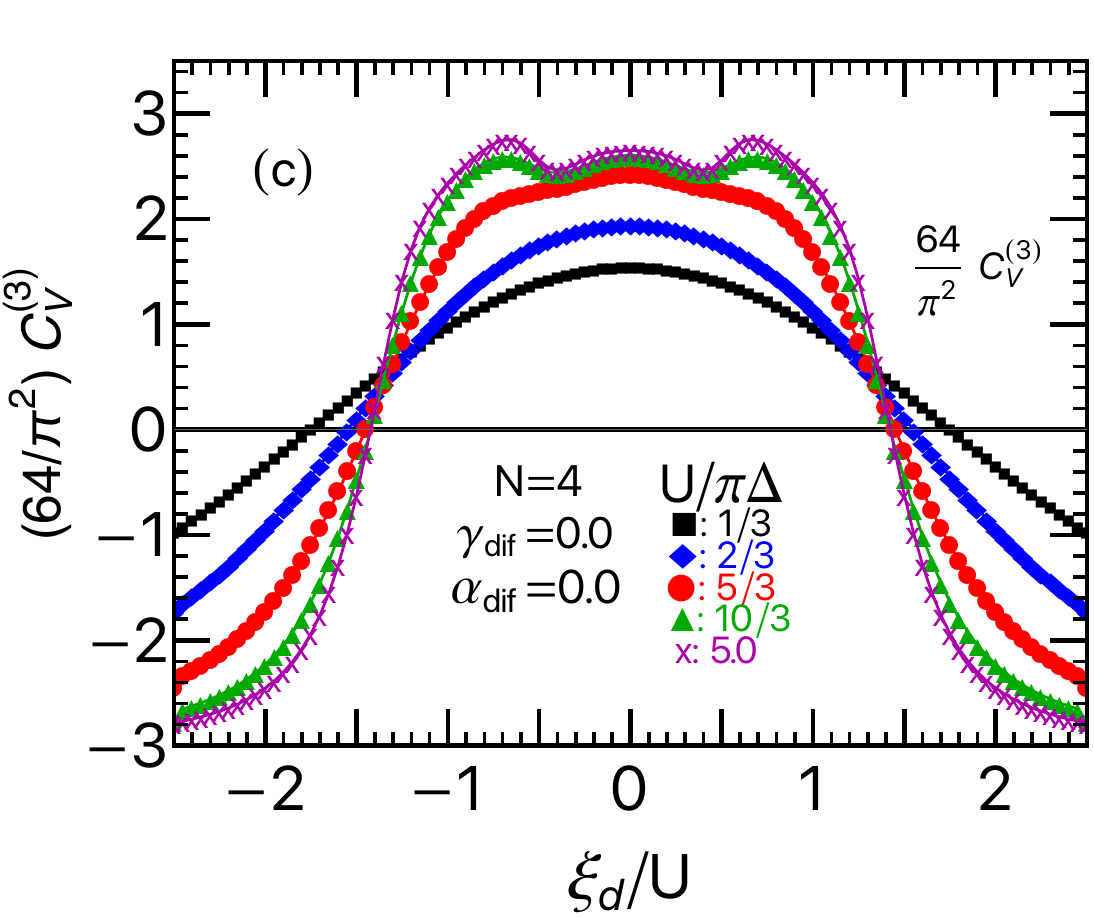}
	\end{minipage}
	\begin{minipage}[r]{0.49\linewidth}
	\centering
	\includegraphics[keepaspectratio,scale=0.23]{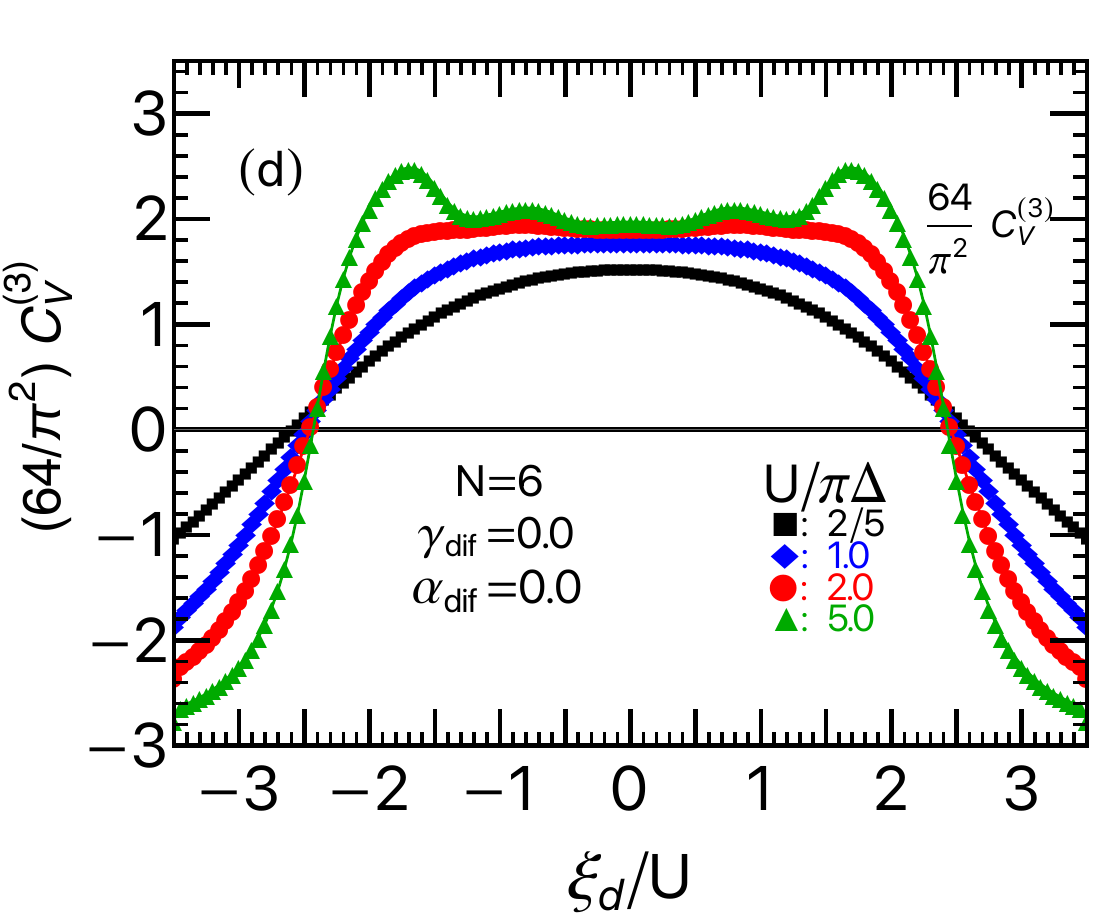}
	\end{minipage}
\caption{ Coefficient $C_V^{(3)}$ 
is plotted vs $\xi_d/U$ 
for the case where both tunnel couplings and bias voltages are symmetric,
$\gamma_\mathrm{dif}^{}=0.0$ and $\alpha_\mathrm{dif}^{}=0.0$.    
Upper panels show the total value of $C_V^{(3)}=(\pi^2/64)(W_V+\Theta_V)$ 
together with the two-body $W_V$ and three-body $\Theta_V$ components,  
(a) for SU(4) and (b) for SU(6) quantum dots 
choosing a large interaction $U/(\pi\Delta) =5.0$. 
Bottom panels show $U$ dependence of $C_V^{(3)}$:    
(c) $U/(\pi\Delta)=1/3$, $2/3$, $5/3$, $10/3$, $5.0$ for SU(4),
and (d) $U/(\pi\Delta)=2/5$, $1.0$, $2.0$,$5.0$ for SU(6).
}
 \label{Cv3ALSymXLSymN4N6}
\end{figure}

\subsection{$C_V^{(3)}$ for symmetric tunnel coupling and 
symmetric bias voltage $\gamma_\mathrm{dif}^{}=\alpha_\mathrm{dif}^{}=0$}

We describe here previous results 
obtained for  symmetric tunnel coupling 
and bias voltage $\gamma_\mathrm{dif}^{}=\alpha_\mathrm{dif}^{}=0$ 
\cite{AO2017_III}
in order to clarify how the breaking of these symmetries affects 
the $C_V^{(3)}$.

Figures \ref{Cv3ALSymXLSymN4N6}(a) and (b) show that $C_V^{(3)}$, 
$W_V$ and $\Theta_V$ 
for $\alpha_\mathrm{dif}^{}=\gamma_\mathrm{dif}^{}=0.0$  
 as functions of the gate voltage $\xi_d^{}$ 
for strong interaction $U/(\pi\Delta)=5.0$ 
 for SU(4) and  SU(6) quantum dots. 
In the strong-coupling limit region $|\xi_d|\lesssim (N-2)U/2$, $C_V^{(3)}$ 
takes plateau structures of the height  $(64/\pi^2)\,C_V^{(3)}\simeq 2.67$ 
 for SU(4),  and $2.0$ for SU(6). 
In particular, the plateau at half filling, i.e., at  $\xi_d^{}=0$, 
is caused by the two-body correlation $W_V$.
In contrast,   the plateau around the Kondo state with 
the fillings of   $\langle n_d^{} \rangle \simeq 1.0$ and $N-1$, 
i.e., corresponding to  the one electron and one hole occupancies, 
  are caused by  the three-body correlation  $\Theta_V^{}$.
Specifically, among three independent components defined  
in Eqs.\ \eqref{eq:ThetaI_def}--\eqref{eq:ThetaIII_def}, $\widetilde{\Theta}_\mathrm{II}$ 
determines the peak structure of $\Theta_V$ 
since 
$\Theta_V \simeq-2\widetilde{\Theta}_\mathrm{II}$ owing to the property  
$\Theta_\mathrm{I}\simeq-\Theta_\mathrm{II}$  in the strongly correlated region. 
In the limit of $|\xi_d^{}|\to\infty$  outside the plateau, 
the coefficient approaches the noninteracting value 
 $(64/\pi^2)\,C_V^{(3)}\to-3$, $W_V \to -1.0$, and $\Theta_V \to -2.0$,  
given by Eqs.\ \eqref{WvNonInt}--\eqref{Cv3NonInt}.

In Figs.\ \ref{Cv3ALSymXLSymN4N6}(c) and (d), 
$C_V^{(3)}$ for $\alpha_\mathrm{dif}^{}=\gamma_\mathrm{dif}^{}=0.0$ 
is plotted as a function of $\xi_d$, 
varying interaction strengths (c)  $U/(\pi\Delta)=1/3$, $2/3$, $5/3$, $10/3$, $5.0$ 
for SU(4), and (d) $U/(\pi\Delta)=2/5$, $1.0$, $2.0$, $5.0$ for SU(6). 
The plateau structure and the peak at the edge of the plateau of $C_V^{(3)}$ as shown above appear as $U$ increases in the strong-coupling limit $|\xi_d|\lesssim (N-2)U/2$.

\subsection{Effects of tunnel asymmetries  $\gamma_\mathrm{dif}^{}\neq 0$
on $C_V^{(3)}$ for symmetric bias voltages $\alpha_\mathrm{dif}^{}= 0$}

We next  consider the effect of tunnel asymmetry 
on $C_V^{(3)}$, setting  bias voltages to be symmetric  
$\alpha_\mathrm{dif}^{}=0$.  
In this case,  among the three types of 
quadratic terms  $\alpha_\mathrm{dif}^{2}$, 
$\alpha_\mathrm{dif}^{}\gamma_\mathrm{dif}^{}$, 
 and $\gamma_\mathrm{dif}^{2}$ 
in Eqs.\ \eqref{Cv3eq}--\eqref{THv},  
the only $\gamma_\mathrm{dif}^{2}$ term 
remains finite  and  contributes to $C_V^{(3)}$: 
\begin{align}
W_V&\xrightarrow{\alpha_\mathrm{dif}^{}=0\,}\,
-\cos2\delta\biggl[1+\biggl\{\frac{5}{N-1}+\frac{3(N-2)}{N-1}\,
\gamma_\mathrm{dif}^{2}\biggr\}\widetilde{K}^2 \biggr],
\label{WvBiasSymEq}
\\
\Theta_V&\xrightarrow{\alpha_\mathrm{dif}^{}=0\,}\,\Theta_{\mathrm{I}}
\,+3\,\widetilde{\Theta}_{\mathrm{II}} 
\,+6\,\gamma_\mathrm{dif}^{2}\,\widetilde{\Theta}_{\mathrm{III}} \,.
\label{THvBiasSymEq}
\end{align}
The  $\gamma_\mathrm{dif}^{2}$ term emerges for $N\geq 3$.   
In particular, it couples to  $\widetilde{\Theta}_\mathrm{III}$,  
 the three-body correlation between electrons in three different levels 
which does not contribute to $C_V^{(3)}$ for symmetric tunnel couplings.

In the strongly correlated region for large $U$,  
the three-body part of $C_V^{(3)}$ takes a simplified form 
$\Theta_V \propto 1-3\gamma_\mathrm{dif}^{2}$  
which does not depend on $\alpha_\mathrm{dif}^{}$, 
as shown in  Eq.\ \eqref{THvStrongLimit}. 
Therefore, 
$\Theta_V$  decreases as tunnel asymmetry 
which enters through $\gamma_\mathrm{dif}^{2}$  
increases,  and vanishes $\Theta_V\simeq 0.0$ 
at  $\gamma_\mathrm{dif}^{}=\pm 1/\sqrt{3}$ ($\simeq \pm 0.577$). 
In particular,
$\widetilde{\Theta}_\mathrm{II}$ has 
a wide  plateau at the fillings $\langle n_d^{} \rangle \simeq 1$ and $N-1$ 
 for large $U$ in Fig.\ \ref{TH123SU4}(b) and Fig.\ \ref{TH123SU6}(b), 
the contribution of the three-body correlations $\Theta_V$ 
becomes significant at these fillings.

\begin{figure}[t]
	\begin{minipage}[t]{0.49\linewidth}
	\centering
	\includegraphics[keepaspectratio,scale=0.23]{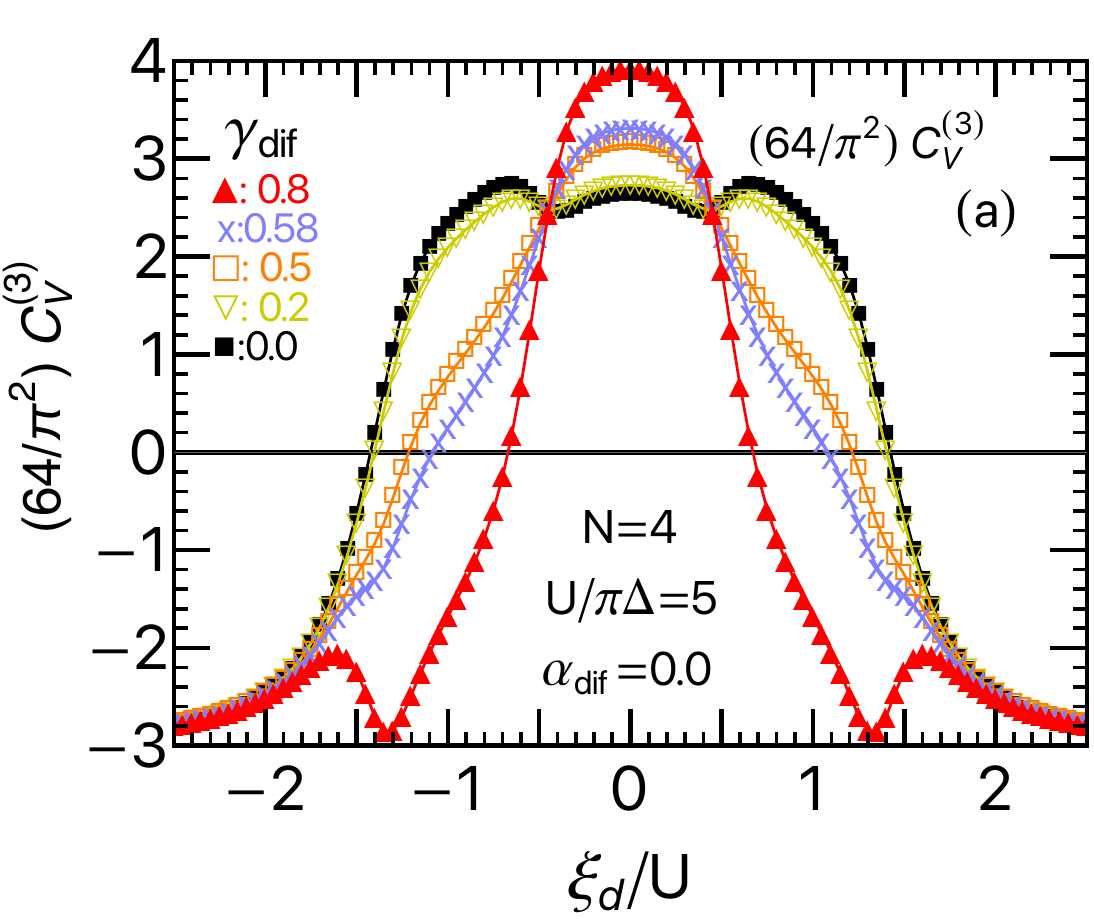}
\end{minipage}
\begin{minipage}[t]{0.49\linewidth}
	\centering
	\includegraphics[keepaspectratio,scale=0.23]{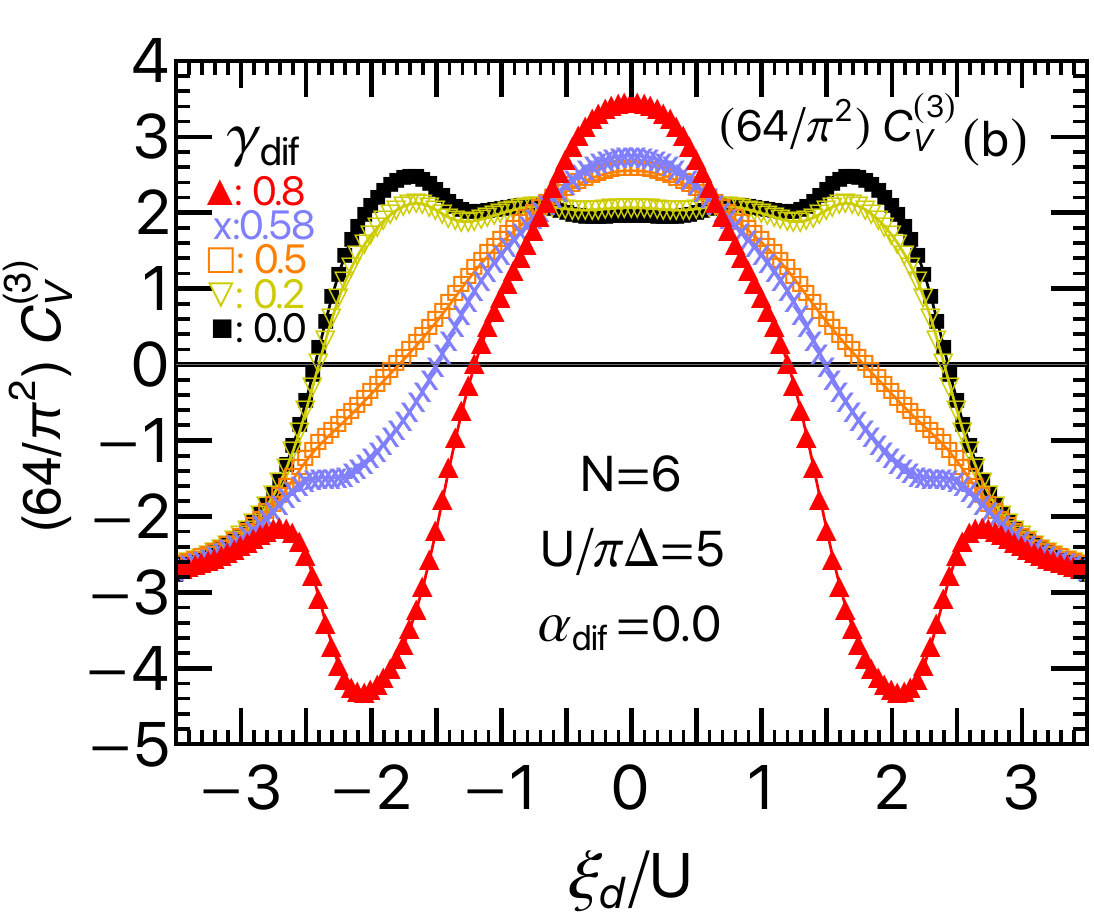}
\end{minipage}\\
\begin{minipage}[t]{0.49\linewidth}
	\centering
	\includegraphics[keepaspectratio,scale=0.23]{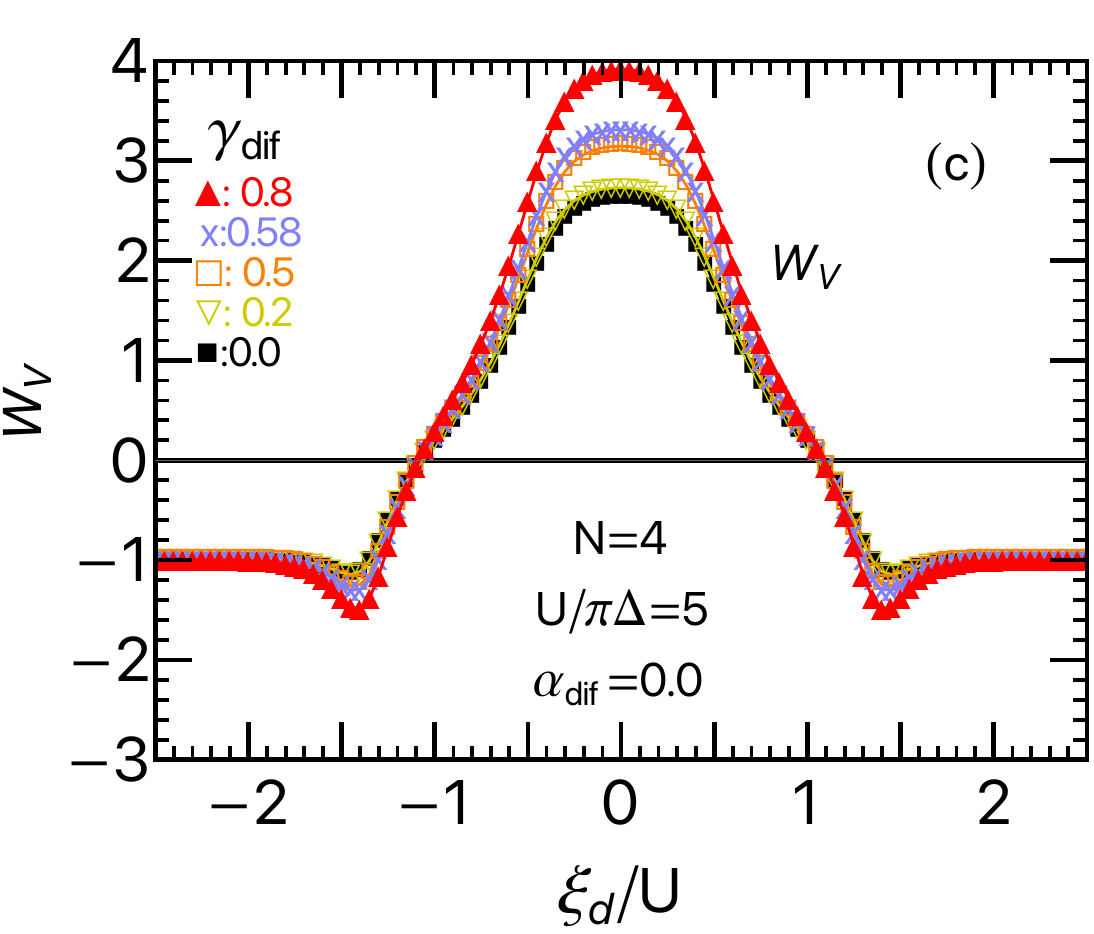}
	\end{minipage}
	\begin{minipage}[t]{0.49\linewidth}
	\centering
	\includegraphics[keepaspectratio,scale=0.23]{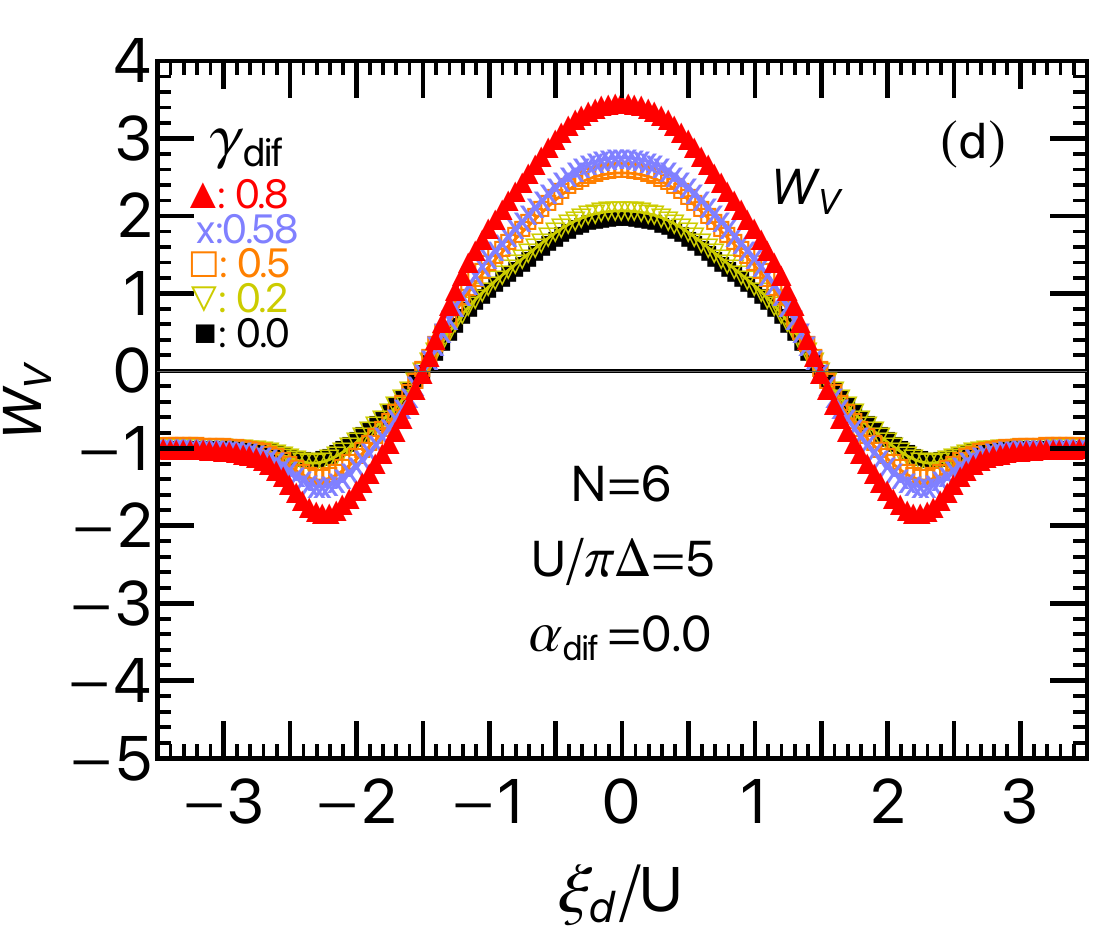}
	\end{minipage}\\
	\begin{minipage}[t]{0.49\linewidth}
	\centering
	\includegraphics[keepaspectratio,scale=0.23]{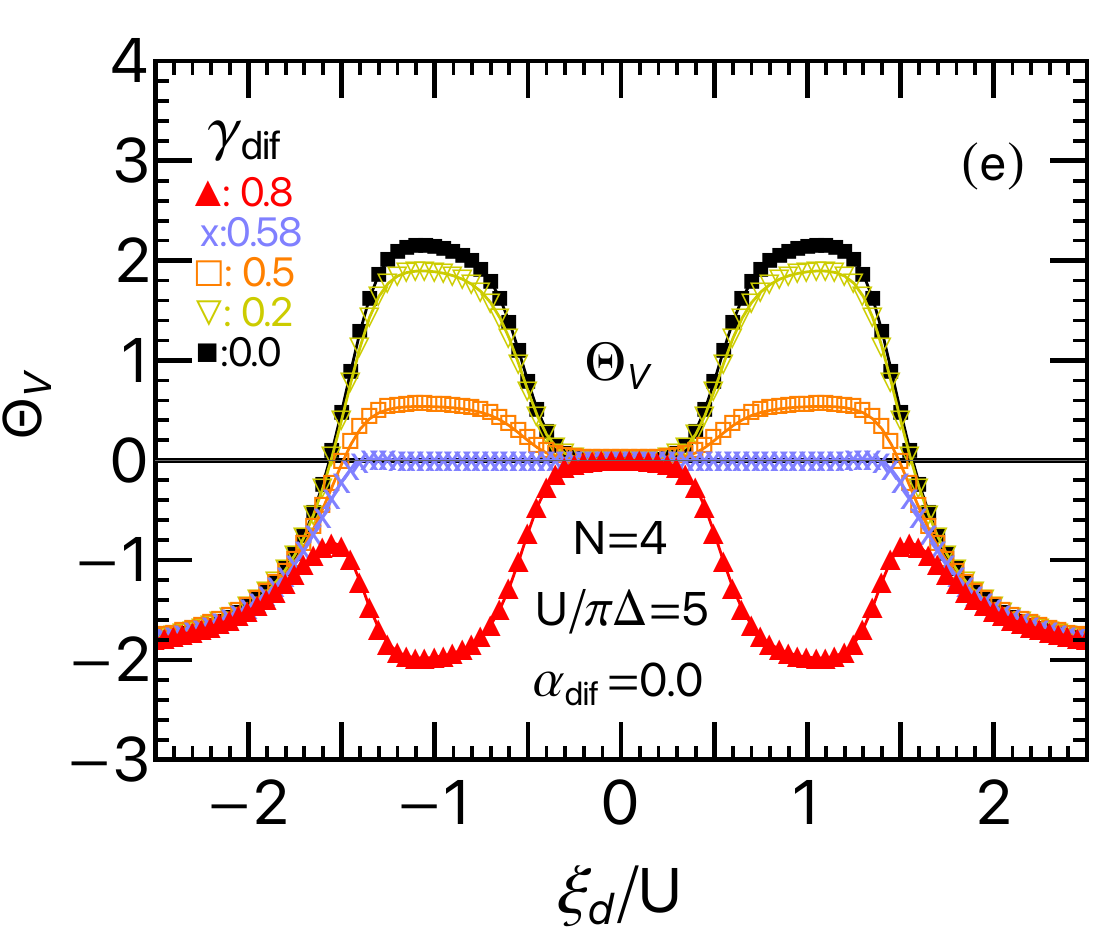}
	\end{minipage}
	\begin{minipage}[t]{0.49\linewidth}
	\centering
	\includegraphics[keepaspectratio,scale=0.23]{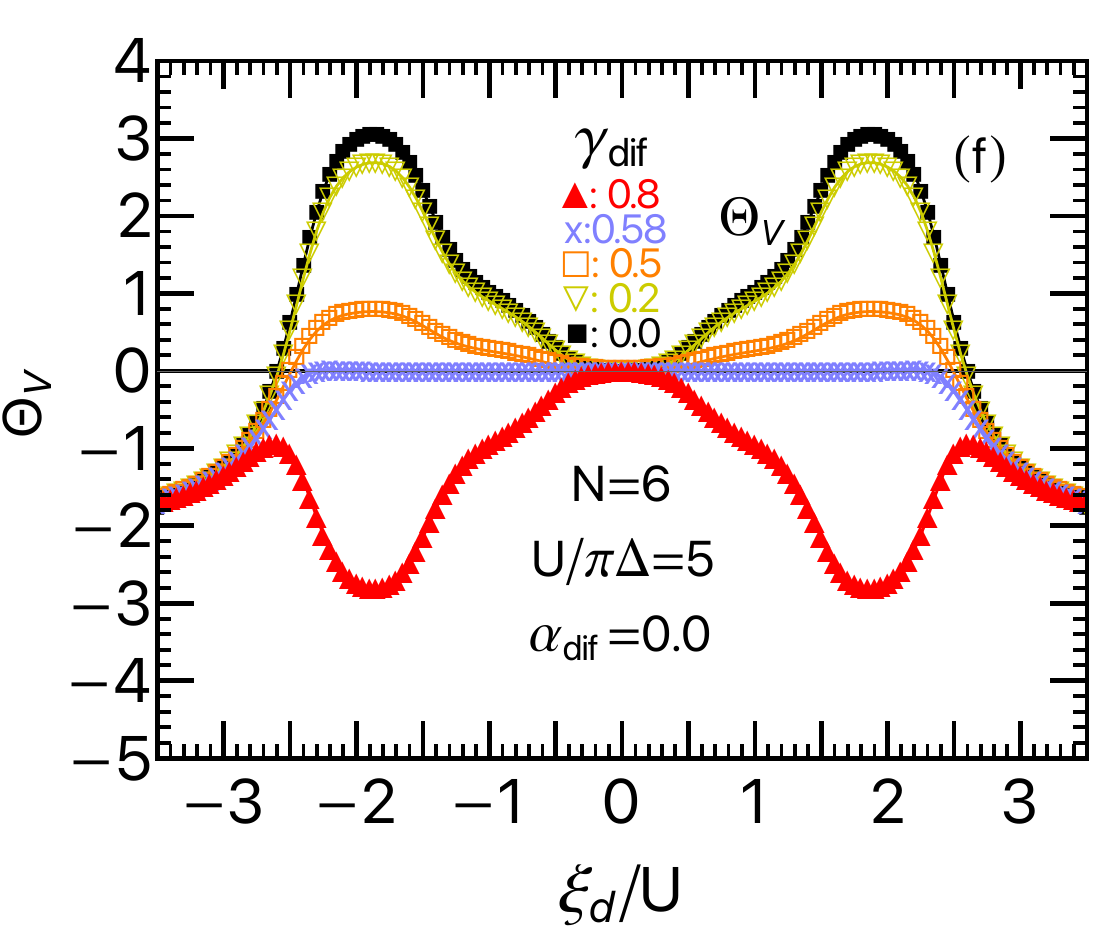}
	\end{minipage}
	\begin{minipage}[t]{0.49\linewidth}
	\centering
	\includegraphics[keepaspectratio,scale=0.23]{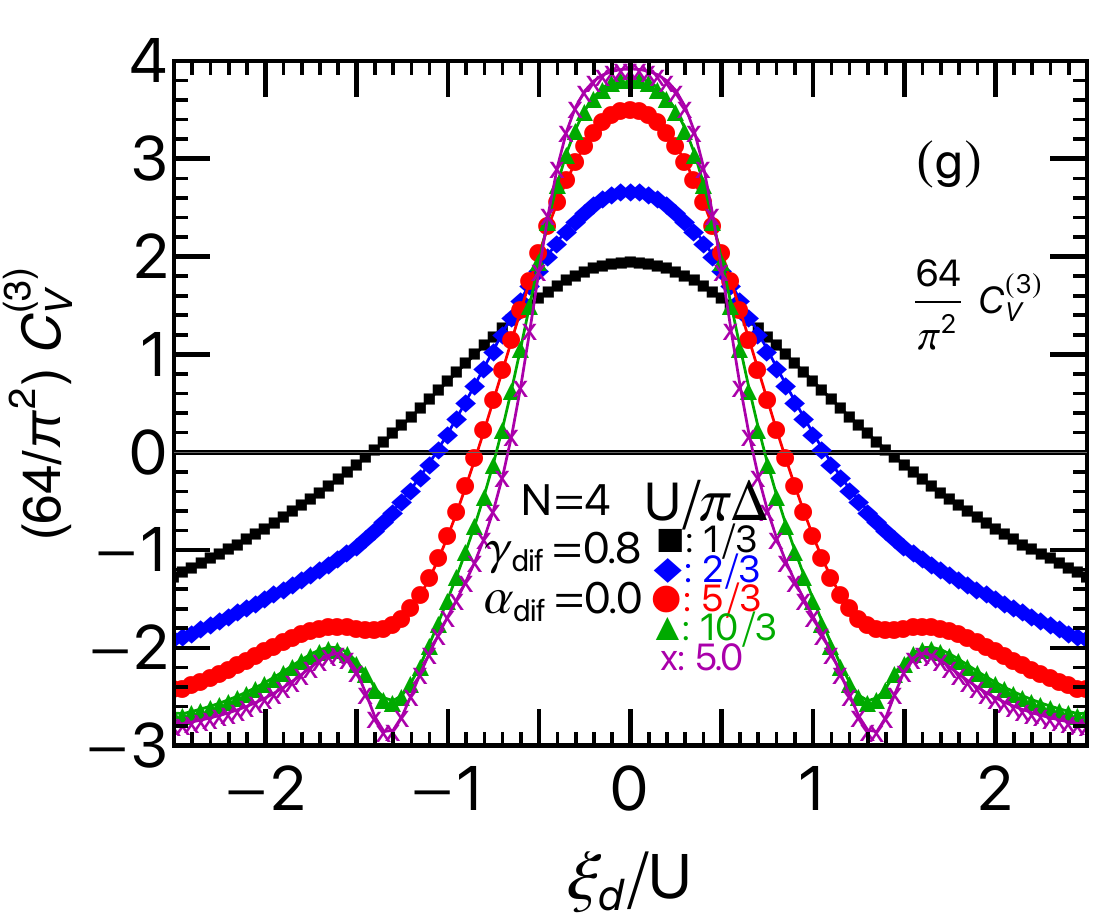}
	\end{minipage}
	\begin{minipage}[t]{0.49\linewidth}
	\centering
	\includegraphics[keepaspectratio,scale=0.23]{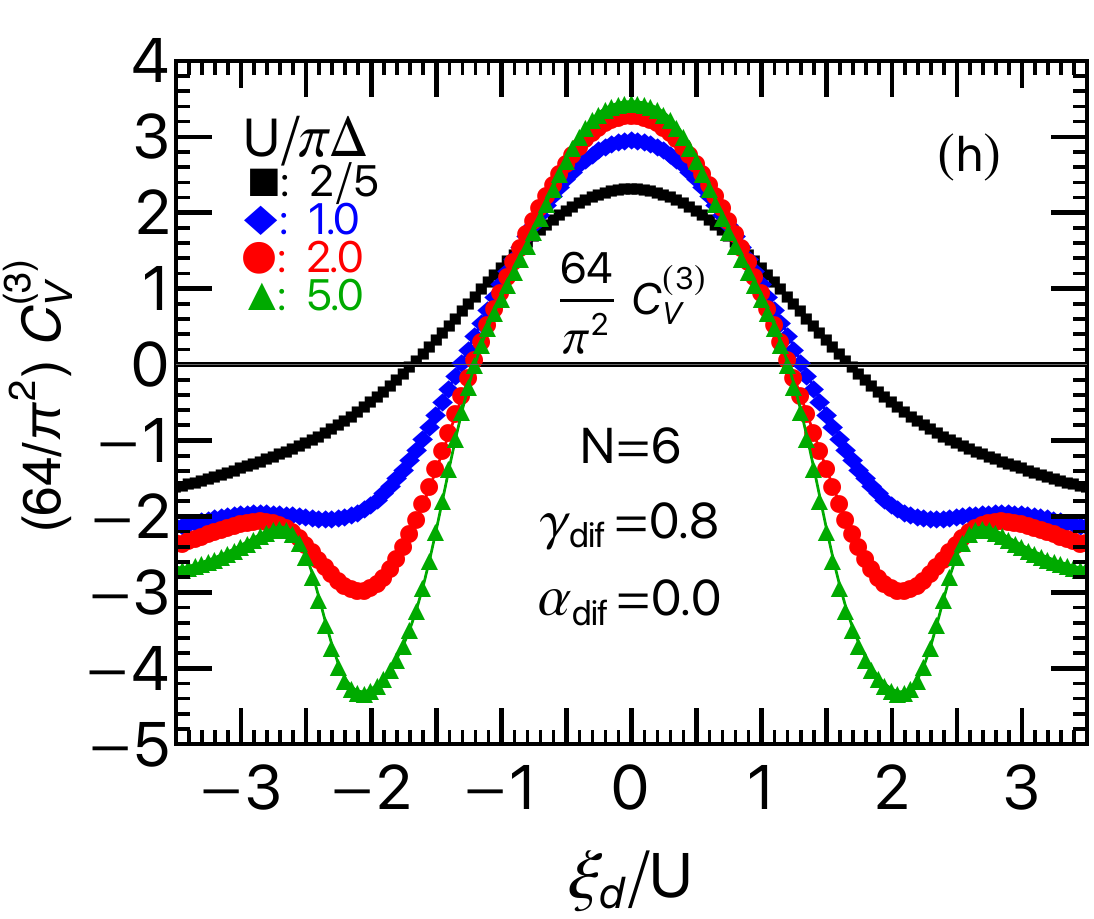}
	\end{minipage}
\caption{Effects of 
tunnel asymmetry $\gamma_\mathrm{dif}^{}$ on order $(eV)^3$ 
nonlinear current: (Top panels) $C_V^{(3)}$, and 
the components (Upper middle panels) $W_V$ and (Lower middle panels) $\Theta_V$ 
defined in  Eqs.\ \eqref{WvBiasSymEq} and \eqref{THvBiasSymEq}
for symmetrical bias voltages  $\alpha_\mathrm{dif}^{}=0.0$ 
are plotted vs $\xi_d^{}/U$, varying tunnel asymmetries, as    
for $\gamma_\mathrm{dif}=0.0$ ($\rule{2mm}{2mm}$),
$0.2$ ($\triangledown$), $0.5$ ($\square$), $0.58$ ($\times$), 
$0.8$ ($\triangle$). 
Interaction strength is chosen to be $U/(\pi\Delta)=5.0$  for (a)--(f):  
  (Left panels)  for SU(4) quantum dots, and (Right panels) for SU(6) quantum dots. 
Bottom panels: $U$ dependence of $C_V^{(3)}$, 
calculated for a fixed large tunnel asymmetry $\gamma_\mathrm{dif}^{}=0.8$,  
(g) $U/(\pi\Delta)=1/3$ ($\rule{2mm}{2mm}$),
$2/3$ ($\diamond$), $5/3$ ($\circ$), $10/3$ ($\triangle$), 
$5.0$ ($\times$) for SU(4), 
and (h) $U/(\pi\Delta)=2/5$ ($\rule{2mm}{2mm}$), 
$1.0$ ($\diamond$), $2.0$ ($\circ$), $5.0$ ($\triangle$) for SU(6). 
}
\label{Cv3BiasSymN4N6}
\end{figure}

The NRG results of  $C_V^{(3)}$,  $W_V$, and  $\Theta_V$, 
for the  bias symmetric case $\alpha_\mathrm{dif}=0.0$ are plotted vs $\xi_d^{}$
 in  the top, upper-middle, and lower-middle 
 panels of Fig.\ \ref{Cv3BiasSymN4N6}, 
for (left panels) SU(4)  and (right panels) SU(6) quantum dots,   
choosing interaction strength to be $U/(\pi\Delta)=5.0$. 
Specifically, Fig.\ \ref{Cv3BiasSymN4N6} (a)--(f)  
are obtained, varying tunnel asymmetries, as   
 $\gamma_\mathrm{dif}=0.0$, $0.2$, $0.5$, $0.58$ and $0.8$. 
The plateau of $C_V^{(3)}$ emerging at 
 $|\xi_d|\lesssim U/2$ is due to 
the half-filling Kondo state  $\langle n_d^{} \rangle \simeq N/2$.
The height of this plateau 
increases  with $\gamma_\mathrm{dif}^{}$, 
and approaches the upper bound 
$(64/\pi^2)\,C_V^{(3)}\to 4+2/(N-1) $ 
in the limit 
 $\gamma_\mathrm{dif}^{2} \to 1$. 
This structure is determined by the two-body part $W_V$ 
as the three-body part $\Theta_V$ vanishes 
around the elelctron-hole symmetric point $\xi_d^{}= 0$. 

We see that the plateau structure of $C_V^{(3)}$ 
at the Kondo states away from half filling 
depend sensitively on tunnel asymmetry. 
The positive plateau  emerging at $|\xi_d^{}|\simeq (N-2)U/2$ 
are due to the Kondo state of the filling  $\langle n_d\rangle\simeq 1$ and $N-1$. 
It disappears as  $\gamma_\mathrm{dif}^{}$ increases, 
and a negative dip develops for  $\gamma_\mathrm{dif}^{2} \geq 1/3$. 
This behavior of $C_V^{(3)}$ 
reflects the evolution of the three-body part $\Theta_V$, 
which is determined by Eq.\ \eqref{THvStrongLimit} in the strongly correlated region. 
Therefore, 
if $C_V^{(3)}$ is measured varying tunneling 
asymmetries $\gamma_\mathrm{dif}^{}$,  
the three-body correlation function $\Theta_V$ and  $\widetilde{\Theta}_\mathrm{III}$
can experimentally be deduced, using 
Eqs.\ \eqref{THvStrongLimit} and \eqref{THvBiasSymEq}.

In the limit of $|\xi_d^{}|\to\infty$, 
the coefficients approach the noninteracting value:
 $W_V \to -1$, 
 $\Theta \to -2$, and 
 $(64/\pi^2)\,C_V^{(3)}\to -3$ 
for symmetric bias $\alpha_\mathrm{dif}^{}=0$. 

Figures  \ref{Cv3BiasSymN4N6}(g) and (h) 
compare  $C_V^{(3)}$ for several different interaction strengths:   
(g) $U/(\pi\Delta)=1/3$, $2/3$, $5/3$, $10/3$, $5.0$ 
for SU(4), and (h) $U/(\pi\Delta)=2/5$, $1.0$, $2.0$, $5.0$ for SU(6), 
choosing a large tunnel asymmetry $\gamma_\mathrm{dif}^{}=0.8$.
The dip structure due to the  Kondo state 
at the filling $\langle n_d^{} \rangle= 1$ and $N-1$ 
becomes clear as $U$ increases, as well as the plateau  
due to the half-filling Kondo state.

\begin{figure}[b]
\begin{tabular}{cc}
\begin{minipage}[r]{0.5\linewidth}
\centering
	\includegraphics[keepaspectratio,scale=0.23]{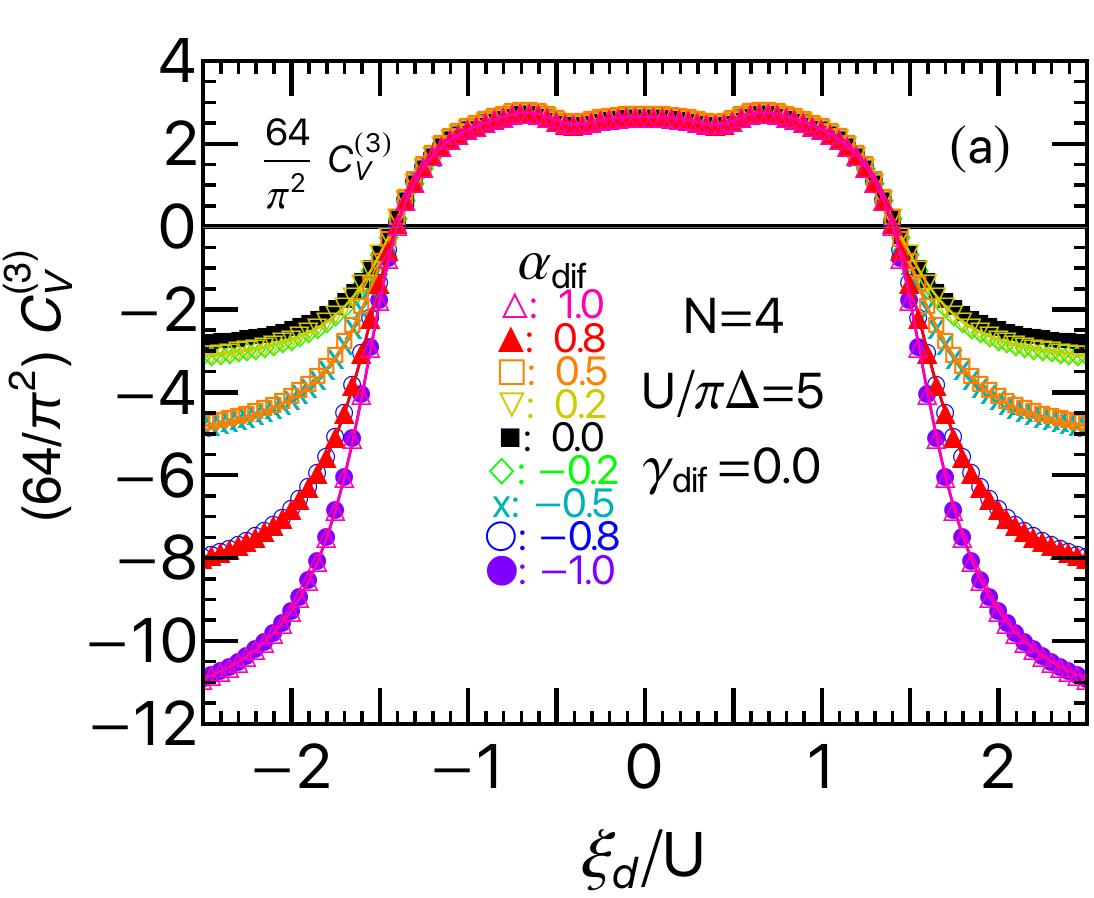}
\end{minipage}
\begin{minipage}[r]{0.5\linewidth}
\centering
	\includegraphics[keepaspectratio,scale=0.23]{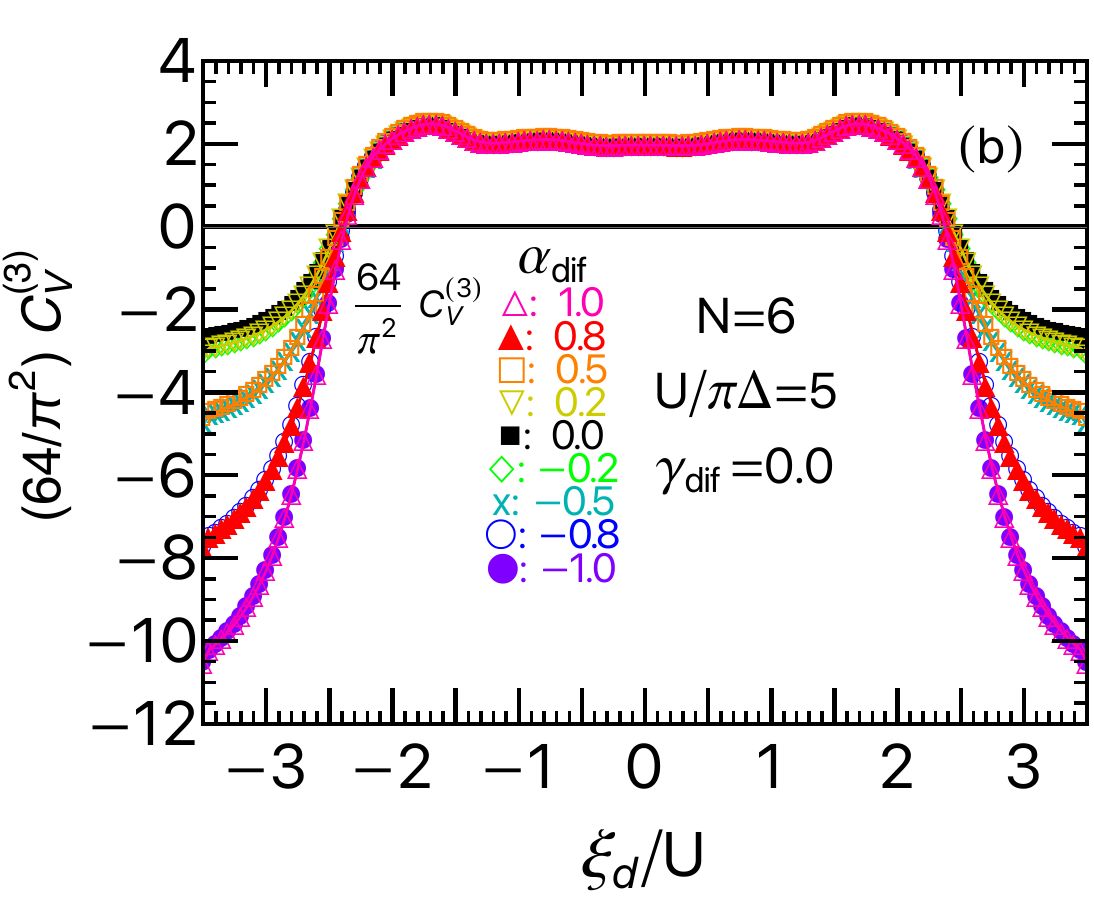}
\end{minipage}\\
\begin{minipage}[r]{0.5\linewidth}
\centering
	\includegraphics[keepaspectratio,scale=0.23]{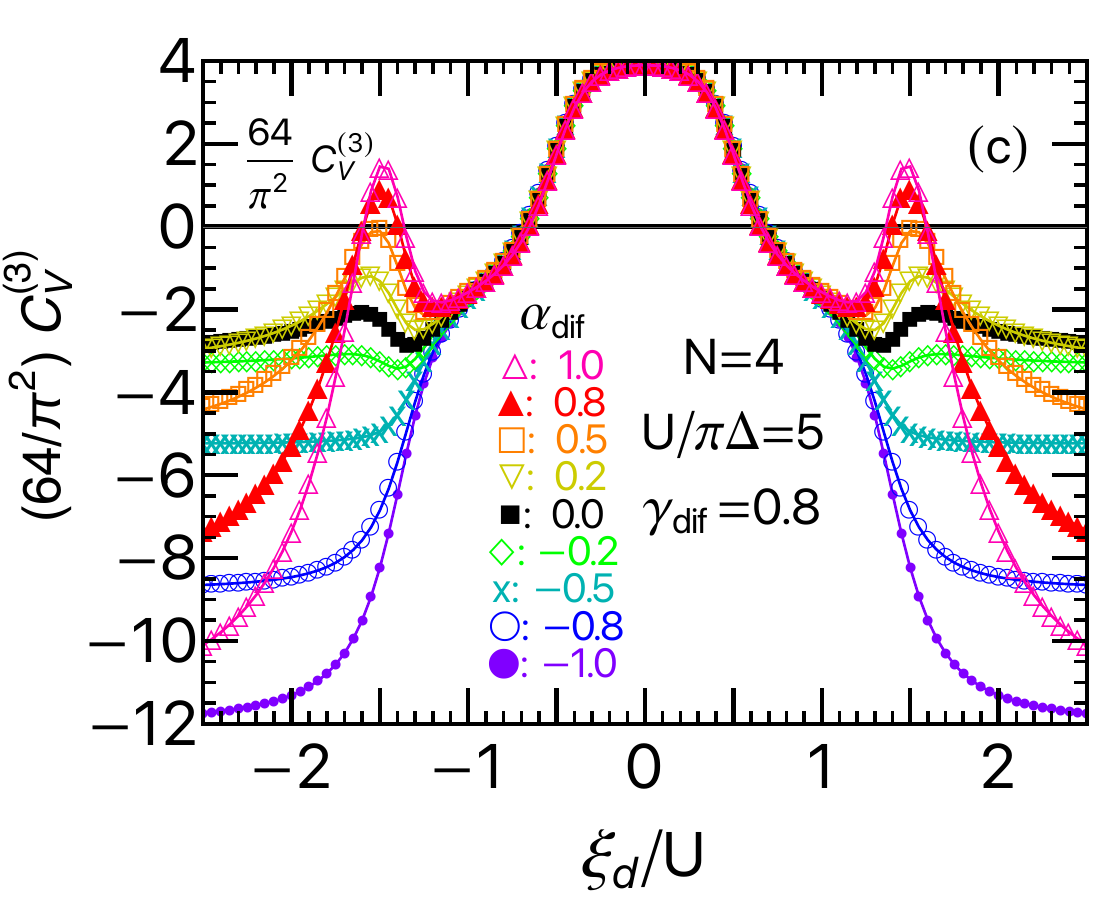}
\end{minipage}
\begin{minipage}[r]{0.5\linewidth}
\centering
	\includegraphics[keepaspectratio,scale=0.23]{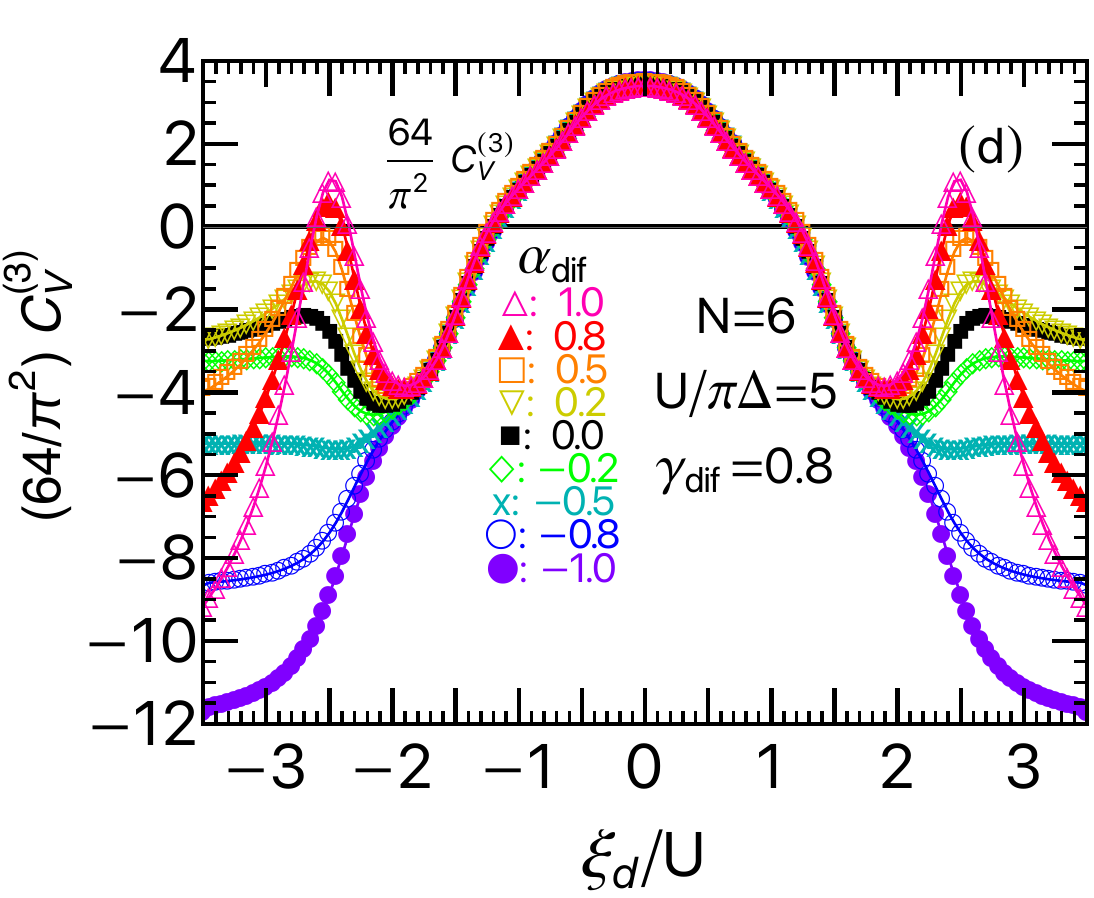}
\end{minipage}
\end{tabular}
\caption{
 Behavior of $C_V^{(3)}$ for two different tunnel couplings, 
 (top panels)  $\gamma_\mathrm{dif}^{}=0.0$ 
and  (bottom panels) $\gamma_\mathrm{dif}^{}=0.8$,
is plotted as a function of $\xi_d^{}$, varying bias asymmetries  
 $\alpha_\mathrm{dif}^{}=-1.0$ ($\bullet$), 
$-0.8$ ($\circ$), 
$-0.5$ ($\times$), $-0.2$ ($\diamond$), $0.0$ ($\blacksquare$), 
$0.2$ ($\triangledown$), $0.5$ ($\square$), $0.8$ ($\blacktriangle$) 
and $1.0$ ($\triangle$).
Interaction strength is fixed at $U/(\pi\Delta)=5.0$,
for both (left panels)  SU(4) and (right panels) SU(6) quantum dots.} 
\label{Cv3tunnelFixN4N6}
\end{figure}

\subsection{$C_V^{(3)}$ under asymmetric bias voltages $\alpha_\mathrm{dif}\neq 0$}

The coefficient $C_V^{(3)}$ depends on the bias and tunnel asymmetries 
through the terms of order $\alpha_\mathrm{dif}^{2}$, 
$\alpha_\mathrm{dif}^{}\gamma_\mathrm{dif}^{}$, 
and $\gamma_\mathrm{dif}^{2}$, 
 in Eqs.\ \eqref{Wv} and \eqref{THv}. 
We have discussed in the role of  
the $\gamma_\mathrm{dif}^{2}$ term of the tunnel asymmetry, 
setting the bias voltages to be symmetric $\alpha_\mathrm{dif}^{}=0$. 
In this section, we examine the contributions of the other two terms,  
 $\alpha_\mathrm{dif}^{2}$ and 
$\alpha_\mathrm{dif}^{}\gamma_\mathrm{dif}^{}$.

\subsubsection{
Effects of bias asymmetry at small 
 $\gamma_\mathrm{dif}^{} = 0$
and large 
 $\gamma_\mathrm{dif}^{} = 0.8$ 
tunnel asymmetries
}

In Fig.\ \ref{Cv3tunnelFixN4N6},  
the coefficient $C_V^{(3)}$ is plotted vs $\xi_d^{}$, 
for (left panels) SU(4) and  (right panels) SU(6) quantum dots,  
varying bias asymmetries 
$\alpha_\mathrm{dif}^{}= 0.0$, $\pm 0.2$, $\pm 0.5$, $\pm 0.8$ 
and $\pm 1.0$. 
Specifically, two different tunnel couplings are examined. 
One  (top panels) is the symmetric coupling $\gamma_\mathrm{dif}^{}=0.0$,  
at which the role of the  $\alpha_\mathrm{dif}^{2}$ term can be clarified.   
The other one (bottom panels) 
is a  large asymmetric coupling $\gamma_\mathrm{dif}^{}=0.8$, 
for which all the quadratic terms,  
 $\alpha_\mathrm{dif}^{2}$, 
$\alpha_\mathrm{dif}^{}\gamma_\mathrm{dif}^{}$, 
and $\gamma_\mathrm{dif}^{2}$ contribute to  $C_V^{(3)}$.

As a large interaction $U/(\pi\Delta)=5.0$ is chosen for each of the panels,  
the coefficient $C_V^{(3)}$ does not depend on bias asymmetry 
 $\alpha_\mathrm{dif}^{}$ 
in the region of $|\xi_d^{}|\lesssim(N-1)U/2$, 
and it takes the values determined by  
$\gamma_\mathrm{dif}^{2}$ with  
Eqs.\ \eqref{WvStrongLimit} and \eqref{THvStrongLimit}.  

In contrast, the coefficient  $C_V^{(3)}$ varies with 
$\alpha_\mathrm{dif}^{}\gamma_\mathrm{dif}^{}$ 
 in the valence fluctuation regime $|\xi_d|\gtrsim(N-1)U/2$.
It approaches the asymptotic value  which is given by Eq.\ \eqref{Cv3NonInt}
in the limit of $|\xi_d^{}|\to\infty$. 
Since the effect of the bias asymmetry 
in the tunnel symmetric case $\gamma_\mathrm{dif}^{}=0.0$  
is determined by the terms of  $\alpha_\mathrm{dif}^{2}$, 
the results  shown in  Figs.\ \ref{Cv3tunnelFixN4N6}(a) and (b) 
do not vary with the sign of the parameter $\alpha_\mathrm{dif}^{}$. 

In Figs.\ \ref{Cv3tunnelFixN4N6}(c) and (d), 
the coefficient $C_V^{(3)}$ has a sharp peak 
in the valence fluctuation region at $\xi_d^{} \simeq \pm  (N-1)U/2$,  
which grows as bias asymmetry $\alpha_\mathrm{dif}^{}$ increases. 
This is caused by the cross term with a positive sign 
 $\alpha_\mathrm{dif}^{}\gamma_\mathrm{dif}^{}>0$:
we will examine  the contributions of this term 
to the peak structures more precisely in the following.

\begin{figure}[b]
	\begin{minipage}[t]{0.49\linewidth}
	\centering
	\includegraphics[keepaspectratio,scale=0.23]{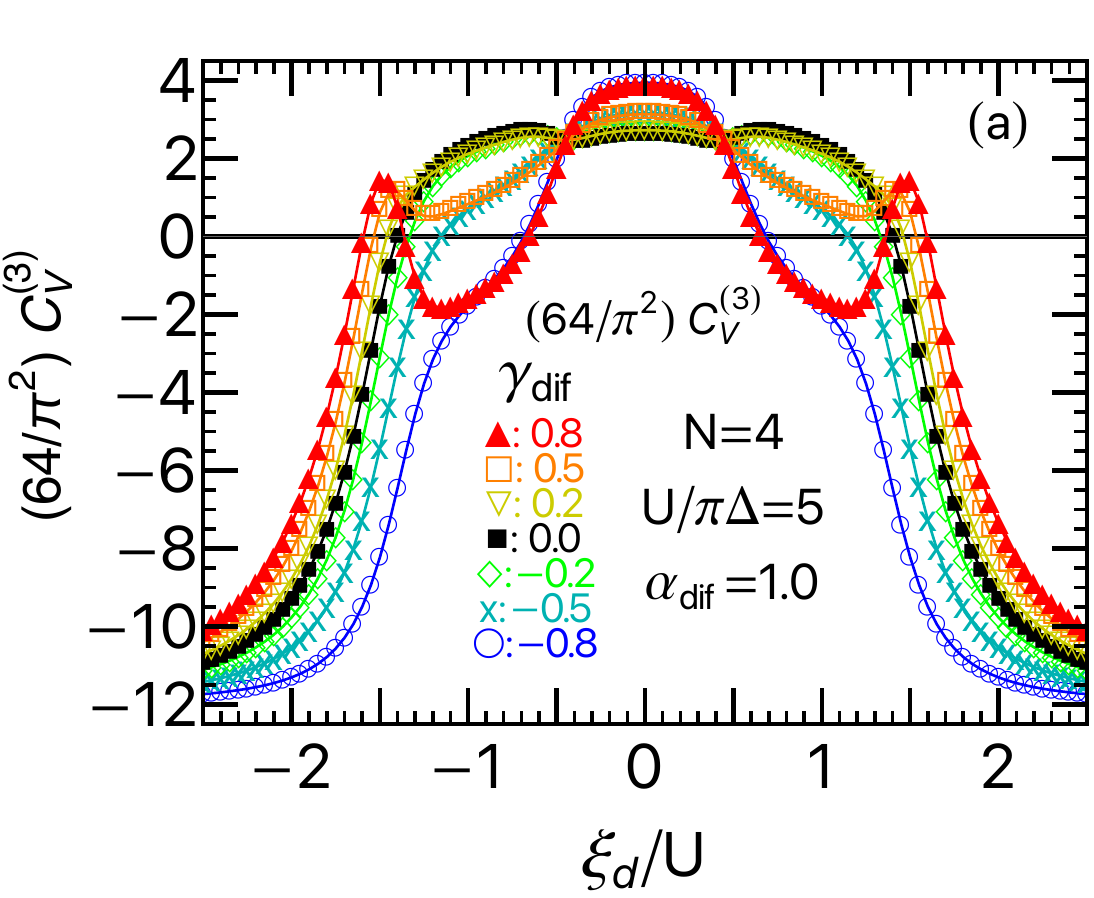}
\end{minipage}
\begin{minipage}[t]{0.49\linewidth}
	\centering
	\includegraphics[keepaspectratio,scale=0.23]{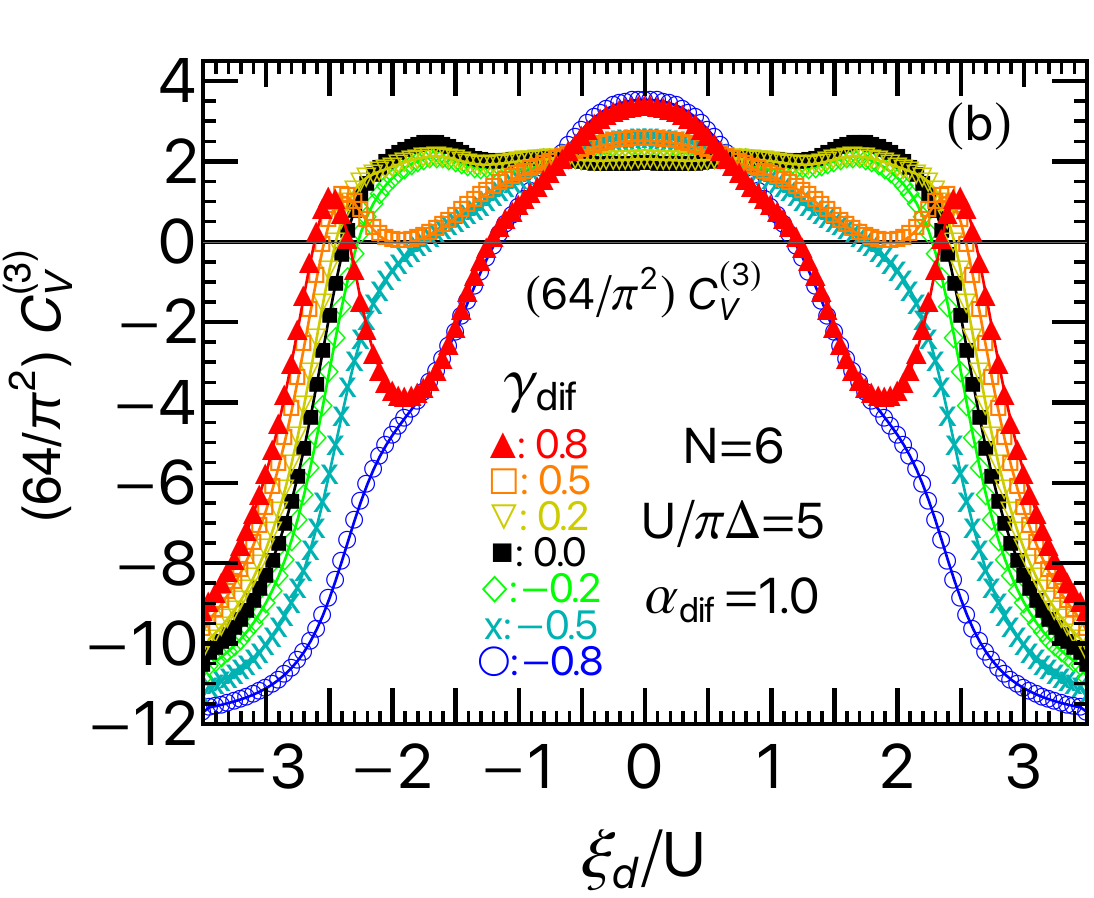}
\end{minipage}\\
\begin{minipage}[t]{0.49\linewidth}
	\centering
	\includegraphics[keepaspectratio,scale=0.23]{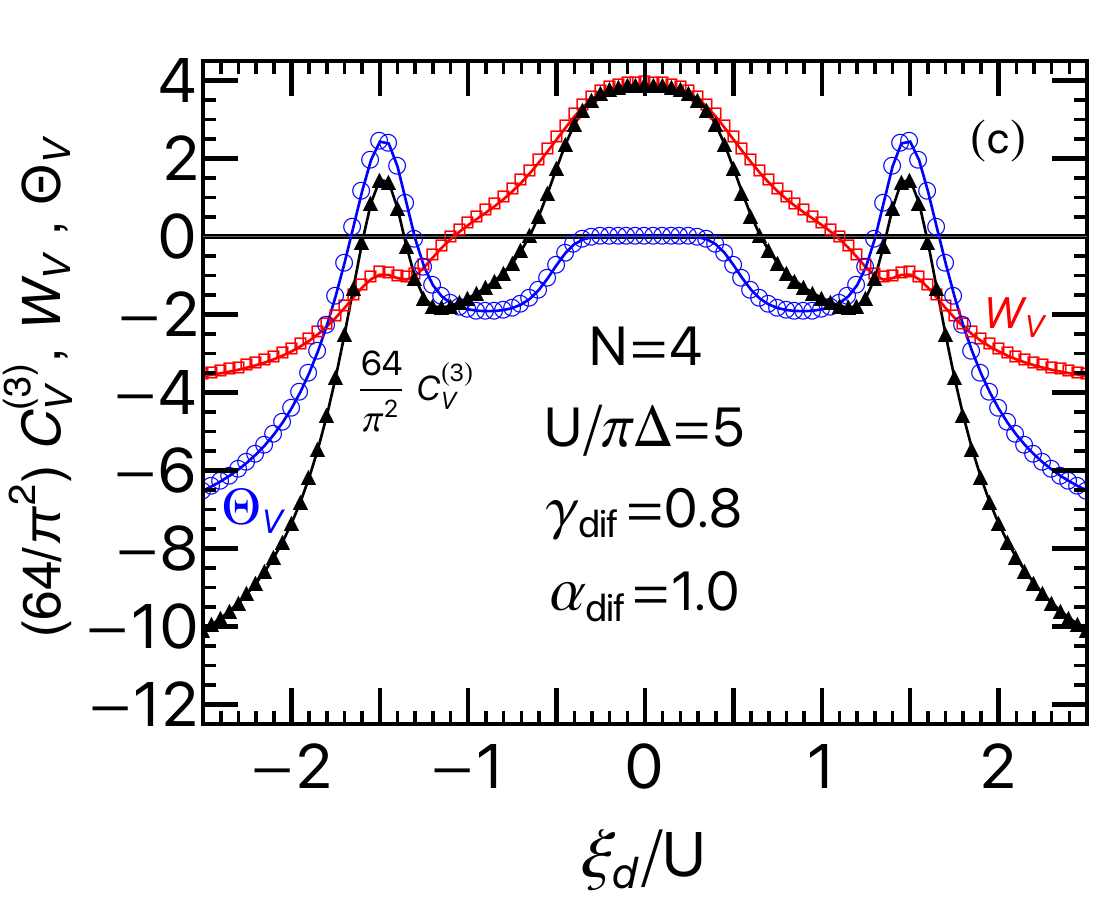}
	\end{minipage}
	\begin{minipage}[t]{0.49\linewidth}
	\centering
	\includegraphics[keepaspectratio,scale=0.23]{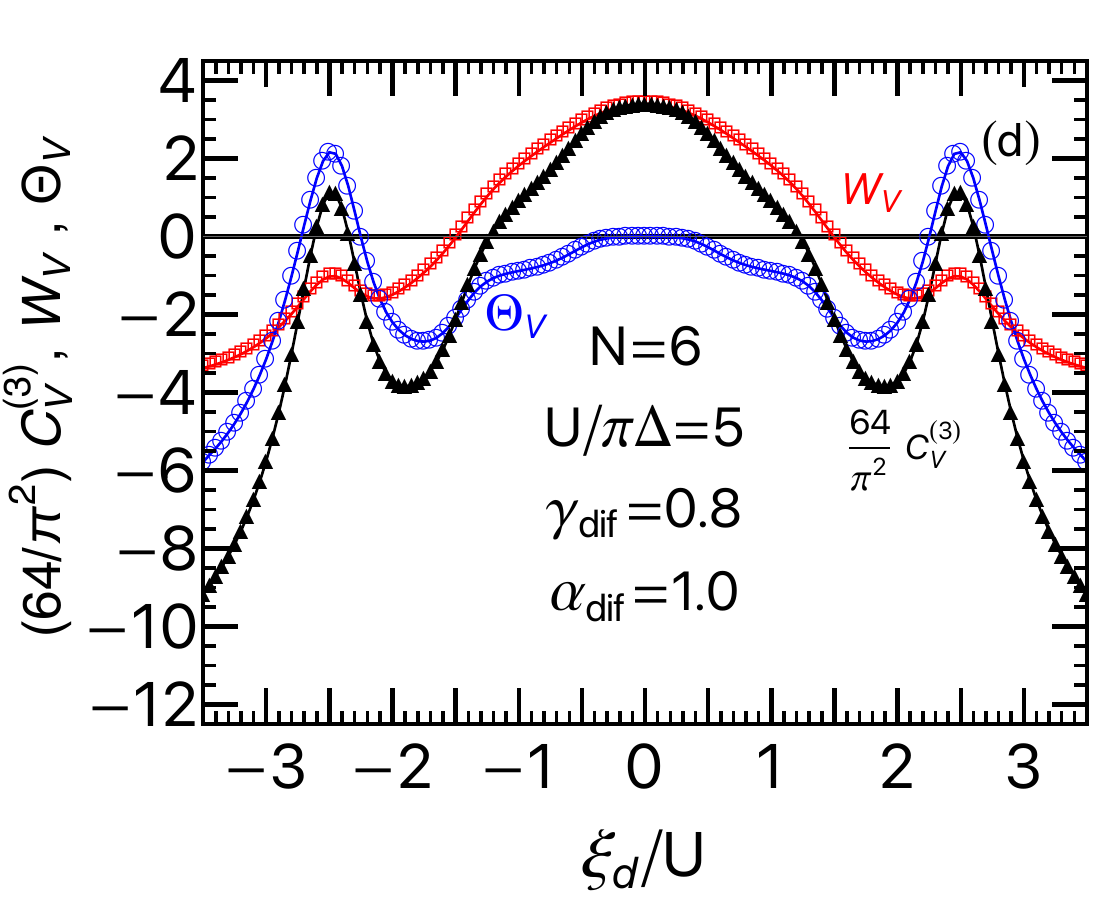}
	\end{minipage}\\
	\begin{minipage}[t]{0.49\linewidth}
	\centering
	\includegraphics[keepaspectratio,scale=0.23]{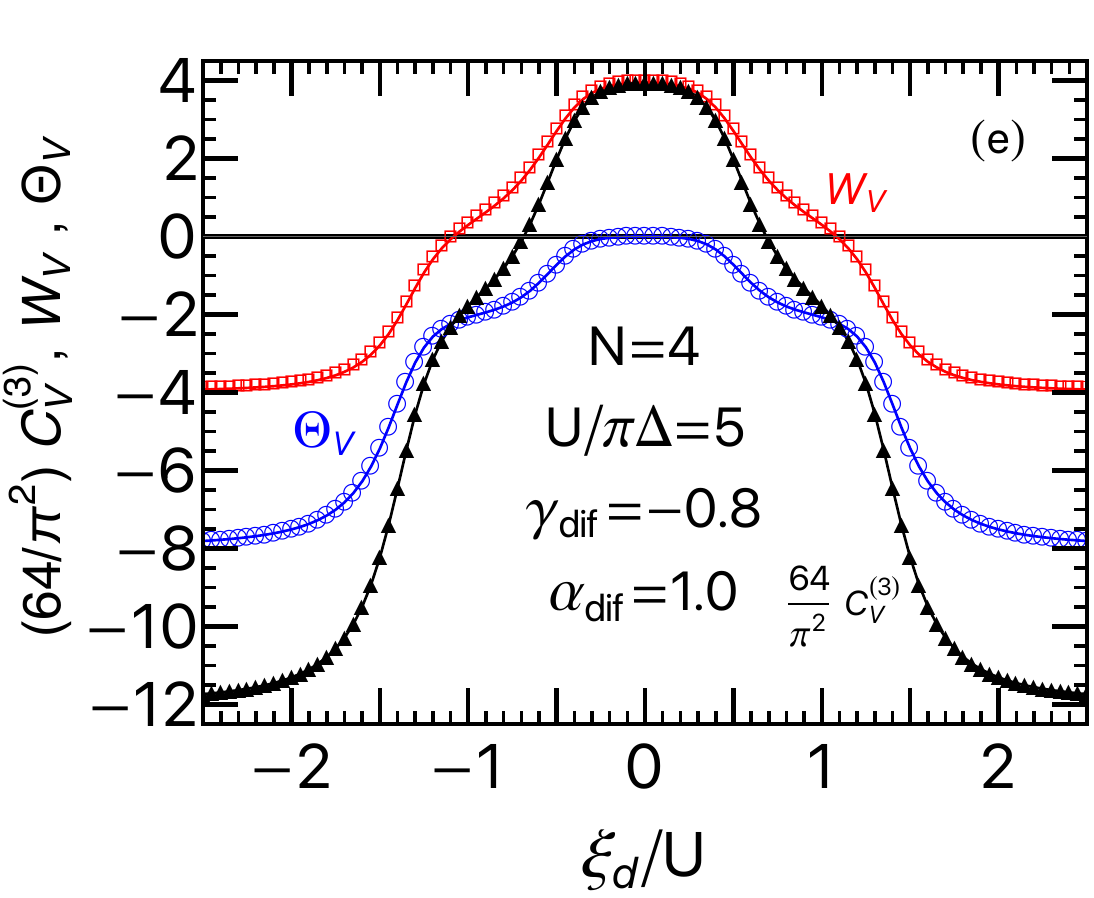}
	\end{minipage}
	\begin{minipage}[t]{0.49\linewidth}
	\centering
	\includegraphics[keepaspectratio,scale=0.23]{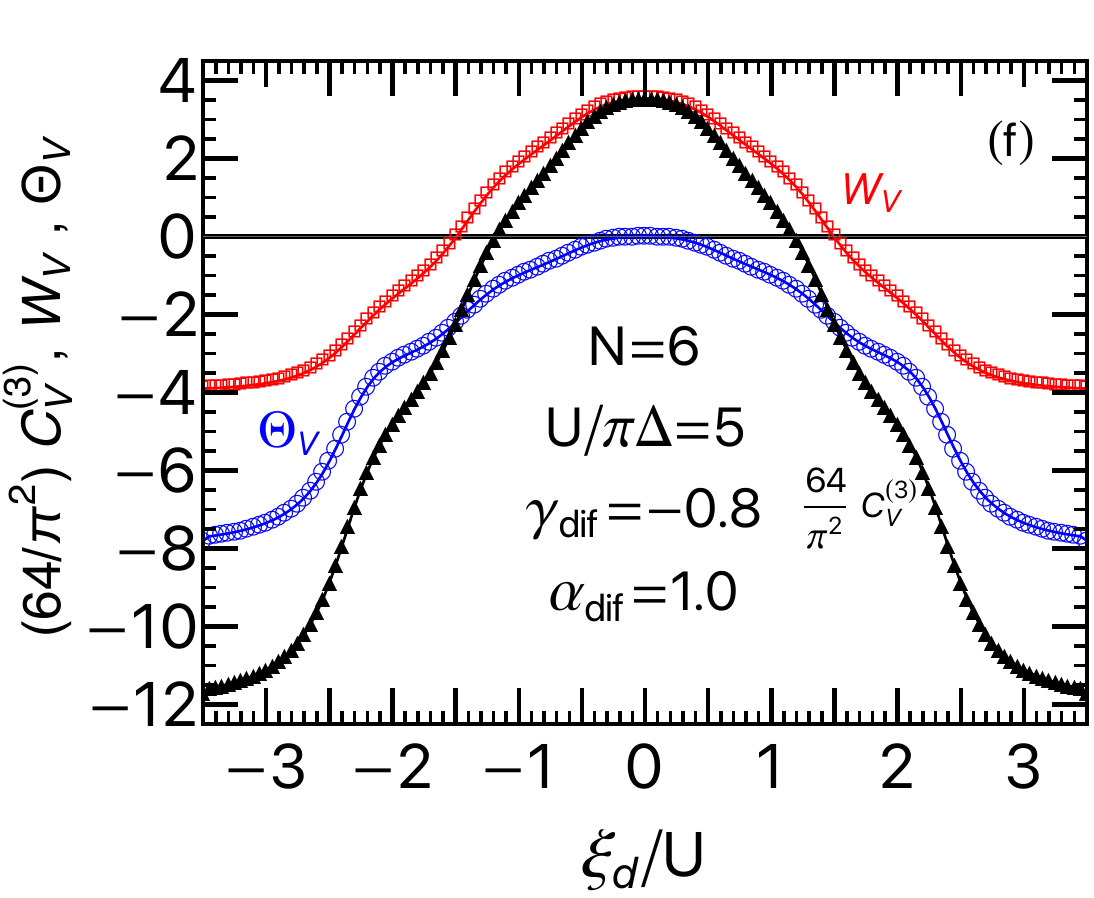}
	\end{minipage}\\
	\begin{minipage}[t]{0.49\linewidth}
	\centering
	\includegraphics[keepaspectratio,scale=0.23]{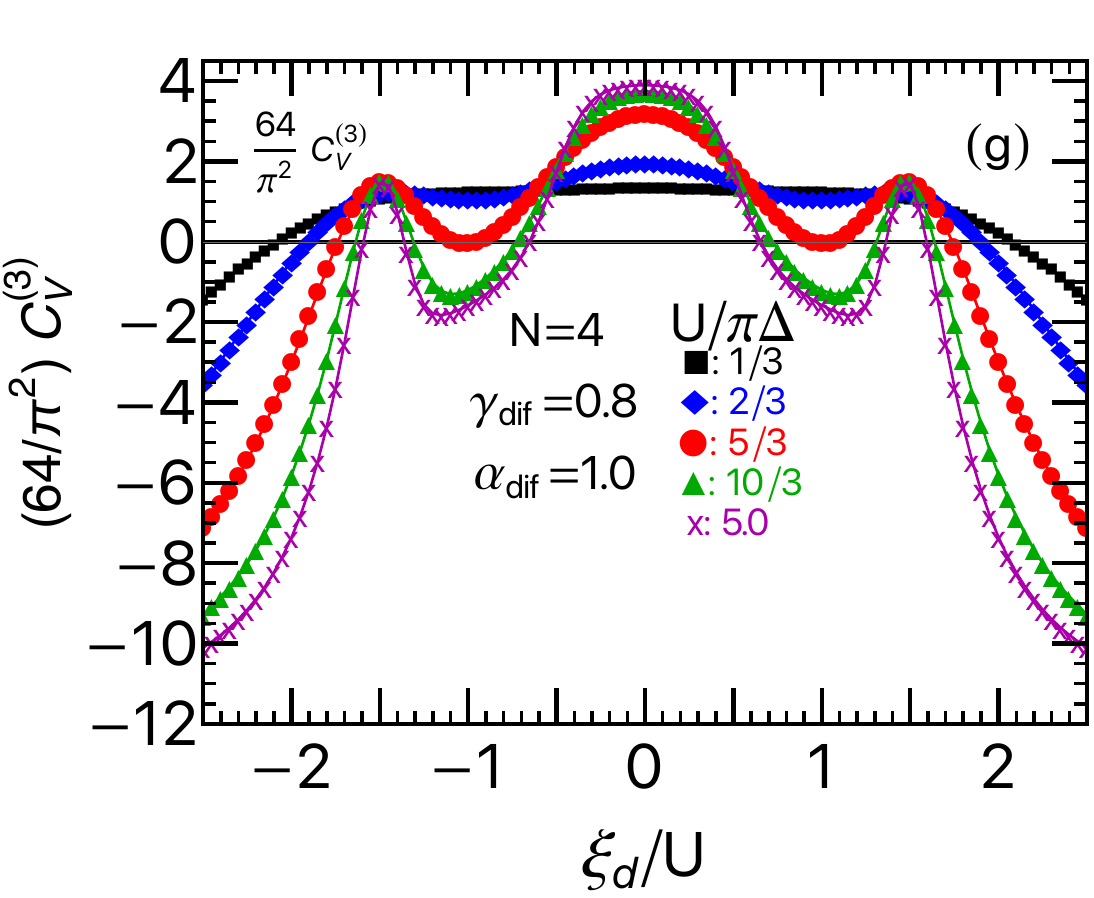}
	\end{minipage}
	\begin{minipage}[t]{0.49\linewidth}
	\centering
	\includegraphics[keepaspectratio,scale=0.23]{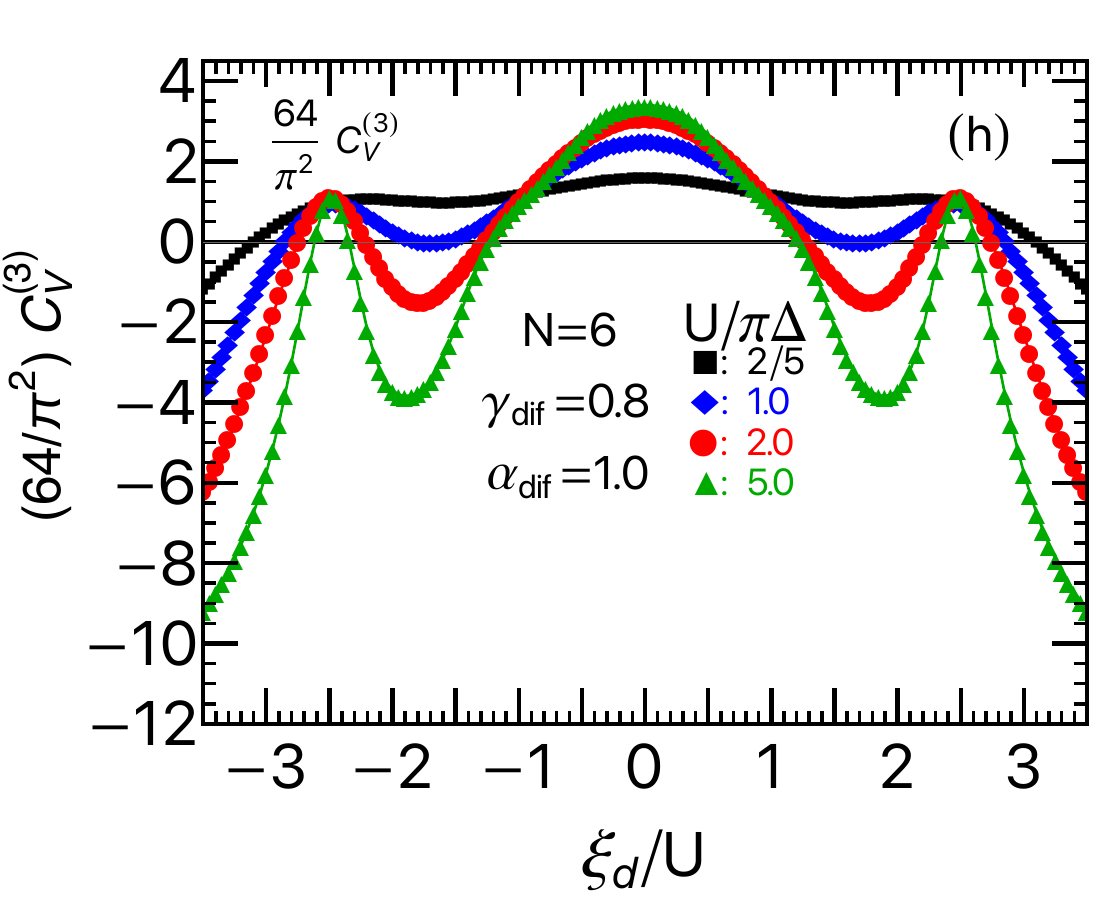}
	\end{minipage}
\caption{
Behavior of $C_V^{(3)}$ 
at large bias asymmetry $\alpha_\mathrm{dif}^{}=1$ 
is plotted vs $\xi_d^{}$ for  
 (left panels) SU(4) and (right panels) SU(6) quantum dots. 
Top panels:  tunnel asymmetry is varied, as   
 $\gamma_\mathrm{dif}^{}=-0.8$ ($\circ$), $-0.5$ ($\times$), 
 $-0.2$ ($\diamond$), $0.0$ ($\rule{2mm}{2mm}$), 
 $0.2$ ($\triangledown$), $0.5$ ($\square$), $0.8$ ($\triangle$). 
In addition,  two-body part  $W_V$ ($\square$) 
and three-body part  $\Theta_V$ ($\circ$) 
are plotted together with $C_V^{(3)}$ ($\blacktriangle$) 
 for two large opposite tunnel asymmetries:    
(upper middle panels) $\gamma_\mathrm{dif}=0.8$,  
and (lower middle panels)  $\gamma_\mathrm{dif}=-0.8$.  
Interaction strength is chosen to be  $U/(\pi\Delta)=5.0$ in (a)--(f). 
Bottom panels show the $U$ dependence of  $C_V^{(3)}$ 
for $\gamma_\mathrm{dif}=0.8$:  
 (g) $U/(\pi\Delta)=1/3$, $2/3$, $5/3$, $10/3$, $5.0$ for SU(4),  
and (h) $U/(\pi\Delta)=2/5$,\,$1.0$,\,$2.0$, $5.0$ for SU(6). 
}
\label{Cv3BiasAL1N4N6}
\end{figure}

\subsubsection{Effects of tunnel asymmetry $\gamma_\mathrm{dif}^{} \neq 0$ on 
$C_V^{(3)}$ at large bias  asymmetry  $\alpha_\mathrm{dif}^{}=1$}

The bias and tunnel asymmetries affect the 
coefficient $C_V^{(3)}$ through 
thus quadratic terms 
 $\alpha_\mathrm{dif}^{2}$, 
$\alpha_\mathrm{dif}^{}\gamma_\mathrm{dif}^{}$, 
 and $\gamma_\mathrm{dif}^{2}$  in Eqs.\ \eqref{Cv3eq}--\eqref{THv}. 
In order to clarify the contribution of the cross term, 
we set the bias asymmetry to be the upper-bound value  $\alpha_\mathrm{dif}^{}=1$,  which describes  the situation where  
 the bias asymmetry is maximized by grounding the right leads. 
Therefore, in this case $W_V$ and  $\Theta_V$ can be expressed in the following form,
\begin{align}
W_V\xrightarrow{\,\alpha_\mathrm{dif}^{}=1\,}&\,
-\cos2\delta\Biggl[4-6\bigl(1+\gamma_\mathrm{dif}^{}\bigr)\widetilde{K}  \nonumber
\\
&+\Biggl\{\frac{3N+2}{N-1}+6\,\gamma_\mathrm{dif}^{}
+\frac{3(N-2)}{N-1}\gamma_\mathrm{dif}^{2}\Biggr\}\widetilde{K}^2\Biggr]\,, 
\end{align}
\begin{align}
\Theta_V\xrightarrow{\,\alpha_\mathrm{dif}^{}=1\,}&\,\,
4\Biggl[\Theta_\mathrm{I}
+6\biggl(\frac{1+\gamma_\mathrm{dif}^{}}{2}\biggr)
\widetilde{\Theta}_\mathrm{II}+6\biggl(\frac{1+\gamma_\mathrm{dif}^{}}{2}
\biggr)^2\widetilde{\Theta}_\mathrm{III}\Biggr]\,. 
\label{THvAL1}
\end{align}
Here, the cross term $\alpha_\mathrm{dif}^{}\gamma_\mathrm{dif}$ 
appears as linear order terms with respect to $\gamma_\mathrm{dif}^{}$. 

Our discussion here is based on  Fig.\ \ref{Cv3BiasAL1N4N6}, 
in which the NRG results of $C_V^{(3)}$ for SU(4) quantum dots 
and those 
 for SU(6) quantum dots 
are presented in the left and right panels, respectively. 
Top panels of Fig.\ \ref{Cv3BiasAL1N4N6} show  
$C_V^{(3)}$ as a function of the gate voltage $\xi_d^{}$, varying
 tunneling asymmetries 
 $\gamma_\mathrm{dif}^{}=0.0$, $\pm 0.2$, $\pm 0.5$, $\pm 0.8$, 
for a strong interaction strength  $U/(\pi\Delta)=5.0$. 
The upper middle panels  show 
 $C_V^{(3)}$, $W_V$,  $\Theta_V$ for large positive tunnel asymmetries
 $\gamma_\mathrm{dif}^{}=0.8$, 
 and correspondingly  lower middle panels show 
the ones for large negative tunnel asymmetries 
 $\gamma_\mathrm{dif}^{}=-0.8$.  
Bottom panels of Fig.\ \ref{Cv3BiasAL1N4N6}
show the results, calculated for several interaction strengths 
$U/(\pi\Delta)=1/3$, $2/3$, $5/3$, $10/3$, $5.0$ for SU(4), 
and $U/(\pi\Delta)=2/5$, $1.0$, $2.0$, $5.0$ for SU(6), 
taking asymmetric tunnel coupling to be $\gamma_\mathrm{dif}^{}=0.8$.

In the strongly correlated  region $|\xi_d^{}|\lesssim(N-1)U/2$ for large $U$, 
the result of  $C_V^{(3)}$ in Figs.\ \ref{Cv3BiasAL1N4N6} (a) and (b) 
almost does not depend on whether or not $\alpha_\mathrm{dif}^{}=1$, 
and the behavior in this region 
 is determined essentially by the $\gamma_\mathrm{dif}^{2}$ term 
in Eqs.\ \eqref{WvStrongLimit} and \eqref{THvStrongLimit}.
The cross term $\alpha_\mathrm{dif}^{}\gamma_\mathrm{dif}^{}$ 
becomes important at $|\xi_d^{}|\gtrsim (N-1)U/2$, 
outside the correlated region.

 In particular,  $C_V^{(3)}$  takes a sharp peak 
near $\xi_d\simeq(N-1)U/2$ in the valence fluctuation regime 
for large positive $\gamma_\mathrm{dif}^{}$, i.e., 
$\alpha_\mathrm{dif}^{}\gamma_\mathrm{dif}^{}>0$ 
 at which bias and tunneling asymmetries 
cooperatively enhance the charge transfer from one of the electrodes.      
The results of $W_V$ and $\Theta_V$, plotted for 
$\gamma_\mathrm{dif}^{}=0.8$ in Figs.\ \ref{Cv3BiasAL1N4N6} (c)-(d), 
 show the sharp peak structure is mainly due to the three-body part $\Theta_V$ 
and two-body part has a small constructive peak at the same position.
In contrast, 
for tunnel asymmetry in the opposite direction 
 $\gamma_\mathrm{dif}^{}=-0.8$, 
each of the two components,  $W_V$ and  $\Theta_V$,  
does not have a peak in Figs.\ \ref{Cv3BiasAL1N4N6} (e)-(f). 

We consider more precisely the peak that emerged 
in the three-body part $\Theta_V$, 
taking the limit of $\gamma_\mathrm{dif}^{}\to 1$ in Eq.\ \eqref{THvAL1}    
\footnote{
This limit of strong tunnel asymmetry $\gamma_\mathrm{dif}^{} \to \pm 1$ 
is meaningful for investigating asymptotic behavior of $C_V^{(3)}$
although it represents the situation where one of the leads is disconnected 
and the current is determined by Eq.\ \eqref{eq:g0_gamma_dif}.},   
\begin{align}
\Theta_V\xrightarrow{\,\alpha_\mathrm{dif}^{}=1,\,
\gamma_\mathrm{dif}^{}\to1\,}
\,4\,\Bigl(\Theta_\mathrm{I}
+6\,\widetilde{\Theta}_\mathrm{II}+6\,\widetilde{\Theta}_\mathrm{III}\Bigr)\,. 
\label{THvAL1XL1}
\end{align}
Note that each of the three-body correlation functions 
of the SU($N$) symmetric quantum dots 
has a definite sign:  $\Theta_\mathrm{I}<0$, 
$\widetilde{\Theta}_\mathrm{II}>0$, 
 and $\widetilde{\Theta}_\mathrm{III}<0$, 
as shown in Figs.\ \ref{TH123SU4} and \ref{TH123SU6}. 
Specifically, in the valence fluctuation regime,  
the positive contribution of $6\,\widetilde{\Theta}_\mathrm{II}$ 
in Eq.\ \eqref{THvAL1XL1}  becomes greater than the negative contribution 
of $\Theta_\mathrm{I}+6\,\widetilde{\Theta}_\mathrm{III}$, 
and the difference between these components yields a sharp peak of  $\Theta_V$ 
at  $|\xi_d^{}|\gtrsim(N-1)U/2$. 
The height,
 which is measured from the 
base value of $\Theta_V$ that is 
defined at the Kondo state in the close vicinity 
of the valence fluctuation region, 
increases with $N$.
Note that in the Kondo state, 
the three-body correlation functions have a property 
$\widetilde{\Theta}_\mathrm{I} \simeq 
-\widetilde{\Theta}_\mathrm{II}\simeq 
\widetilde{\Theta}_\mathrm{III}, $
and  $\Theta_V$ in this limit   $\gamma_\mathrm{dif} \to 1$ 
 takes  a negative  value $\Theta_V\to 4\Theta_\mathrm{I}<0$.
In the opposite limit of the tunnel asymmetries  $\gamma_\mathrm{dif}^{}\to-1$, 
the cross term is negative 
$\alpha_\mathrm{dif}^{}\,\gamma_\mathrm{dif}^{}<0$. 
Therefore,  the three-body contribution 
also becomes negative $\Theta_V\to 4\Theta_\mathrm{I}<0$,  
and $C_V^{(3)}$ monotonically 
decreases at $|\xi_d^{}|\gtrsim(N-1)U/2$.

We also examine how the peak structure evolves with $U$  
in Figs.\ \ref{Cv3BiasAL1N4N6}(g) and (h). 
The results show that 
the dip structure in the SU($N$) Kondo regime 
near $|\xi_d^{}|\simeq(N-1)U/2$ 
and the sharp peak structure in the valence fluctuation 
regime $|\xi_d^{}|\gtrsim(N-1)U/2$ clearer 
as interaction $U$ increases.

\section{Conclusion}
\label{Conclusion}

We have derived the exact low-bias expansion formula of  
the differential conductance $dI/dV$ through 
a multilevel Anderson impurity up to terms of order $(eV)^2$,  
without assuming symmetries in tunnel couplings or bias voltages. 
It is applicable to a wide class of quantum dots 
with arbitrary energy level  $\epsilon_{d\sigma}^{}$. 
The expansion coefficients are expressed in terms of  
the phase shift $\delta_\sigma^{}$,
 linear susceptibilities $\chi_{\sigma\sigma'}^{}$, 
and three-body correlation functions 
$\chi_{\sigma\sigma'\sigma''}^{[3]}$, 
defined with respect to the equilibrium ground state.

The tunnel and bias asymmetries enter the transport coefficients 
through the parameters  
$\gamma_\mathrm{dif}^{}=(\Gamma_L - \Gamma_R)/(\Gamma_L+\Gamma_R)$ and
$\alpha_\mathrm{dif}^{}=(\mu_L+\mu_R-2E_F)/(\mu_L-\mu_R)$.
In contrast to the linear conductance which depends only 
on the tunnel asymmetry through 
the prefactor 
$g_0^{}=\frac{e^2}{h}\,4\Gamma_L \Gamma_R/(\Gamma_L+\Gamma_R)^2 $, 
the nonlinear terms 
$c_{V,\sigma}^{(2)}$ and $c_{V,\sigma}^{(3)}$,  
given in Eqs.\  (B.5)  and (B.6),  depend on both asymmetries. 
The  $\gamma_\mathrm{dif}^{}$-dependence  
enters additionally through the self-energy corrections due to the Coulomb interactions, 
and the $\alpha_\mathrm{dif}^{}$-dependence arises also 
through a shift of bias window. 
In particular, $c_{V,\sigma}^{(2)}$ emerges 
 when tunnel couplings and/or bias voltages are asymmetrical, 
and is given by a linear combination of $\gamma_\mathrm{dif}^{}$ 
and  $\alpha_\mathrm{dif}^{}$.

We have explored the behaviors of these coefficients of SU($N$) quantum dots 
of $N=4$ and $6$ with the NRG, in a wide range  of parameter space, 
the varying gate voltage $\xi_d^{}$ and interaction $U$ 
as well as  $\gamma_\mathrm{dif}^{}$ and  $\alpha_\mathrm{dif}^{}$.
In particular, for large $U$, transport exhibits quite different behaviors 
depending on electron fillings, especially in the two regions:
one is the strongly-correlated region $1 \lesssim \langle n_d^{}\rangle \lesssim N-1$,  
and the other is the valence fluctuation region in which 
 $0 \lesssim\langle n_d^{}\rangle \lesssim 1$ or
 $N-1 \lesssim\langle n_d^{}\rangle \lesssim N$. 
The coefficient $C_V^{(2)}$ of the first nonlinear term of $dI/dV$ 
for SU($N$) quantum dots 
   becomes almost independent of 
bias asymmetries in the strongly-correlated region 
as charge fluctuation is suppressed in this region.
Conversely,  it becomes less sensitive to tunnel asymmetries 
in the valence fluctuation region 
since interaction effects are suppressed as the filling approaches 
 $\langle n_d^{}\rangle \to 0$ or $N$.

The three-body correlation functions contribute to  
the order $(eV)^2$ nonlinear term of $dI/dV$, 
especially in the SU($N$) Kondo regime other than the half-filled one 
 occurring in electron-hole asymmetric cases,  and also in the valence fluctuation region.
The coefficient $C_V^{(3)}$ of the order $(eV)^2$ term shows 
the quadratic  $\alpha_\mathrm{dif}^{2}$, 
$\alpha_\mathrm{dif}^{}\gamma_\mathrm{dif}^{}$ 
 and $\gamma_\mathrm{dif}^{2}$ dependences   
on the bias and tunnel asymmetries.

In particular, the $\gamma_\mathrm{dif}^{2}$ term, 
which is absent in the SU(2) case,
emerges for multilevel quantum dots with $N \geq 3$, 
and it couples to 
a three-body correlation 
between electrons occupying three different local levels:  
$\chi_{\sigma\sigma'\sigma''}^{[3]}$ 
for $\sigma\neq\sigma'\neq\sigma''\neq\sigma$. 
We have found that,  as $\gamma_\mathrm{dif}^{2}$ increases, 
the structure of SU($N$) Kondo plateaus of $C_V^{(3)}$ at  
$\langle n_d^{}\rangle \simeq 1$ and $N-1$ fillings  
 vary significantly from the one 
for symmetric junctions with $\gamma_\mathrm{dif}^{}=\alpha_\mathrm{dif}^{}=0$. 
It suggests that the tunnel asymmetries could be used 
as a sensitive probe for observing three-body correlations 
in the SU($N$) Kondo states.

The cross term $\alpha_\mathrm{dif}^{}\gamma_\mathrm{dif}^{}$ 
plays an important role, especially in the valence fluctuation region. 
It yields a sharp peak of $C_V^{(3)}$  
 when  $\alpha_\mathrm{dif}^{} \gamma_\mathrm{dif}^{}>0$,
i.e., in the case at which the tunneling and bias asymmetries 
cooperatively enhance the charge transfer 
from one of the electrodes.   
This is caused by a constructive enhancement 
of the three-independent components 
of the three-body correlation function of SU($N$) quantum dots: 
$\chi_{\sigma\sigma\sigma}^{[3]}$,  
$\chi_{\sigma\sigma'\sigma'}^{[3]}$, and  
$\chi_{\sigma\sigma'\sigma''}^{[3]}$. 
Our results indicate that 
these three-independent components can separately be deduced 
if $C_V^{(3)}$ is measured varying tunneling 
asymmetries $\gamma_\mathrm{dif}^{}$ 
and bias asymmetries $\alpha_\mathrm{dif}^{}$.

\begin{acknowledgments}

This work was supported by JSPS KAKENHI
 Grant Nos.\ JP18K03495,  JP18J10205,  JP21K03415, 
and JP23K03284,  
and by JST CREST Grant No.\ JPMJCR1876.
KM was supported by JST, the establishment of university
fellowships towards the creation of science technology
innovation, Grant Number JPMJFS2138.
\end{acknowledgments}

\appendix

\section{Fermi-liquid relations}
\label{sec:FL_relations}

The retarded Green's function 
defined in Eq.\ \eqref{eG} can be expressed in the form, 
\begin{align}
G_\sigma^r(\omega)\,=& 
 \ \frac{1}{\omega-\epsilon_{d\sigma}+i\Delta-\Sigma_\sigma^r(\omega)}\,.
\label{eG_selfEG}
\end{align}
Specifically, the ground-state properties 
and the leading Fermi-liquid corrections are  determined by 
 the low-frequency behavior of  the equilibrium self-energy 
$\Sigma_{\mathrm{eq},\sigma}^r(\omega)
\equiv \left. \Sigma_\sigma^r(\omega) \right|_{T=eV=0}^{}$, 
or  the Green's function: 
\begin{align}
G_{\sigma}^{r}(\omega)\,\simeq\,
\frac{z_\sigma^{}}{\omega-\widetilde{\epsilon}_{d\sigma}^{}+i\widetilde{\Delta}_{\sigma}}\,. 
\end{align}
Here, the renormalized parameters are defined by 
\begin{align}
\widetilde{\epsilon}_{d\sigma}^{}
\,\equiv & \ z_\sigma^{}
\left[\epsilon_{d\sigma}+\Sigma_{\mathrm{eq},\sigma}^r(0)\right] 
\, = \,  
\widetilde{\Delta}_{\sigma}  \cot \delta_\sigma\,,
\nonumber 
\\
\widetilde{\Delta}_{\sigma} \,\equiv & \  z_\sigma\Delta, 
\qquad 
\frac{1}{z_\sigma^{}}\, \equiv \ 
 1-\left.\frac{\partial \Sigma_{\mathrm{eq},\sigma}^r(\omega)}
{\partial \omega}\right|_{\omega=0}^{} . 
\rule{0cm}{0.6cm}
\label{zdefeddef}
\end{align}
Furthermore, 
the renormalization factor $z_\sigma$ and 
the derivative of $\Sigma_{\mathrm{eq},\sigma}^r(0)$ 
with respect to the impurity level $\epsilon_{d\sigma'}$ 
are related to each other 
through the Ward identity \cite{Yamada1975II,Yoshimori1976}:
\begin{align}
\frac{1}{z_\sigma^{}}
\, = \, 
\widetilde{\chi}_{\sigma\sigma}
\,, 
\qquad \quad 
\widetilde{\chi}_{\sigma\sigma'} 
\,\equiv\,
\delta_{\sigma\sigma'}+\frac{\partial \Sigma_{\mathrm{eq},\sigma}^r(0)}
{\partial \epsilon_{d\sigma'}} \,. 
\end{align}
Note that $\widetilde{\chi}_{\sigma\sigma'}$ corresponds to 
an enhancement factor for the linear susceptibilities 
defined in Eq.\ \eqref{eq:linear_susceptibility}, i.e., 
$\chi_{\sigma\sigma'} =
 - \partial \bigl\langle n_{d\sigma}\bigr\rangle/\partial\epsilon_{d\sigma'}
= \rho_{d\sigma}^{} \widetilde{\chi}_{\sigma\sigma'}$ 
 at $T=0$.  

Recently, the Ward identity
for the  order $\omega^2$ real part  of  the self-energy has also been obtained, as \cite{FMvDM2018,AO2017_I,AO2017_II,AO2017_III}
\begin{align}
\left.
\frac{\partial^2}{\partial \omega^2}
\mathrm{Re}\,\Sigma_{\mathrm{eq},\sigma}^{r}(\omega)
\right|_{\omega \to 0}^{}
\,=\, \frac{\partial^2 \Sigma_{\mathrm{eq},\sigma}^{r}(0)}
{\partial \epsilon_{d\sigma}^{2}}\,
\ = \
 \frac{\partial \widetilde{\chi}_{\sigma\sigma}}{\partial \epsilon_{d\sigma}^{}} \,
\,.
\label{eq:self_w2}
\end{align}
This identity shows that the $\omega^2$ real part  is determined by
the intra-level component of the three-body correlation function
 $\chi_{\sigma\sigma\sigma}^{[3]}$.
Physically,  this term of the self-energy induces
 higher-order energy shifts for single-quasiparticle excitations.

\section{Derivation of  $c_{V,\sigma}^{(2)}$ and  $c_{V,\sigma}^{(3)}$ }
\label{sec:asymptotic_form_of_A}

We describe here the derivation of  the coefficients 
 $c_{V,\sigma}^{(2)}$ and  $c_{V,\sigma}^{(3)}$  
which appeared in  Eq.\ \eqref{dif_cond_first}. 
 The low-energy asymptotic form of  
the retarded self-energy $\Sigma_\sigma^r(\omega)$ 
for multi-orbital Anderson impurity model 
was derived up to terms of order $\omega^2$, $T^2$, and $(eV)^2$ 
in previous work \cite{AO2017_III,Oguri2022}, 
\begin{align}
& 
\mathrm{Im} \, \Sigma_\sigma^r (\omega)\,= \,  
-\frac{\pi}{2} 
\frac{1}{\rho_{d\sigma}}\sum_{\sigma'(\neq \sigma)} 
\chi_{\sigma \sigma'}^2 
\nonumber
\\
& \qquad 
\times \Biggl[ (\omega-\alpha eV)^2 
+  \frac{3\Gamma_L\Gamma_R}{(\Gamma_L+\Gamma_R)^2} (eV)^2 
 +  (\pi T)^2 \Biggr]+\cdots   \! ,  
\label{ImSelf}
\end{align}
\begin{align}
&\mathrm{Re} \, \Sigma_\sigma^r(\omega) \, =  \    
\Sigma_{\mathrm{eq},\sigma}^r(0)
-\sum_{\sigma'(\neq \sigma)} \widetilde{\chi}_{\sigma \sigma'}\,\alpha \, eV 
+ (1-\widetilde{\chi}_{\sigma \sigma})\, \omega
\nonumber
\\
& \qquad \ 
+\frac{1}{6}\frac{1}{\rho_{d\sigma}}\sum_{\sigma'(\neq \sigma)}
\frac{\partial \chi_{\sigma \sigma'}}{\partial \epsilon_{d\sigma'}}
\Biggl[  \frac{3\Gamma_L\Gamma_R}{(\Gamma_L+\Gamma_R)^2} (eV)^2
+(\pi T)^2 \Biggr] 
\nonumber
\\ 
& \qquad \ 
+\frac{1}{2}
\frac{\partial \widetilde{\chi}_{\sigma \sigma}}{\partial \epsilon_{d\sigma}}
\,\omega^2 
+\sum_{\sigma'(\neq \sigma)}
\frac{\partial \widetilde{\chi}_{\sigma \sigma'}}{\partial \epsilon_{d\sigma}} 
\,\alpha \, eV  \omega \,
\nonumber
\\
& \qquad \ 
+\frac{1}{2}\sum_{\sigma'(\neq \sigma)}\sum_{\sigma''(\neq \sigma)}
\frac{\partial \widetilde{\chi}_{\sigma \sigma'}}{\partial \epsilon_{d\sigma''}}
 \,\alpha^2(eV)^2
+\cdots\,.  
\label{ReSelf}
\end{align}
Here, $\alpha$ is 
a parameter defined in a way such that  $\alpha\, eV 
\equiv  (\Gamma_L \,\mu_L+\Gamma_R\, \mu_R)/ (\Gamma_L+\Gamma_R) $, 
and thus 
\begin{align}
\alpha\,\equiv  & \ 
\frac{\alpha_L\Gamma_L-\alpha_R\Gamma_R}{\Gamma_L+\Gamma_R}
\,=\, \frac{1}{2}\Bigl(\alpha_\mathrm{dif}^{}+\gamma_\mathrm{dif}^{}\Bigr)\,. 
\end{align}

The spectral function can be deduced exactly  
up to terms of order $\omega^2$, $T^2$, and $(eV)^2$ 
from the above results of $\Sigma_\sigma^r(\omega)$ 
using Eq.\ \eqref{eG_selfEG}: 
\begin{widetext}

\begin{align}
\pi\Delta A_\sigma(\omega) \,= & \  
\sin^2\delta_\sigma+\pi\sin2\delta_\sigma\,
\biggl[\, \chi_{\sigma\sigma}\,\omega
+\sum_{\sigma' (\neq\sigma)}\chi_{\sigma\sigma'}
\  \frac{1}{2}\bigl(\alpha_{\mathrm{dif}}+\gamma_\mathrm{dif}\bigr) \,eV\,\biggr] 
\nonumber
\\
&+\pi^2\biggl[\,\cos2\delta_\sigma
\biggl(\chi_{\sigma\sigma}^2
+\frac{1}{2}\sum_{\sigma'(\neq\sigma)}\chi_{\sigma\sigma'}^2\biggr)
-\frac{\sin2\delta_\sigma}{2\pi}\chi_{\sigma\sigma\sigma}^{[3]}\,\biggr]
\,\omega^2 
\nonumber
\\
&+\pi^2\biggl[\,\cos2\delta_\sigma\sum_{\sigma'(\neq\sigma)}
\biggl(\chi_{\sigma\sigma}\chi_{\sigma\sigma'}
-\frac{1}{2}\chi_{\sigma\sigma'}^2\biggr)
-\frac{\sin2\delta_\sigma}{2\pi}\sum_{\sigma'(\neq\sigma)}
\chi_{\sigma\sigma\sigma'}^{[3]}\,\biggr]
\bigl(\alpha_{\mathrm{dif}}+\gamma_\mathrm{dif}\bigr)\,\omega\,eV
\nonumber
\\
&+\frac{\pi^2}{3}\sum_{\sigma'(\neq\sigma)}
\biggl[\, \frac{3}{2}\cos2\delta_\sigma\chi_{\sigma\sigma'}^2
-\frac{\sin2\delta_\sigma}{2\pi}\chi_{\sigma\sigma'\sigma'}^{[3]}\,\biggr]
\biggl[\, \frac{3}{4}
\biggl(1+2\alpha_\mathrm{dif}\gamma_\mathrm{dif}+\alpha_{\mathrm{dif}}^2
\biggr)(eV)^2+(\pi T)^2\,\biggr]
\nonumber
\\ 
&+\frac{\pi^2}{3}\sum_{\sigma'(\neq\sigma)}
\sum_{\sigma''(\neq\sigma,\sigma')}
\biggl[\,\cos2\delta_\sigma\chi_{\sigma\sigma'}\chi_{\sigma\sigma''}
-\frac{\sin2\delta_\sigma}{2\pi}\chi_{\sigma\sigma'\sigma''}^{[3]}\,\biggr]
\ \frac{3}{4}
\biggl(\gamma_\mathrm{dif}^2\!+2\alpha_\mathrm{dif}\gamma_\mathrm{dif}+\alpha_{\mathrm{dif}}^2\biggr)(eV)^2 
\ + \ \cdots \,.
\label{spectral_expansion}
\end{align}
\end{widetext}
Substituting this low-energy asymptotic form 
 into the  the Landauer-type formula in Eq.\ \eqref{eq:MWFormula}, we obtain the exact expression for the coefficients  $c_{V,\sigma}^{(2)}$ and  $c_{V,\sigma}^{(3)}$ for the differential conductance at $T=0$ defined in Eq.\ \eqref{dif_cond_first}:  
\begin{widetext}
\begin{align}
 c_{V,\sigma}^{(2)}\,= & \ 
\pi \sin 2\delta_\sigma
\left[ \, \alpha_\mathrm{dif}\,\chi_{\sigma\sigma}\,
+\,\bigl(\alpha_\mathrm{dif}+\gamma_\mathrm{dif}^{}\bigr)
\sum_{\sigma'(\neq\sigma)}
\chi_{\sigma\sigma'}\right]\,,
\label{cv2_general} \\ 
  c_{V,\sigma}^{(3)}\,=& \ \frac{\pi^2}{4}
\Biggl[\, 
-\cos2\delta_\sigma\Biggl\{ \bigl(1+3\alpha_\mathrm{dif}^2\bigr)\,
\chi_{\sigma\sigma}^2 
\,+\, \left(5-3\gamma_\mathrm{dif}^{2} \right)
\sum_{\sigma'(\neq\sigma)}\chi_{\sigma\sigma'}^2 
\,+\,6\alpha_\mathrm{dif}\,
\bigl(\alpha_\mathrm{dif}+\gamma_\mathrm{dif}^{}\bigr)
\,  \chi_{\sigma\sigma} \! 
\sum_{\sigma'(\neq\sigma)}\chi_{\sigma\sigma'} 
\nonumber
\\
& \qquad \qquad \qquad \quad
 +3\bigl(\alpha_\mathrm{dif}+\gamma_\mathrm{dif}^{}\bigr)^2
\sum_{\sigma'(\neq\sigma)}\sum_{\sigma''(\neq\sigma)}
\chi_{\sigma\sigma'}\chi_{\sigma\sigma''}\Biggr\} 
\nonumber
\\ 
& \qquad
+\,\frac{\sin2\delta_\sigma}{2\pi}
\Biggl\{ \bigl(1+3\alpha_\mathrm{dif}^2\bigr)\,\chi_{\sigma\sigma\sigma}^{[3]}
\,+\,6\alpha_\mathrm{dif}\,
\left(\alpha_\mathrm{dif}+\gamma_\mathrm{dif}^{}\right)\,
\sum_{\sigma'(\neq\sigma)}\chi_{\sigma\sigma\,\sigma'}^{[3]} 
\,+\,3\left(1-\gamma_\mathrm{dif}^{2}\right)
\sum_{\sigma'(\neq\sigma)}\chi_{\sigma\sigma'\sigma'}^{[3]}
\nonumber
\\ 
& \qquad \qquad \qquad \quad \ 
+3\left(\alpha_\mathrm{dif}+\gamma_\mathrm{dif}^{}\right)^2
\sum_{\sigma'(\neq\sigma)}\sum_{\sigma''(\neq\sigma)}
\chi_{\sigma\sigma'\sigma''}^{[3]}
\Biggr\}\, 
\Biggr]\,.
\label{cv3_general}
\end{align}
\end{widetext}
Note that $c_{V,\sigma}^{(2)}$ is proportional to the derivative of the density of state as  $\sin2\delta_\sigma^{} =  \Delta \rho_{d\sigma}'  /\chi_{\sigma\sigma}$ 
from Eq.\ \eqref{DerivativeDOS}. 
The coefficients $c_{V,\sigma}^{(2)}$ and $c_{V,\sigma}^{(3)}$ 
have odd and even inversion-symmetrical properties, respectively.
Namely, if  the left and right leads are inverted 
together with  their tunnel couplings and  chemical potentials, 
sign of $c_{V,\sigma}^{(2)}$ changes while $c_{V,\sigma}^{(3)}$ does not: 
\begin{align}
&c_{V,\sigma}^{(2)}(\alpha_\mathrm{dif}^{},\,\gamma_\mathrm{dif}^{}) 
\,=\,
-c_{V,\sigma}^{(2)}(-\alpha_\mathrm{dif}^{},\,-\gamma_\mathrm{dif}^{})\,, 
\label{AntiSymCv2}
\\
&c_{V,\sigma}^{(3)}(\alpha_\mathrm{dif}^{},\,\gamma_\mathrm{dif}^{}) 
\,=\, 
c_{V,\sigma}^{(3)}(-\alpha_\mathrm{dif}^{},\,-\gamma_\mathrm{dif}^{})\,.
\label{AntiSymCv3}
\end{align}

\section{Behavior of three-body correlation functions for large $U$}
\label{ThreeBodySame}

\begin{figure}[b]
	\rule{1mm}{0mm}
	\begin{minipage}[r]{\linewidth}
	\centering
	\includegraphics[keepaspectratio,scale=0.22]{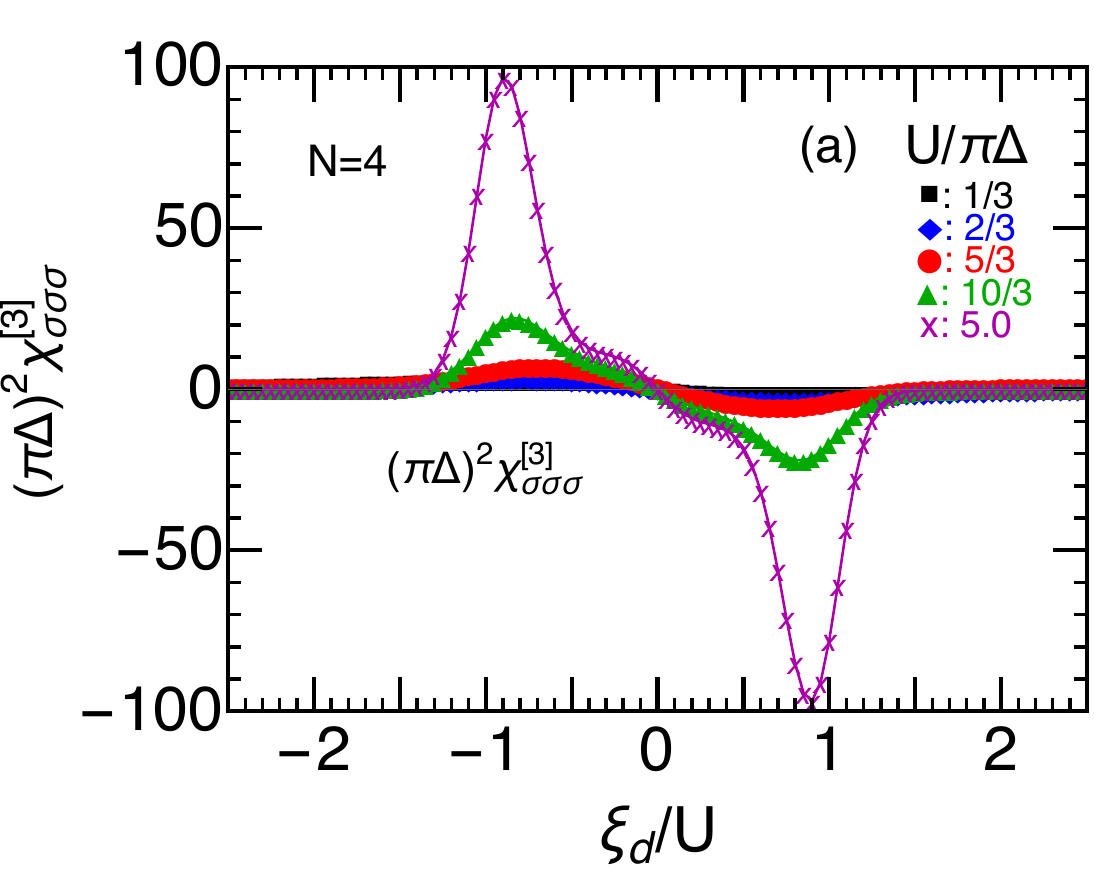}
	\centering
	\includegraphics[keepaspectratio,scale=0.22]{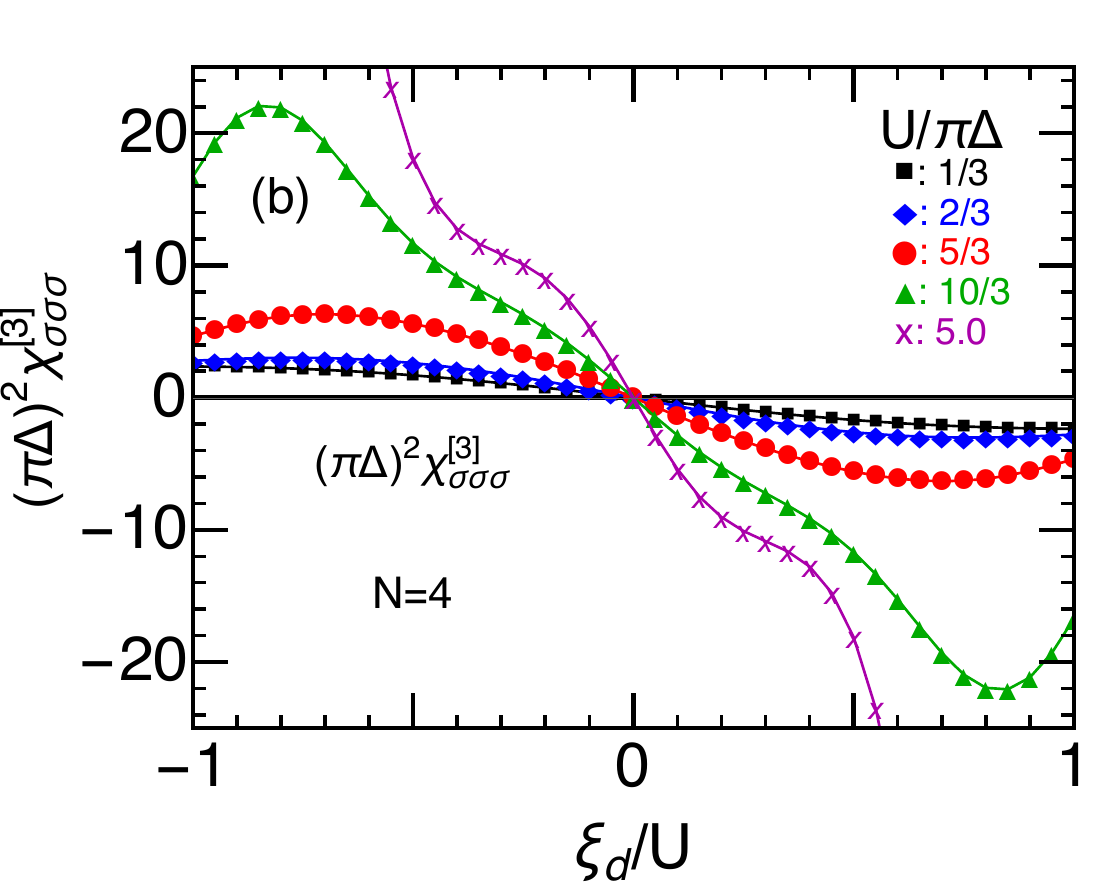}
	\centering
	\includegraphics[keepaspectratio,scale=0.22]{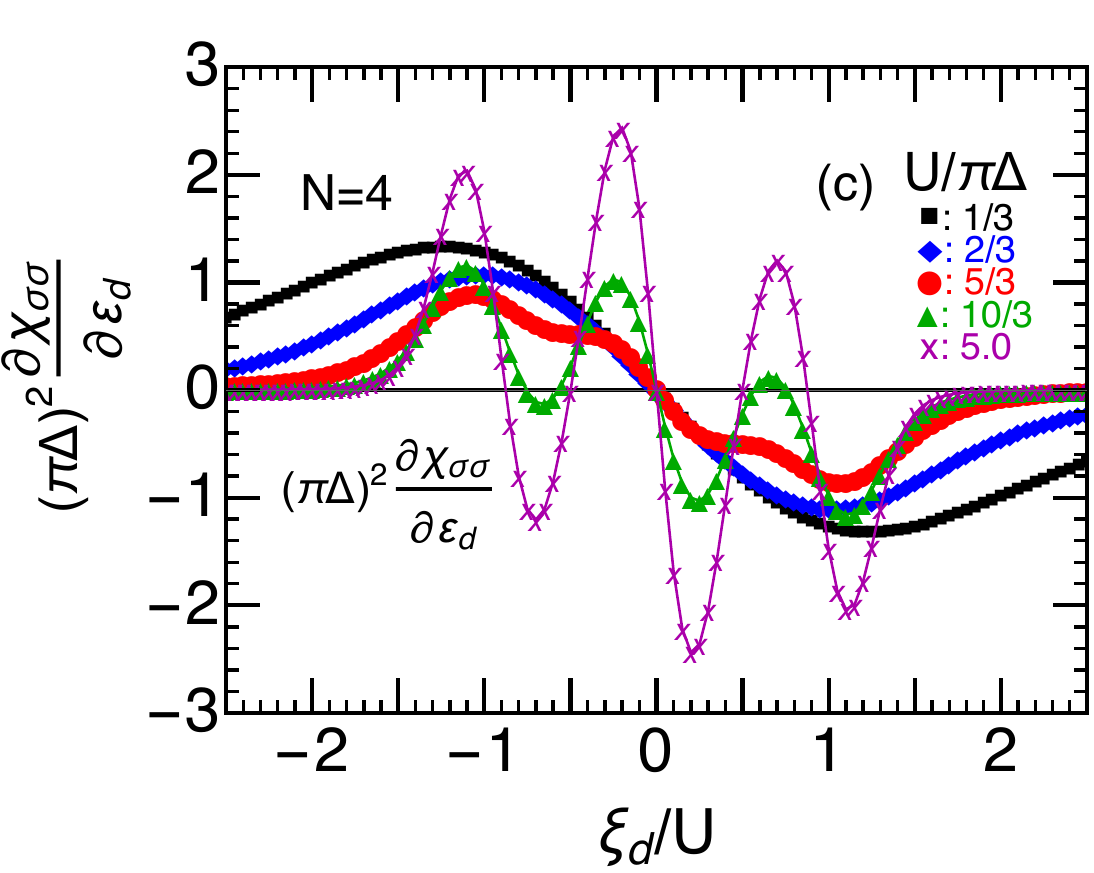}
	\centering
	\includegraphics[keepaspectratio,scale=0.22]{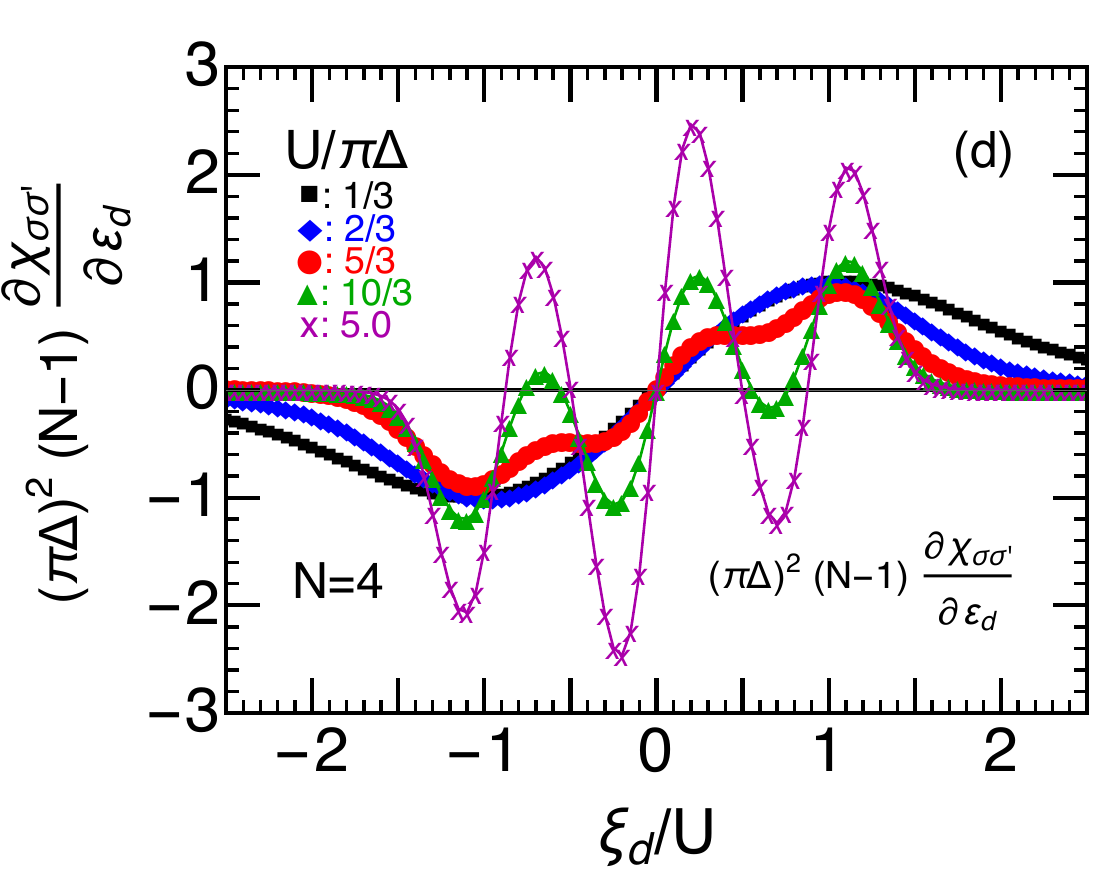}
	\end{minipage}
\caption{
Correlation functions emerging on the right-hand side 
of Eqs.\ \eqref{THIIDefEq} and \eqref{THIIIDefEq} 
are plotted vs $\xi_d/U$,  for $N=4$.  
(a)-(b): the diagonal component $(\pi\Delta)^2\chi_{\sigma\sigma\sigma}^{[3]}$.   
(c):  $(\pi\Delta)^2\frac{\partial \chi_{\sigma\sigma}}{\partial \epsilon_d}$.
(d):  $(\pi\Delta)^2\, (N-1)\, 
\frac{\partial \chi_{\sigma\sigma'}}{\partial \epsilon_d}$. 
(b) represents an enlarged view of (a). 
In each of these panels, interaction strength is chosen to be 
 $U/(\pi\Delta)=1/3$, $2/3$, $5/3$, $10/3$, $5.0$. 
}
  \label{Chi_TH}
\end{figure}

In the SU($N$) symmetric case, 
the two different off-diagonal components 
 $\chi_{\sigma\sigma'\sigma'}^{[3]}$ 
 and $\chi_{\sigma\sigma'\sigma''}^{[3]}$
of the three-body correlation functions 
for $\sigma\neq\sigma'\neq\sigma''\neq\sigma$  
can be expressed 
as a linear combination of the diagonal one 
$\chi_{\sigma\sigma\sigma}^{[3]}$ 
 and the derivative of the linear susceptibilities \cite{Teratani2020PRL}:   
\begin{align}
(N-1)\chi_{\sigma\sigma'\sigma'}^{[3]}&=-\chi_{\sigma\sigma\sigma}^{[3]}
+\frac{\partial \chi_{\sigma\sigma}}{\partial \epsilon_d}\,, 
\label{THIIDefEq}
\\
\frac{(N-1)(N-2)}{2}\chi_{\sigma\sigma'\sigma''}^{[3]}& 
=\chi_{\sigma\sigma\sigma}^{[3]}
-\frac{\partial \chi_{\sigma\sigma}}{\partial \epsilon_d}
+\frac{N-1}{2}\frac{\partial \chi_{\sigma\sigma'}}{\partial \epsilon_d}\,. 
\label{THIIIDefEq}
\end{align}
Figure \ref{Chi_TH} compares the components emerging 
on the right-hand side for $N=4$, i.e.,  
$\chi_{\sigma\sigma\sigma}^{[3]}$, 
 $\frac{\partial \chi_{\sigma\sigma}}{\partial \epsilon_d}$, and  
$(N-1)\frac{\partial \chi_{\sigma\sigma'}}{\partial \epsilon_d}$,
varying interaction strength $U$. 
We can see that 
the diagonal component $|\chi_{\sigma\sigma\sigma}^{[3]}|$ 
becomes much larger than the derivative terms 
$|\frac{\partial \chi_{\sigma\sigma}}{\partial \epsilon_d}|$ 
and $(N-1)\,|\frac{\partial \chi_{\sigma\sigma'}}{\partial \epsilon_d}|$, 
for strong interactions $U/(\pi\Delta)\gtrsim 2.0$, 
over a wide parameter range  $-(N-1)U/2\lesssim\xi_d^{}\lesssim (N-1)U/2$.
Therefore, $\chi_{\sigma\sigma\sigma}^{[3]}$ 
dominates on the right hand side of 
Eq.\ \eqref{THIIDefEq} and also on  Eq.\ \eqref{THIIIDefEq},
and thus the corresponding dimensionless parameters  show 
the property  
$\Theta_\mathrm{I}\simeq-\widetilde{\Theta}_\mathrm{II}
\simeq\widetilde{\Theta}_\mathrm{III}$ 
described in  Eq.\ \eqref{TH1simTH2simTH3} 
for the strong interaction region. 
We have confirmed that the same behavior occurs in the SU(6) symmetric case.

\end{document}